\documentstyle[preprint, aps, epsf, floats]{revtex}
\renewcommand{\baselinestretch}{1.5}

\newcommand{\mysection}[1]{\section{#1}\setcounter{equation}{0}}
\def\lb{\mbox{$\bar{\lambda}$}}

\def\cm{\mbox{${\cal M}$}}

\def\cz{\mbox{${\cal Z}$}}
\def\gb{\bar{g}}
\def\mn{{\mu\nu}}

\def\M{M}

\def\y{y}

\newcommand{\be}{\begin{equation}}
\newcommand{\ee}{\end{equation}}
\newcommand{\bea}{\begin{eqnarray}}
\newcommand{\eea}{\end{eqnarray}}

\newcommand{\beann}{\begin{eqnarray*}}
\newcommand{\eeann}{\end{eqnarray*}}
\newcommand{\benn}{\begin{displaymath}}
\newcommand{\eenn}{\end{displaymath}}
\newcommand{\Tr}{\mbox{Tr}}

\newcommand{\kh}{\breve{k}}
\newcommand{\lh}{\breve{\lambda}}
\newcommand{\Gh}{\breve{G}}

\newcommand{\p}[4]{\Phi^{#1}_{#2}{(#3)}^{\mbox{\rm \scriptsize {#4}}}}
\newcommand{\pt}[4]{\widetilde{\Phi}^{#1}_{#2}{(#3)}^{\mbox{\rm \scriptsize {#4}}}}

\newcommand{\Fbeta}{\mbox{\boldmath $\beta$}}

\newcommand{\rf}[1]{(\ref{#1})}

\def\M{M}

\begin{document}
\begin{titlepage}
\renewcommand{\baselinestretch}{1}
\small\normalsize
\begin{flushright}
hep-th/0110054\\
MZ-TH/01-27
\end{flushright}

\vspace{1cm}

\begin{center}
{\LARGE \sc Renormalization Group Flow of \\[1mm]
Quantum Gravity \\[3mm]
in the Einstein-Hilbert Truncation}

\vspace{1cm}
{\large M. Reuter and F. Saueressig}\\

\vspace{1cm}
\noindent
{\it Institute of Physics, University of Mainz\\
Staudingerweg 7, D-55099 Mainz, Germany}\\

\end{center}

\vspace*{1.5cm}
\begin{abstract}
The exact renormalization group equation for pure quantum gravity is used to derive the non-perturbative $\Fbeta$-functions for the dimensionless Newton constant and cosmological constant on the theory space spanned by the Einstein-Hilbert truncation. The resulting coupled differential equations are evaluated for a sharp cutoff function. The features of these flow equations are compared to those found when using a smooth cutoff. The system of equations with sharp cutoff is then solved numerically, deriving the complete renormalization group flow of the Einstein-Hilbert truncation in $d=4$. The resulting renormalization group trajectories are classified and their physical relevance is discussed. The non-trivial fixed point which, if present in the exact theory, might render Quantum Einstein Gravity nonperturbatively renormalizable is investigated for various spacetime dimensionalities.
\end{abstract}
\end{titlepage}
\mysection{Introduction}
Exact renormalization group (RG) equations \cite{bag} provide a powerful tool for the nonperturbative investigation of both fundamental (renormalizable) and effective quantum field theories. In particular the RG equation of the effective average action \cite{ber} has been applied to a variety of matter field theories as well as to Euclidean quantum gravity \cite{ERGE,ERGE3a}.

The effective average action $\Gamma_k$ is a Wilsonian (coarse grained) free energy functional defined on a smooth spacetime manifold. Its construction is based upon a modified version of the standard path integral for the generating functional which has a built-in infrared (IR) cutoff at the mass scale $k$. All quantum fluctuations with momenta $p^2>k^2$ are integrated out as usual, while the contributions from modes with $p^2<k^2$ are suppressed by the cutoff. This cutoff is implemented by giving a momentum dependent (squared) mass $R_k(p^2)$ to the modes with covariant momentum $p$ \cite{ber}.

The $k$-dependence of $\Gamma_k$ is governed by an exact functional RG equation. In any realistic theory it is impossible to solve this equation exactly, but by appropriately truncating the space of action functionals (``theory space'') one can obtain nonperturbative approximate solutions which do not rely upon small expansion parameters. The truncation is carried out by making an ansatz for $\Gamma_k$ which contains finitely or infinitely many free, $k$-dependent parameters ${\rm g}_i(k)$. Upon inserting this ansatz into the functional RG equation and projecting the RG flow onto the truncation subspace one obtains a coupled system of ordinary differential equations for the ${\rm g}_i$.

For the case of Euclidean quantum gravity, the effective average action and its RG equation have been constructed in \cite{ERGE} where also the flow equations for the ``Einstein-Hilbert truncation'' were derived. The Einstein-Hilbert truncation retains only the invariants $\int \, d^dx \sqrt{g} R$ and $\int d^dx \sqrt{g}$, the associated coupling constants being $G_k$, the running Newton constant, and $\lb_k$, the running cosmological constant, respectively. The original construction of ref. \cite{ERGE} employs a cutoff of ``TYPE A'' which is formulated in terms of the complete metric fluctuation $h_{\mn}$. In refs. \cite{oliver,oliver2} a new cutoff of ``TYPE B'' was introduced which is formulated in terms of the component fields of $h_{\mn}$ appearing in its transverse-traceless decomposition \cite{tt}.

One of the interesting predictions of the Einstein-Hilbert truncation is that the high energy behavior of 4-dimensional quantum gravity is governed by a non-trivial ultraviolet (UV)-attractive fixed point for the dimensionless Newton constant and cosmological constant \cite{ERGE,oliver,oliver2,souma}. (In $d$ dimensions, they are defined as $g_k \equiv k^{d-2} G_k$ and $\lambda_k \equiv \lb_k / k^2$, respectively.) If this fixed point is present also in the exact theory, Quantum Einstein Gravity is likely to be renormalizable at the nonperturbative level \cite{weinbergcc}. Despite its perturbative nonrenormalizability it could then be a predictive, fundamental theory valid at arbitrarily small distances. An UV fixed point $g^*$ for $g_k$ entails that $G_k \equiv g^*/k^{d-2}$ vanishes for $k \rightarrow \infty$ (provided $d>2$) so that the theory becomes asymptotically free.

Also the ``phenomenological'' implications of the running gravitational constants for black hole physics \cite{bh1,BHST} and for cosmology \cite{horizon,astro} have been investigated. In particular, it has been argued \cite{horizon} that the UV fixed point reflects itself in the cosmology of the Planck era and that it might lead to a solution of the flatness and the horizon problems of classical Friedmann-Robertson-Walker cosmology.

Previous investigations and applications of the Einstein-Hilbert truncated flow equations have mainly been based on approximate solutions of the RG-equations in the far IR $(k \rightarrow 0)$ and far UV $(k \rightarrow \infty)$. Besides these solutions only little is known about the complete RG flow given by the Einstein-Hilbert truncation.

In this paper we shall therefore study the numerical solutions to the flow equation for the dimensionless cosmological constant $\lambda_k$ and the dimensionless Newton constant $g_k$, analyzing the RG flow in the complete $\lambda$-$g-$plane. For a generic cutoff, the pertinent $\Fbeta$-functions $\Fbeta_{\lambda}$ and $\Fbeta_{g}$ contain rather complicated ``threshold functions'' which are functions of $\lambda_k$ and $g_k$, and functionals of $R_k(p^2)$. In order to make a numerical solution of the flow equation feasible we introduce a sharp cutoff \cite{ber,sc} for which the integrals defining the threshold functions can be evaluated analytically. In contrary to many standard matter field theories where the singular nature of the sharp cutoff often leads to ill defined or divergent $\Fbeta$-functions, we shall see that applying the sharp cutoff to the $\Fbeta$-functions arising form the Einstein-Hilbert truncation leads not only to well defined functions $\Fbeta_{\lambda}, \Fbeta_g$, but in addition yields the same characteristic behavior of the RG trajectories. Moreover, for universal quantities, even the quantitative results are very similar to those found when using a smooth cutoff function.

The $\Fbeta$-functions employed in the present investigation are those derived in the original paper \cite{ERGE}, i.e. they arise from a cutoff of ``TYPE A'' and a constant gauge fixing parameter $\alpha=1$.

The remaining sections of this paper are organized as follows. In Section II we give a brief summary of the original derivation \cite{ERGE} of the ``TYPE A'' flow equation for $\bar{\lambda}_k$ and $G_k$, introducing all the definitions needed for the later analysis. In Section III we investigate the properties of the threshold functions as an important ingredient of the flow equation. In this course we introduce a sharp cutoff function which allows the analytical evaluation of the integrals appearing in these functions. The Sections IV and V contain the results of our numerical analysis. In Section IV we investigate the renormalization group flow below the Planck scale and compare the results obtained with the sharp and a smooth exponential cutoff function. In Section V we use the sharp cutoff to derive the full RG flow on the $g$-$\lambda$--phase space and classify all possible solutions to the flow equation in $d=4$. Here we also investigate the reliability of the Einstein-Hilbert truncation in $d=4$ and its limitations for $d>4$ by investigating the behavior of the non-trivial fixed point in various other dimensions.
%
%
\mysection{The flow equation of the Einstein-Hilbert truncation}
\setcounter{equation}{0} 
In order to derive the non-perturbative flow equation for the dynamical Newton constant $G_k$ and the cosmological constant $\lb_k$ on the theory space spanned by the Einstein-Hilbert truncation we use the effective average action approach to pure quantum gravity. For the details of the following discussion we refer to the original paper \cite{ERGE}.

The main ingredient of this method is the exact evolution equation for the effective average action $\Gamma_k[g_{\mn}]$ for gravity. The derivation of this evolution equation parallels the approach already successfully tested for Yang-Mills theories \cite{ym,ym2}. In principle it is also possible to include the additional renormalization effects of $\bar{\lambda}_k$ and $G_k$ coming from matter fields \cite{percacci,odintsov}, but these are not included in the present derivation. 

In the construction of $\Gamma_k[g]$ one starts out with the usual path integral of $d$-dimensional Euclidean gravity. This is gauge fixed by  using the background field method \cite{abb,adl} and employing a background gauge fixing condition. Due to the introduction of the background metric $\gb$ the effective average action $\Gamma_k[g; \gb]$ now depends on both the full metric $g$ and the background metric $\gb$. The conventional effective action $\Gamma[g]$ is regained as the $k\rightarrow 0$ limit of $\Gamma_k[g] \equiv \Gamma_k[g;\gb=g]$, where the two metrics have been identified. In this manner $\Gamma_k[g]$ becomes invariant  under general coordinate transformations. 

The crucial new component in the construction of $\Gamma_k[g, \gb]$ is the $k$-dependent IR-cutoff term $\Delta_k S$ in the action under the path integral. This term discriminates between the high $(p^2>k^2)$ and low-momentum modes $(p^2<k^2)$. It suppresses the contribution of the low-momentum modes to the path integral by adding a momentum dependent mass term
\be\label{2.5}
\Delta_k S[h,C,\bar{C};\gb]
= \frac{1}{2}\kappa^2\int d^dx \, \sqrt{\gb}\, h_\mn R^{\rm grav}_k[\gb]^{\mn\rho\sigma}h_{\rho\sigma} +\sqrt{2}\int d^dx\, \sqrt{\gb}\, \bar{C}_\mu R^{\rm gh}_k[\gb]C^\mu
\ee
Here $\kappa^2$ is a constant, and the first and second term on the RHS provide the cutoff for the fluctuations of the metric $h_{\mn} = g_{\mn} - \gb_{\mn}$ and the ghost fields $\bar{C}_{\mu}, C^{\mu}$, respectively. In this work we choose the following form of the cutoff 
operators $R^{\rm grav}_k$ and $R^{\rm gh}_k$: 
\be\label{2.6}
 R^{\rm grav}_k[\gb] = \cz^{\rm grav}_k k^2 R^{(0)}(-\bar{D}/k^2), \qquad R^{\rm gh}_k[\gb] = k^2 R^{(0)}(-\bar{D}/k^2)
\ee
Here $(\cz^{\rm grav}_k)^{\mn\rho\sigma} = \left[ (I-P_{\phi})^{\mn\rho\sigma} - (d-2)/2 P_{\phi}^{\mn\rho\sigma}  \right] Z_{Nk}$ is a matrix acting on $h_{\mn}$. In this expression $(P_\phi)^{\mn\rho\sigma} \equiv d^{-1} \, \gb^{\mn} \, \gb^{\rho\sigma}$ projects $h_{\mn}$ onto its trace part $\phi$. In the terminology of \cite{oliver}, this form of the cutoff function defines the ``cutoff of TYPE A''. The so-called shape function $R^{(0)}$ is arbitrary except that it has to satisfy the conditions
\be\label{2.7}
R^{(0)}(0) = 1, \qquad R^{(0)}(z \rightarrow \infty) = 0
\ee

Neglecting the evolution of the ghost sector which corresponds to a first truncation of the general structure of $\Gamma_k$, one finds that $\Gamma_k[g, \gb]$ satisfies the following flow equation:
\bea\label{2.13}
\nonumber \partial_t \Gamma_k\left[g, \gb \right] &=& \frac{1}{2} \Tr \left[ (\kappa^{-2} \Gamma^{(2)}_k +  R^{\rm grav}_k\left[ \bar{g} \right])^{-1} \partial_t R^{\rm grav}_k\left[ \gb \right] \right] \\
&& - \Tr\left[ \left( -\cm \left[ g, \gb \right] + R^{\rm gh}_k\left[ \gb \right] \right)^{-1} \partial_t R^{\rm gh}_k \left[ \gb \right] \right]
\eea
Here $\Gamma_k^{(2)}[g, \gb]$ denotes the Hessian of $\Gamma_k\left[g, \gb \right]$ with respect to $g_{\mn}$ at fixed background field $\gb_{\mn}$ and $t \equiv \ln(k/\hat{k})$ is the  ``renormalization group time'' with respect to the reference scale $\hat{k}$. Furthermore, $\cm$ represents the Faddeev-Popov ghost operator. 

In order to obtain the nonperturbative flow equation for the running Newton constant $G_k$ and the cosmological constant $\lb_k$ we now approximate $\Gamma_k[g, \gb]$ by the Einstein-Hilbert truncation, considering the subspace spanned by the operators $\int d^dx \sqrt{g}$ and $\int d^dx \sqrt{g}\,R$ only: 
\be\label{2.15}
\Gamma_k[g\,,\,\bar{g}]\, = (16 \pi G_k)^{-1} \int  d^dx \,\sqrt{g} \, \left\{ -R + 2 \bar{\lambda}_k \right\}  + \mbox{classical gauge fixing}
\ee
Substituting this ansatz into the evolution equation \rf{2.13} and projecting the resulting flow onto the subspace given by the Einstein-Hilbert truncation then leads to the flow equation for $\bar{\lambda}_k$ and $G_k$. This equation is conveniently written down using the dimensionless Newton constant $g_k$ and cosmological constant $\lambda_k$: 
\be\label{2.16}
g_k \equiv k^{d-2} G_k, \qquad \lambda_k \equiv \lb_k k^{-2}
\ee 
For these couplings the flow equation reads
\be\label{2.17}
\nonumber \partial_t \lambda_k = \Fbeta_{\lambda}(\lambda_k, g_k), \qquad
\partial_t g_k = \Fbeta_g(\lambda_k, g_k) 
\ee
where the $\Fbeta$-functions are given by
\bea\label{2.18}
\nonumber \Fbeta_{\lambda}(\lambda, g) &=& -(2-\eta_N)\, \lambda + \frac{1}{2}\, (4 \pi)^{1-d/2}  \, g \cdot \\
\nonumber  && \! \! \cdot \left[ 2 \, d(d+1) \, \Phi^1_{d/2}(-2\lambda)- 8 \, d \, \Phi^1_{d/2}(0) - d(d+1) \, \eta_N \, \widetilde{\Phi}^1_{d/2}(-2 \lambda) \right]  \\[1.5ex]
\Fbeta_g(\lambda, g) &=& \left(d-2+\eta_N \right) \, g  
\eea
Here $\eta_N$ is the anomalous dimension of the operator $\int d^dx \sqrt{g}R$,
\be\label{2.19}
\eta_N(g, \lambda) = \frac{g \, B_1(\lambda)}{1-g \, B_2(\lambda) },
\ee
and the functions $B_1(\lambda)$ and $B_2(\lambda)$ have the following definition:
\bea\label{2.20}
\nonumber B_1(\lambda) &\equiv& \frac{1}{3} \,(4 \pi)^{1-d/2} \bigg[ d(d+1) \, \Phi^1_{d/2-1}(-2\lambda) \, - 6d(d-1)\,\Phi^2_{d/2}(-2\lambda) \hspace*{2cm}\\
\nonumber &&  \hspace{2cm}-4d \,\Phi^1_{d/2-1}(0) - 24 \Phi^2_{d/2}(0) \bigg] \\[2ex]
B_2(\lambda) &\equiv& -\frac{1}{6} \,(4 \pi)^{1-d/2} \left[d(d+1) \, \widetilde{\Phi}^1_{d/2-1}(-2\lambda)-6d(d-1)\widetilde{\Phi}^2_{d/2}(-2\lambda)  \right] 
\eea
Furthermore, we introduced the $R^{(0)}$-dependent threshold functions $\Phi^{p}_{n}$ and $\widetilde{\Phi}^{p}_{n}$ $(p=1,2,\ldots)$ as
\bea\label{2.21}
\nonumber \Phi^p_n(w) &=& \frac{1}{\Gamma(n)} \int^{\infty}_{0} \!\! dz z^{n-1} \frac{ R^{(0)}(z) \, - \, z\, R^{(0)\prime}(z)}{\left[ z +  R^{(0)}(z) + w\right]^p}\\[2ex]
\widetilde{\Phi}^p_n(w) &=& \frac{1}{\Gamma(n)} \int^{\infty}_{0} \!\! dz z^{n-1} \frac{ R^{(0)}(z) }{\left[ z +  R^{(0)}(z) + w\right]^p}
\eea

In order to investigate the renormalization group flow below the Planck scale it is more convenient to introduce a second set of coupling constants which are made dimensionless by using an arbitrary but fixed scale $\M$ rather than $k$:
\be\label{2.22}
\Gh(\kh) \equiv G_k \, \M^{d-2}, \qquad \lh(\kh)  \equiv  \lb_k \, \M^{-2} 
\ee
For convenience we also introduce $\kh \equiv k M^{-1}$ as a dimensionless scale parameter.

Substituting the dimensionless coupling constants \rf{2.22} into the $\Fbeta$-functions  \rf{2.18} one finds the following set of differential equations
\be\label{2.23}
\nonumber \kh \, \frac{d}{d \kh} \Gh(\kh) = \eta_N \Gh(\kh)
\ee
\bea\label{2.23a} 
\nonumber \kh \, \frac{d}{d \kh} \lh(\kh) &=& \eta_N \, \lh + \frac{1}{2} \, (4 \pi)^{1-d/2}   \, \kh^d \, \Gh \bigg[2 \, d(d+1) \, \Phi^1_{d/2}(-2 \lh/\kh^2)\\
&& \qquad  -8 \, d \, \Phi^1_{d/2}(0) - d(d+1) \, \eta_N \, \widetilde{\Phi}^1_{d/2}(-2 \lh/\kh^2) \bigg] \hspace*{2cm}
\eea
with
\be\label{2.23b}
\eta_N = \frac{\kh^{d-2} \, \Gh \, B_1(\lh/\kh^2)}{1-\kh^{d-2} \, \Gh \, B_2(\lh/\kh^2) }
\ee
Here $\partial_t = k \, d/dk = \kh \, d/d\kh$ has been used. The equations \rf{2.23}, \rf{2.23a} and \rf{2.23b} will be our starting point for investigating the renormalization group flow below the scale $\M$ in Section IV.
%
%
\mysection{threshold functions}
Before we can turn to the numerical solution of the flow equations \rf{2.18} and \rf{2.23}, \rf{2.23a} it is important to understand the characteristics of the threshold functions $\Phi^{p}_{n}$ and $\widetilde{\Phi}^{p}_{n}$ which depend on the form of the cutoff function $R^{(0)}$ chosen.
\begin{subsection}{Smooth Cutoff Functions}
For the purpose of studying the general properties of the threshold functions, we first assume that $R^{(0)}$ satisfies \rf{2.7} and is a smooth function which does not vanish too quickly for small values of its argument: $z+R^{(0)}(z) \ge 1$, $\forall \, z \ge 0$. Under this condition the definition of the threshold functions \rf{2.21} immediately shows that they vanish for $w \rightarrow \infty$. 

Furthermore one notes by examining the denominator of the integrands in \rf{2.21} that the functions $\p{p}{n}{w}{}$ and $\pt{p}{n}{w}{}$ are well defined only for arguments $w>-1$. For $w<-1$ the denominator vanishes for some $z$ in the region of integration, yielding a non-integrable singularity. Therefore $\p{p}{n}{w}{}$ and $\pt{p}{n}{w}{}$ are finite and well defined for $w \in (-1, \infty)$ only.

One can derive a recursion formula which connects the threshold functions with different $p$-values. Interchanging the derivative with respect to $w$ and the $z$-integration yields:
\be\label{3.2}
\frac{d}{d\,w}\, \Phi^p_n(w) = -p \,  \Phi^{p+1}_n(w), \qquad  
\frac{d}{d\,w}\, \widetilde{\Phi}^p_n(w) = -p \,  \widetilde{\Phi}^{p+1}_n(w)
\ee
\end{subsection} 
\begin{subsection}{The Exponential Cutoff}
For practical computations it is necessary to have an explicit form of the cutoff function $R^{(0)}$. In the calculations done in \cite{souma,BHST,horizon} the one-parameter family of exponential cutoffs
\be\label{3.3}
R^{(0)}(z; s)^{\rm \scriptsize Exp} = \frac{s \, z}{\exp (s \, z) -1} \quad {\rm for} \quad s>0
\ee
has been used. Here $s$ is a ``shape parameter''. For these cutoffs the threshold functions \rf{2.21} are seen to be positive definite:
\be\label{3.4}
\p{p}{n}{w}{Exp} \ge 0,  \quad  \pt{p}{n}{w}{Exp} \ge 0, \qquad p = 1,2, \ldots \; {\rm ,} \; n = 1,2, \ldots \; \; \mbox{and} \; w \; \in (-1, \infty) 
\ee
The integral defining the threshold functions with exponential cutoff can be carried out analytically for vanishing argument. Using the integral representations for polylogarithms and the Riemann $\zeta$-function, ${\rm Li}_{\nu}(s) = \frac{1}{\Gamma(\nu)}\, \int_0^\infty dz \, \frac{s \, z^{\nu-1}}{e^z-s}$ and $\zeta(\nu) = \frac{1}{\Gamma(\nu)}\, \int_0^\infty dz \, \frac{z^{\nu-1}}{e^z-1}$ \cite{Le82}, one easily verifies
\bea\label{3.5}
\nonumber \p{1}{n}{0}{Exp} &=& \frac{n}{s^n} \left\{ \, \zeta(n+1) - {\rm Li}_{n+1}(1-s)\right\}\\[1.5ex]
\p{2}{n}{0}{Exp} &=& \frac{1}{s^{n-2}(1-s)} \, {\rm Li}_{n-1}(1-s)
\eea
But for non-vanishing arguments $w$ an analytic solution to these integrals is unknown.   
\end{subsection}
\begin{subsection}{The Sharp Cutoff}
In order to be able to evaluate the threshold functions for any argument $w$ we introduce a different cutoff function, the sharp cutoff. On the level of the dimensionful function $R_k(p^2) \equiv k^2 R^{(0)}(p^2/k^2)$ the sharp cutoff is defined as
\be\label{3.6}
R_k(p^2)^{\rm sc} \equiv \widehat{R} \, \Theta(1-p^2/k^2)
\ee
where the limit $\widehat{R} \rightarrow \infty$ is to be taken {\it after} the integration over $p$, i.e. after substituting \rf{3.6} into the threshold functions. In the path integral this choice of cutoff leads to a complete suppression of modes with momentum $p^2<k^2$ while all modes with $p^2 \ge k^2$ are completely integrated out without the additional mass term.

The evaluation of the integrals in the threshold functions then proceeds as follows. In the first step one substitutes $z=p^2/k^2$ and $w=v/k^2$ into the definition of $\p{p}{n}{w}{}$:
\be\label{3.7}
\Phi^p_n(v/k^2) = \frac{1}{\Gamma(n)}\,(k^2)^{p-n} \int^\infty_0 dp^2 \, (p^2)^{n-1} \, \frac{R^{(0)}(p^2/k^2)- (p^2/k^2) R^{(0)\prime}(p^2/k^2)}{(p^2 + R_k(p^2) + v)^p} 
\ee
Here the prime denotes the derivative of $R^{(0)}$ with respect to its argument. The integrand of equation \rf{3.7} can then be written as a total derivative with respect to $k$, for any  $p>1$:
\bea\label{3.8}
\nonumber \Phi^p_n(v/k^2) &=& -\frac{1}{2 \, \Gamma(n) \, (p-1)} \, (k^2)^{p-n-1} \int_0^\infty dp^2 \, k \frac{D}{D \, k} \, \frac{(p^2)^{n-1}}{(p^2 + R_k(p^2) + v)^{p-1}} \\
&& = -\frac{1}{2 \, \Gamma(n) \, (p-1)} \, (k^2)^{p-n-1} \int_0^\infty dp^2 \, k \frac{\partial}{\partial \, k} \, \frac{(p^2)^{n-1}}{(p^2 + R_k(p^2) + \tilde{v})^{p-1}} \bigg|_{\tilde{v}=v}  
\eea
Here the derivative $D/Dk$ acts by definition only on the $k$-dependence of $R_k$, but not on $v$. In order to rewrite it in terms of $\partial/\partial k$ we introduced the constant $\tilde{v}$ which is strictly independent of $k$. Only at the end of the calculation it is identified with $v \equiv w k^2$. (For $p=1$ the formula \rf{3.8} breaks down since its right hand side is no longer well defined.) Assuming $p>1$, we interchange the $z$-integration and  the derivative with respect to $k$. This is allowed since the integral in \rf{3.8} is absolutely convergent. If one now substitutes the sharp cutoff $R_k(p^2)^{\rm sc} \equiv \widehat{R} \, \Theta(1-p^2/k^2)$ one finds:
\be\label{3.9}
\p{p}{n}{v/k^2}{\rm sc} = -\frac{(k^2)^{p-n-1} }{2 \, \Gamma(n) \, (p-1)} \; k \frac{\partial}{\partial \, k} \, \int_0^\infty dp^2 \, \frac{(p^2)^{n-1} }{(p^2 + \widehat{R} \, \Theta(1-p^2/k^2)  + \tilde{v})^{p-1}} \bigg|_{\tilde{v}=v} 
\ee
Taking the limit $\widehat{R} \rightarrow \infty$ restricts the momentum integration to $p^2<k^2$: 
\be\label{3.10}
\p{p}{n}{v/k^2}{sc} = -\frac{(k^2)^{p-n-1}}{2 \, \Gamma(n) \, (p-1)} \; k \frac{\partial}{\partial \, k} \, \int_{k^2}^{\infty} \, dp^2 \, \frac{(p^2)^{n-1}}{(p^2 +  \tilde{v})^{p-1}} \bigg|_{\tilde{v}=v} 
\ee
The resulting integral is trivially evaluated by acting with the $k$-derivative on the lower integration limit. This yields our final result for the threshold functions with a sharp cutoff:
\be\label{3.11}
\p{p}{n}{w}{sc} = \frac{1}{\Gamma(n)} \, \frac{1}{p-1} \, \frac{1}{(1+w)^{p-1}} \quad {\rm for} \quad p>1
\ee

One easily verifies that these threshold functions also satisfy the recursion relation \rf{3.2} derived for a smooth cutoff. This relation can be used to {\it define} $\p{1}{n}{w}{\rm sc}$ as the solution to the differential equation $\frac{d}{d\,w} \p{1}{n}{w}{sc} = - \frac{1}{\Gamma(n)} \, \frac{1}{1+w} $ which arises from substituting $\p{2}{n}{w}{\rm sc}$ into \rf{3.2}. $\p{1}{n}{w}{\rm sc}$ is then determined up to a constant of integration:
\be\label{3.12}
\p{1}{n}{w}{sc} = - \, \frac{1}{\Gamma(n)} \, \ln(1+w) + \p{1}{n}{0}{sc}
\ee
The constants $\p{1}{n}{0}{sc} \equiv \varphi_n$ are undefined a priori; they parameterize the residual cutoff scheme dependence which is still present after having opted for a sharp cutoff.

For numerical calculations we need to fix the $\varphi_n$'s. Usually we shall assume them equal to the corresponding constants arising from the exponential cutoff with the shape parameter $s=1$: 
\be\label{3.13}
\varphi_n \equiv \p{1}{n}{0}{sc} = \p{1}{n}{0}{Exp~($s$=1)}
\ee
This choice leads to a very good agreement between the  threshold functions evaluated by using the exponential and the sharp cutoff, at least for values of $w$ in the ``confidence interval'' $[-0.7, 1]$. This is shown in FIG. \ref{eins}, where the numerical values of those $\Phi^{p}_{n}$-functions which appear in the $\Fbeta$-functions for $d=4$ are compared for the two types of cutoffs.
\begin{figure}[t]
\renewcommand{\baselinestretch}{1}
\epsfxsize=0.49\textwidth
\begin{center}
\leavevmode
\epsffile{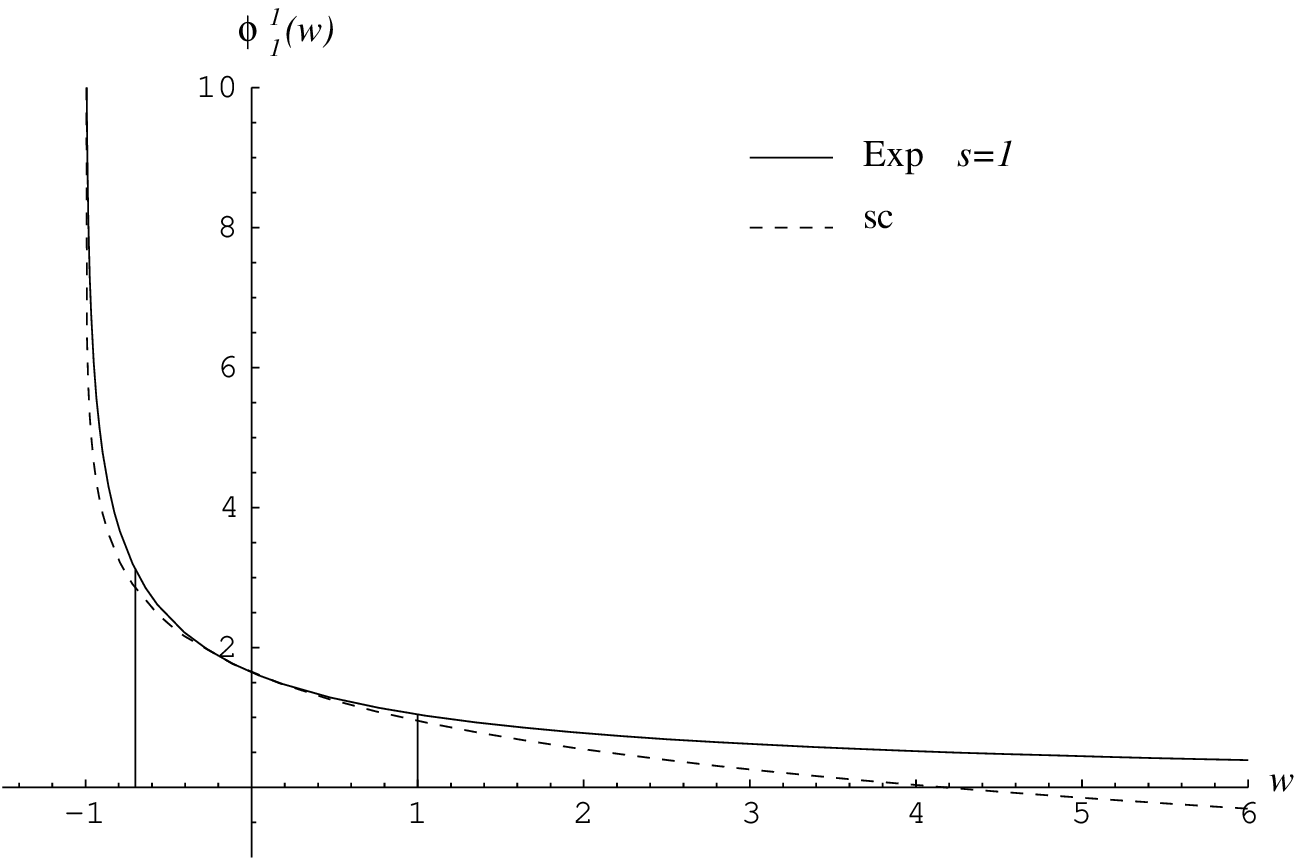}
\epsfxsize=0.48\textwidth
\epsffile{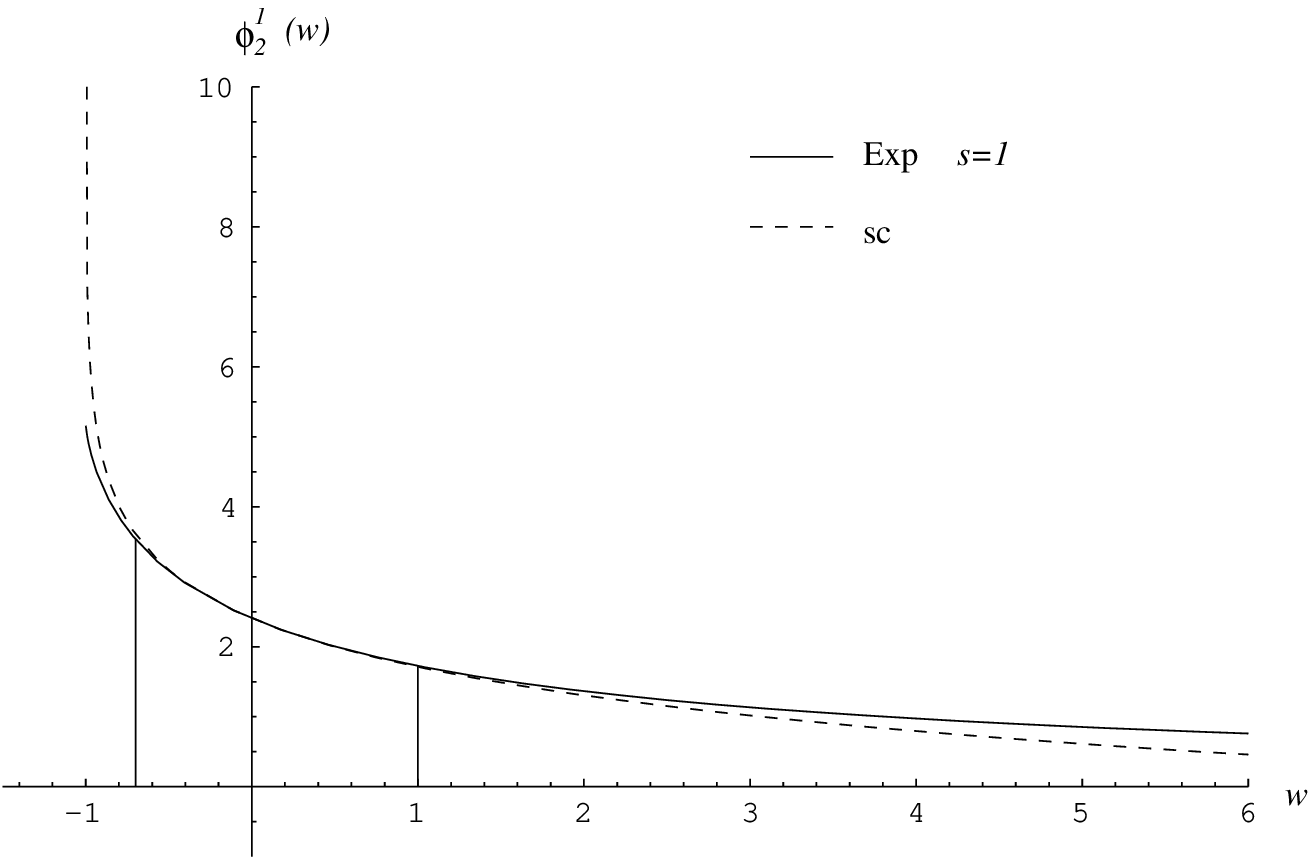}
\\
\leavevmode
\epsfxsize=0.49\textwidth 
\epsffile{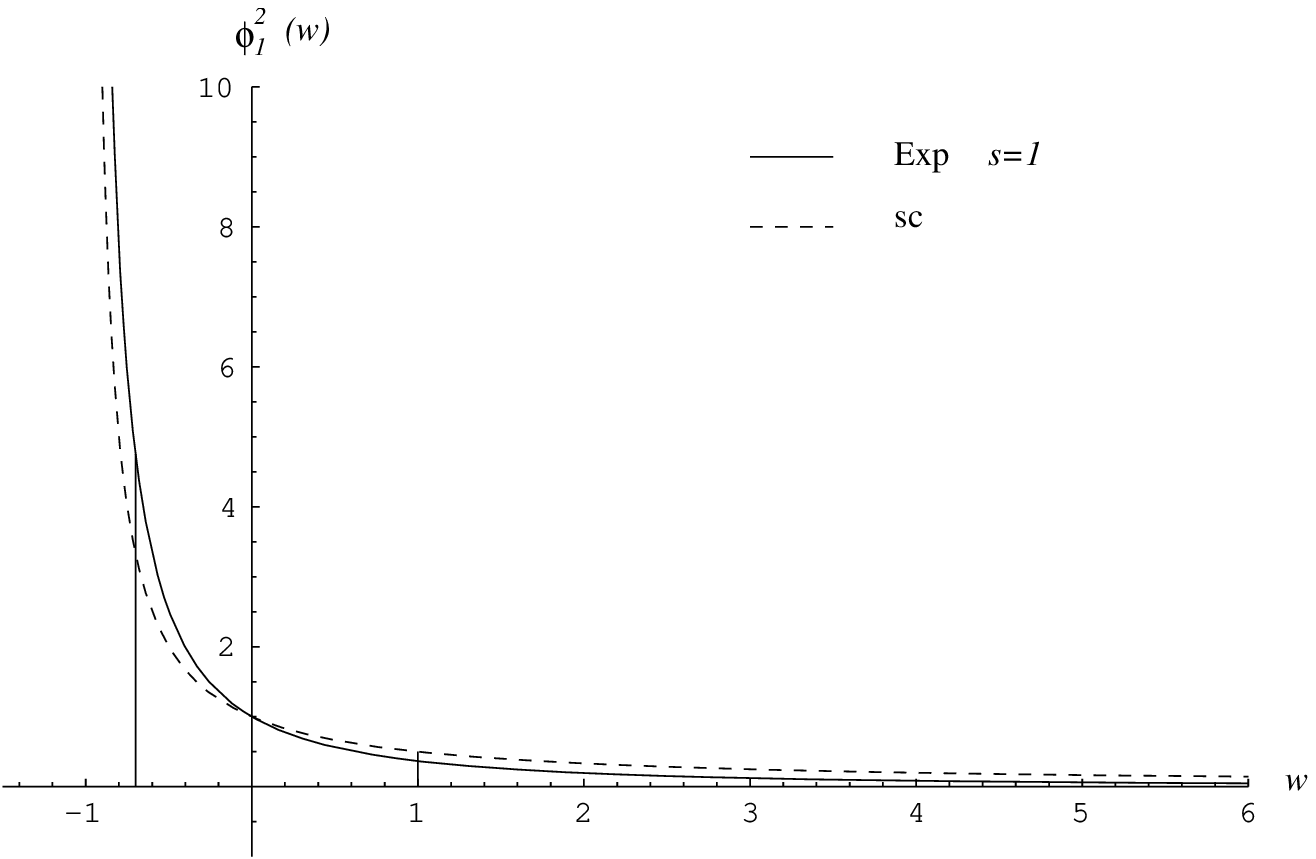}
\epsfxsize=0.48\textwidth
\epsffile{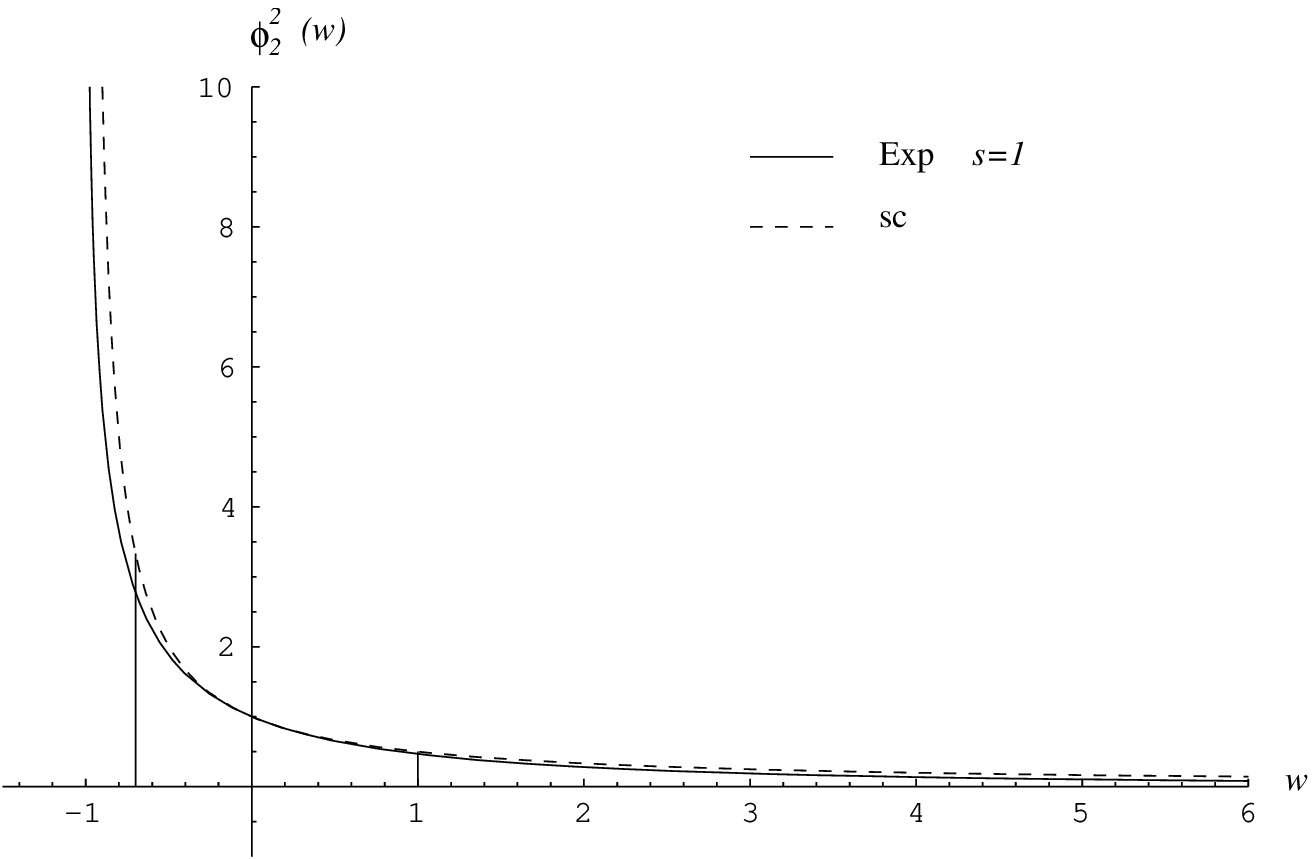}
\end{center}
\parbox[c]{\textwidth}{\caption{\label{eins}{\footnotesize Comparison between the threshold functions in the $\Fbeta$-functions for $d=4$ for exponential and the sharp cutoff. The horizontal lines indicate the ``confidence interval'' in which the two cutoff types yield similar results.}}}
\end{figure}
\begin{figure}
\epsfxsize=0.48\textwidth
\renewcommand{\baselinestretch}{1}
\begin{center}
\leavevmode
\epsffile{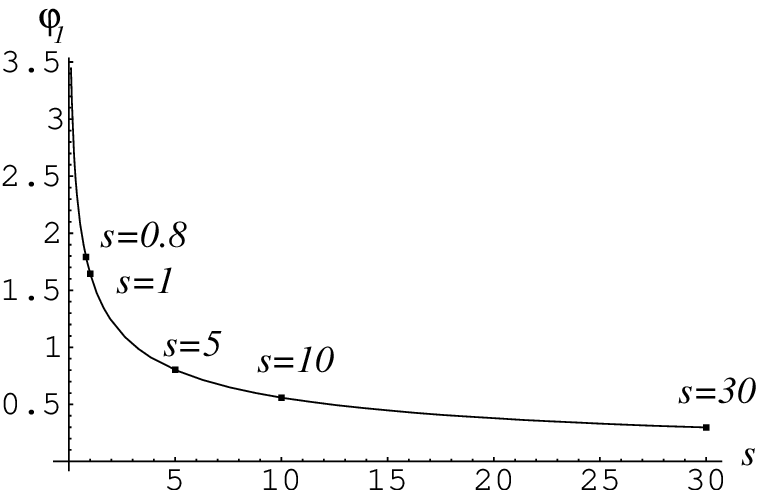}
\epsfxsize=0.48\textwidth
\epsffile{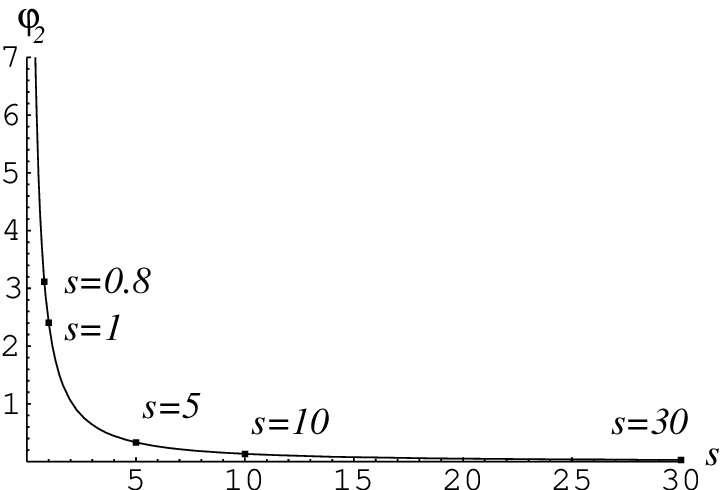}
\end{center}
\parbox[c]{\textwidth}{\caption{\label{f"unf}{\footnotesize Dependence of the constants of integration $\varphi_1$ and $\varphi_2$ on the shape parameter $s$.}}}
\end{figure}

In order to investigate the residual cutoff scheme dependence of the results obtained from the sharp cutoff it is useful to define a one-parameter family of $\varphi_n$'s, analogous to the one-parameter family of exponential cutoff functions \rf{3.3}. This is done by generalizing the relation \rf{3.13} to arbitrary values of the shape parameter $s$:  
\be\label{5.14b} 
\varphi_n(s) \equiv \p{1}{n}{0}{Exp (s)}
\ee
The resulting functions $\varphi_1(s)$ and $\varphi_2(s)$, which are the only constants of  integration appearing in $d=4$, are shown in FIG. \ref{f"unf}. Here one  finds that for $s>15$ the numerical values of $\varphi_1(s)$ and $\varphi_2(s)$ are not subject  to relevant changes any more. Therefore we will limit further investigations of the residual cutoff scheme dependence to the region $s<15$.

For the $\widetilde{\Phi}^{p}_{n}$-functions the integral appearing in \rf{2.21} can also be evaluated explicitly by substituting the sharp cutoff. In a calculation similar to the one shown above one obtains
\bea\label{3.14}
\nonumber \pt{1}{n}{w}{sc} &=& \frac{1}{\Gamma(n+1)} \qquad {\rm for} \qquad p=1 \\
\pt{p}{n}{w}{sc} &=& 0  \qquad {\rm for} \qquad p>1
\eea

Having at hand the explicit form of the threshold functions, equations \rf{3.11}, \rf{3.12} and \rf{3.14}, we are now in a position to write down the $\Fbeta$-functions \rf{2.18} with all integrations carried out. The result reads
\bea\label{3.15}
\nonumber  \Fbeta_g(\lambda, g)^{\rm sc} &=& \left(d-2+\eta_N^{\rm sc} \right) \, g \\[1.5ex]
\nonumber \Fbeta_{\lambda}(\lambda, g)^{\rm sc} &=& -(2-\eta_N^{\rm sc})\, \lambda + \frac{1}{2} \, (4 \pi)^{1-d/2} \, g \cdot \\
&& \! \! \cdot \left[ - \, \frac{2 d (d+1)}{\Gamma(d/2)} \, \ln(1 - 2 \lambda) + 2 d (d-3) \varphi_{d/2} - \frac{d(d+1)}{\Gamma(d/2+1)} \, \eta_N^{\rm sc} \right] 
\eea
with
\be\label{3.16}
\eta_N^{\rm sc}(\lambda, g) = \frac{g \, B_1(\lambda)^{\rm sc} }{1-g \, B_2(\lambda)^{\rm sc} }
\ee
Here $B_1(\lambda)^{\rm sc}$ and $B_2(\lambda)^{\rm sc}$ are given by
\bea\label{3.17}
\nonumber B_1(\lambda)^{\rm sc} &=& \frac{1}{3} \, (4 \pi)^{1-d/2}  \bigg\{ -\frac{d \, (d+1)}{\Gamma(d/2-1)} \, \ln(1-2\lambda) + d \, (d-3) \varphi_{d/2-1} \\
\nonumber && \qquad - \frac{6\,d\,(d-1) }{\Gamma(d/2)} \, \frac{1}{1-2 \lambda} - \frac{24}{\Gamma(d/2)} \bigg\}\\[1.5ex]
B_2(\lambda)^{\rm sc} &=& - \frac{1}{6} \, (4 \pi)^{1-d/2} \, \frac{d \, (d+1)}{\Gamma(d/2) }
\eea
Thus the flow equation \rf{2.17} has now boiled down to a coupled system of ordinary differential equations whose RHS is known explicitly and which is easily solved numerically.
\end{subsection}
%
%
\mysection{RG trajectories below the Planck scale}
In this section we discuss numerical solutions to the ``$M$-scaled'' differential equations \rf{2.23}, \rf{2.23a} for $\Gh$ and $\lh$ in $d=4$ dimensions. We specify initial conditions at some fixed scale $k = \hat{k}$ and identify the mass scale $M$, which was used in \rf{2.22} for defining the dimensionless variables $\Gh$ and $\lh$, with this initial point $M \equiv \hat{k}$. It will prove convenient to rewrite the flow equations in terms of the scale parameter
\be\label{4.1a}
y \equiv \kh^2 \equiv k^2/\M^2
\ee
Hence the initial point $\hat{k} \equiv M$ corresponds to $\hat{y} = 1$.

We shall see that some RG trajectories can be continued down to $k=0$, while others terminate in a singularity at a nonzero value of $k$. For every trajectory which can be extended to $k=0$ we define a Planck mass in terms of the final, i.e. infrared, value of Newton's constant:
\be\label{4.1b}
m_{\rm Pl} \equiv G_0^{-1/2} \equiv G(k=0)^{-1/2}
\ee
As a consequence of this definition,
\be\label{4.1c}
\Gh(0) \equiv G_0 M^2 = M^2/m_{\rm Pl}^2
\ee
After having solved the RG equation and having found the final value $\Gh(0)$ one could in principle use equation \rf{4.1c} in order to express $M$ in terms of the more physical Planck mass.

In order to disentangle the various effects which contribute to the running of $\Gh$ and $\lh$ it is helpful to start the discussion by analyzing two approximate forms of the system of equations \rf{2.23}, \rf{2.23a}. At a first level of approximation, we neglect the running of $\Gh$ by setting $\eta_N = 0$ and focus on the scale dependence of $\lh$ alone. The motivation is that according to canonical dimensional analysis the running of $\lh$ is much more ``relevant'' than that of Newton's constant. At a second level of approximation, we further simplify the remaining equation, \rf{2.23a} with $\eta_N$ set to zero, by neglecting the backreaction which the changing $\lh$ has on the flow via the threshold functions; to this end we set $\p{1}{n}{-2 \lh / \kh^2}{} \approx \p{1}{n}{0}{}$.
\begin{subsection}{The Naive Renormalization Group Flow}
Let us start with the ``naive'' RG flow which is defined by the following two approximations:
\bea\label{4.1}
\nonumber && \Gh(\kh) \approx \Gh(0) = \mbox{const}, \quad \mbox{i.e.} \quad \eta_N \approx 0, \quad \mbox{and}\\
&& \p{p}{n}{- 2 \lh(y)/y}{} \approx \p{p}{n}{0}{} 
\eea
The remaining differential equation $\frac{d \, \lh(\y)}{d \, \y} = \frac{\y}{2 \, \pi} \, \Gh(0) \, \p{1}{2}{0}{}$ is easily solved:
\bea\label{4.2}
\lh(\y) = \lh{}(\hat{\y}) + \frac{1}{4 \, \pi} \, \p{1}{2}{0}{} \, \Gh(0) \, \left( \y^2 - \hat{\y}^2 \right)
\eea
\begin{figure}[t]
\renewcommand{\baselinestretch}{1}
\epsfxsize=8.5cm
\centerline{\epsfbox{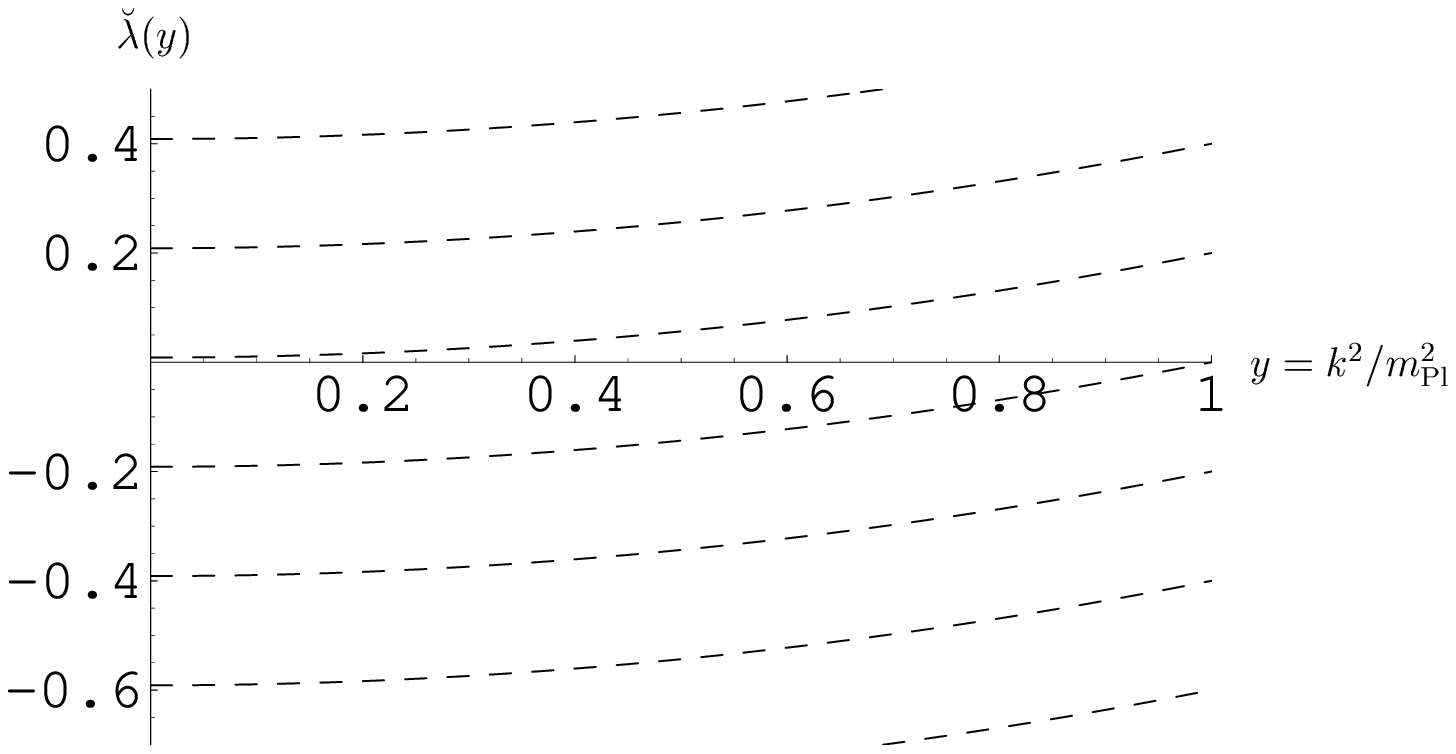}}
\parbox[c]{\textwidth}{\caption{\label{zwei}{\footnotesize Solution \rf{4.2} to the naive flow equation for different initial values $\lh(\hat{\y})$ and $\Gh(0)=1$.}}}
\end{figure}
This corresponds to the result of a naive one loop calculation with an IR cutoff which also yields a running of $\lb_k$ proportional to $y^2 \propto k^4$. 

The solutions \rf{4.2} are plotted in FIG. \ref{zwei} where the dimensionless Newton constant was chosen to be $\Gh(0)=1$. By equation \rf{4.1c} this choice implies that $M=m_{\rm Pl}$, i.e. that the initial conditions are imposed at $\hat{k}=m_{\rm Pl}$, and that the cosmological constant is measured in Planckian units.

As FIG. \ref{zwei} shows, the solutions \rf{4.2} allow for any value of the ``renormalized'' cosmological constant $\lh(0)$ in the limit $k\rightarrow 0$, depending on the initial value set at the scale $\widehat{y}$. We could ``fine-tune'' this initial value so that the renormalized cosmological constant becomes $\lh(0)=0$. But this fine-tuning depends on the numerical value of $\p{2}{1}{0}{}$ which in turn depends on the non-physical cutoff function $R^{(0)}$ chosen. Therefore it cannot have a universal meaning.   
\end{subsection}
\begin{subsection}{The Impact of the Threshold Functions}
\begin{figure}[t]
\renewcommand{\baselinestretch}{1}
\epsfxsize=0.49\textwidth
\begin{center}
\leavevmode
\epsffile{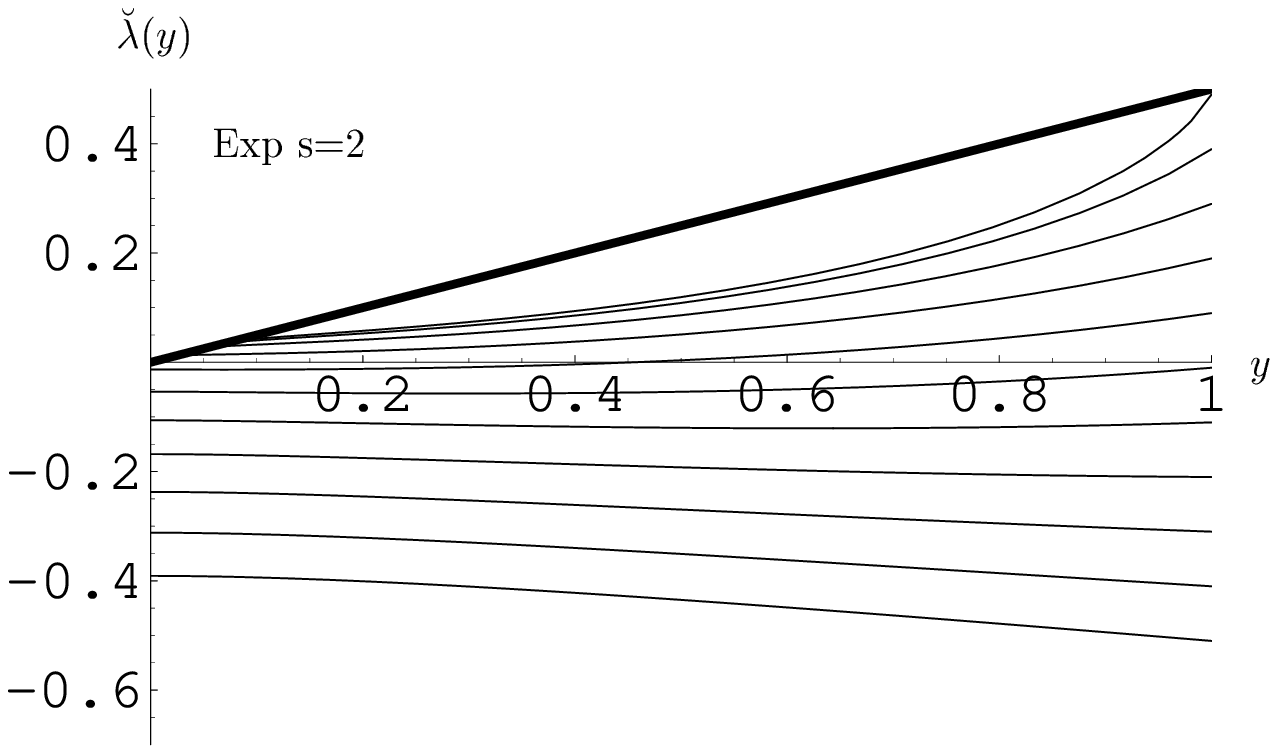}
\epsfxsize=0.48\textwidth
\epsffile{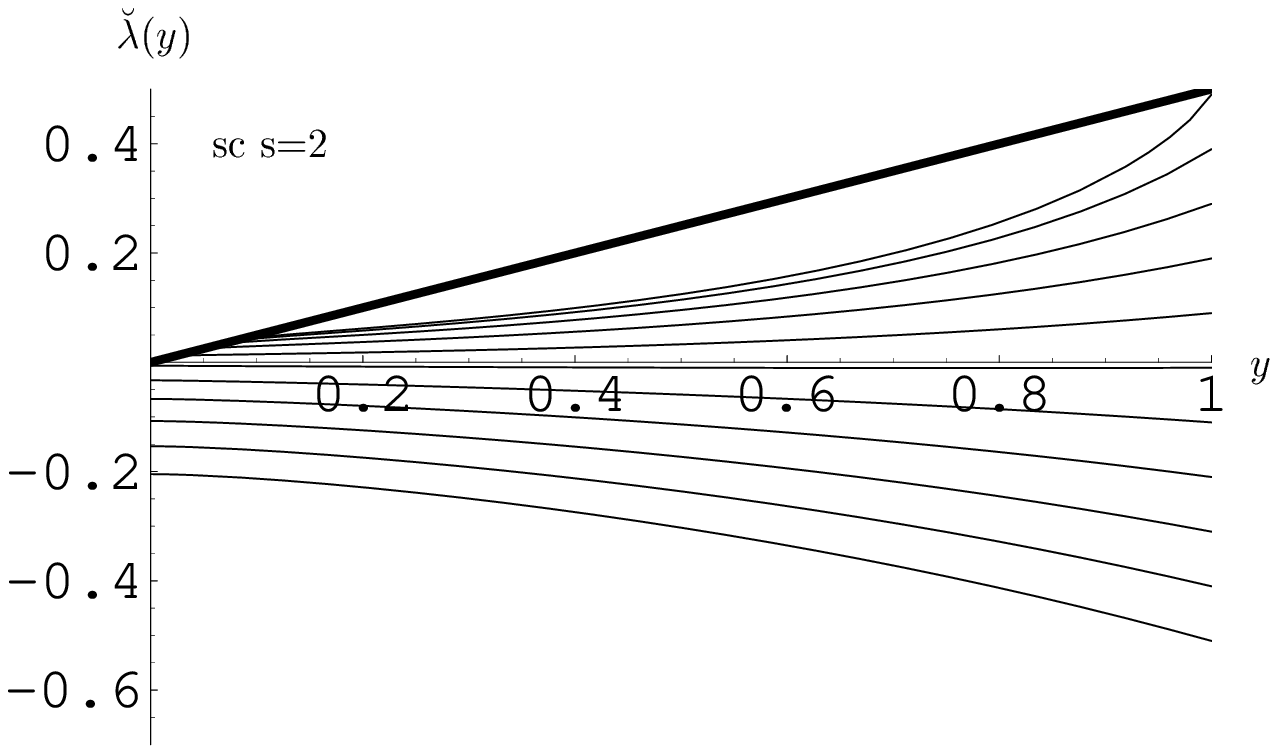}
\end{center}
\parbox[c]{\textwidth}{\caption{\label{dreia}{\footnotesize Numerical solutions to the flow equations \rf{4.3} and \rf{4.3a} for the exponential and the sharp cutoff for $\Gh(0)=1$ and shape parameter $s=2$. The straight line indicates the boundary $\lh=y/2$ introduced by the threshold functions. Trajectories leading to negative values $\lh(0)$ as well as trajectories terminating in the singularity are found.}}}
\end{figure}
Now we drop the second assumption made in \rf{4.1} and allow for a non-trivial argument of the threshold function $\Phi^{1}_{2}$, while keeping $\eta_N = 0$. The resulting flow equation for a generic smooth cutoff reads
\be\label{4.3}
\frac{d \, \lh{}}{d \, \y} = \frac{\y}{2 \pi} \, \Gh(0) \, \left[ 5 \, \p{1}{2}{-2 \lh{}/ \y}{} - 4 \, \p{1}{2}{0}{} \right] \\
\ee
For the sharp cutoff it becomes:
\be\label{4.3a}
\frac{d \, \lh{}}{d \, \y} = \frac{\y}{2 \pi} \, \Gh(0) \, \left[ -5 \, \ln(1-2 \lh{}/ \y) +  \varphi_2 \right]
\ee
We shall solve eqs. \rf{4.3} and \rf{4.3a} numerically and compare the resulting trajectories. We specify initial values $\lh(\hat{y})$ at the initial point $y = \hat{y} \equiv 1$. In order to visualize the generic solution to these flow equations we have to abandon our choice  $s=1$ for the shape parameter because for this special value the RG trajectories have certain properties which are not typical. Let us start with $s=2$ instead which illustrates the general situation. (For the sharp cutoff this change of $s$ leads to a different numerical value of the $\varphi_n$'s, whose $s$-dependence has been defined in equation \rf{5.14b}.) 

The resulting trajectories for $s=2$ are shown in FIG. \ref{dreia}. The special case arising from $s=1$ is displayed in FIG. \ref{drei}. Again we chose $\Gh(0)=1$, implying that $M = m_{\rm Pl}$ for those trajectories for which $G(k=0)$ is defined. 

Comparing the ``Exp'' to the ``sc'' diagrams in FIGs. \ref{dreia} and \ref{drei} clearly indicates that the sharp and the smooth exponential cutoff yield essentially the same RG trajectories. 

The most striking feature of FIG. \ref{dreia} is that some trajectories cannot be continued below a certain finite value $y=y_{\rm term}$. Trajectories which, in FIG. \ref{zwei}, have led to positive IR values $\lh(0)>0$ now terminate because they hit the singular line $\lh = y/2$. It is due to the singularities of the $\Phi$- and $\widetilde{\Phi}$-functions at $w \equiv -2 \lh/y = -1$. Looking at the flow equations \rf{4.3} and \rf{4.3a} we see that their right hand sides diverge, $\Fbeta_{\lh} \rightarrow +\infty$, if $\lh$ approaches $y/2$. 
\begin{figure}[t]
\epsfxsize=0.48\textwidth
\renewcommand{\baselinestretch}{1}
\begin{center}
\leavevmode
\epsffile{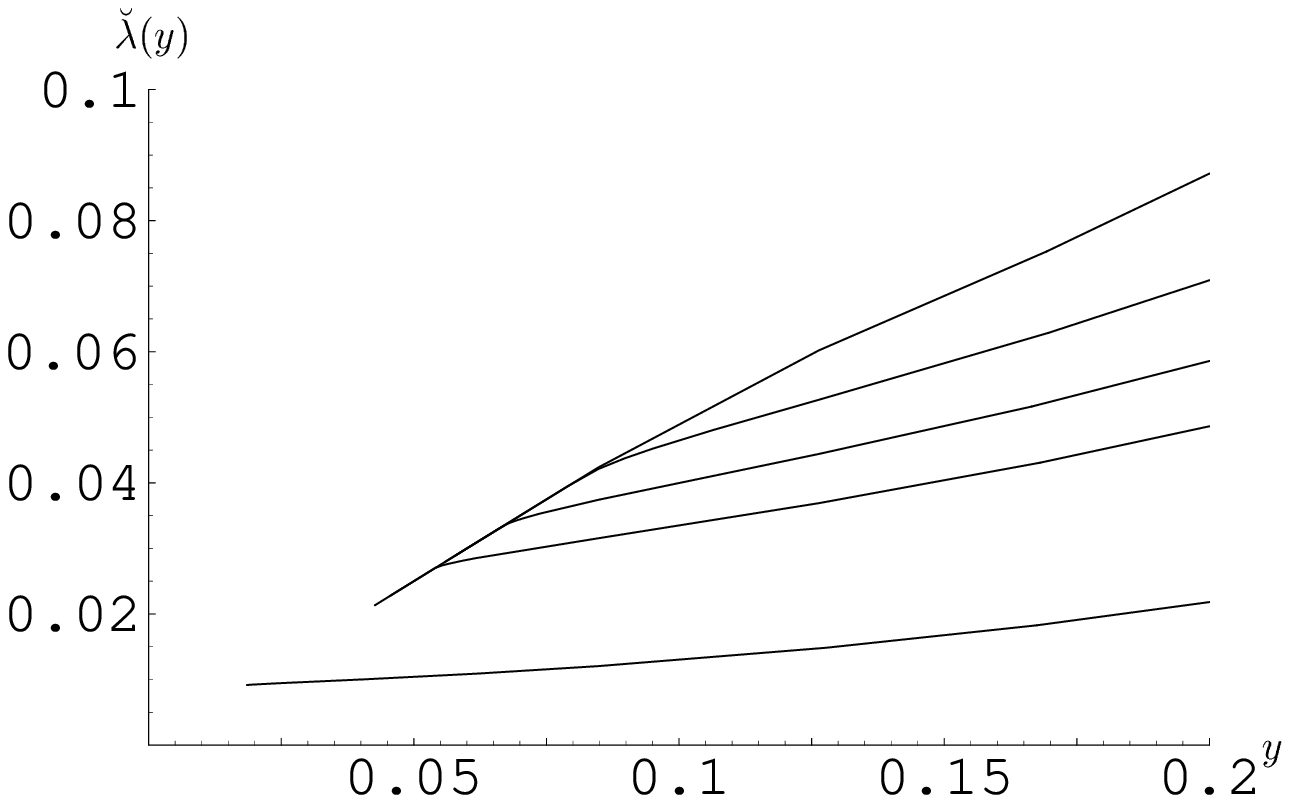}
\end{center}
\parbox[c]{\textwidth}{\caption{\label{dreib}{\footnotesize Behavior of trajectories close to the boundary $\lh = y/2$.}}}
\end{figure}
Some solutions to eq. \rf{4.3a} in the vicinity of $\lh=y/2$ are shown in FIG. \ref{dreib}. The trajectories approach the singularity with a slope $\Fbeta_{\lh}$ smaller than $1/2$. Only directly on the boundary the slope would jump to $\+\infty$ discontinuously. Equations \rf{4.3} and \rf{4.3a}  show that the singularity of $\Fbeta_{\lh}$ also extends to 
\be\label{4.4}
\lh{} \ge \frac{\y}{2}
\ee
Hence the RG flow is defined only for $\lh < y/2$. This leads to the following consequences for all cutoffs: 
\begin{enumerate}
\item It is not possible to choose initial values $\lh(\hat{\y})$ larger than $\hat{y}/2$. 
\item There are no trajectories in the region $\lh > y/2$. Therefore we obtain only negative values for the cosmological constant $\lb_0$ in the limit $k \rightarrow 0$. Trajectories that in FIG. \ref{zwei} have led to positive values $\lh(0)>0$ now either lead to negative values of the cosmological constant, $\lh(0) < 0$, or run into the boundary line $\lh = y/2$ and terminate at finite values $\y_{\rm term} > 0$. 
\end{enumerate} 

Comparing those trajectories that in FIG. \ref{zwei} und FIG. \ref{dreia} yield negative IR-values $\lh(0) < 0$ we see that including the nontrivial argument of the threshold functions leads to a (rather weak) focusing of the cosmological constant towards zero, in the sense that the trajectories in FIG. \ref{dreia} curve upward for $y \rightarrow 0$.
\begin{figure}[t]
\renewcommand{\baselinestretch}{1}
\epsfxsize=0.49\textwidth
\begin{center}
\leavevmode
\epsffile{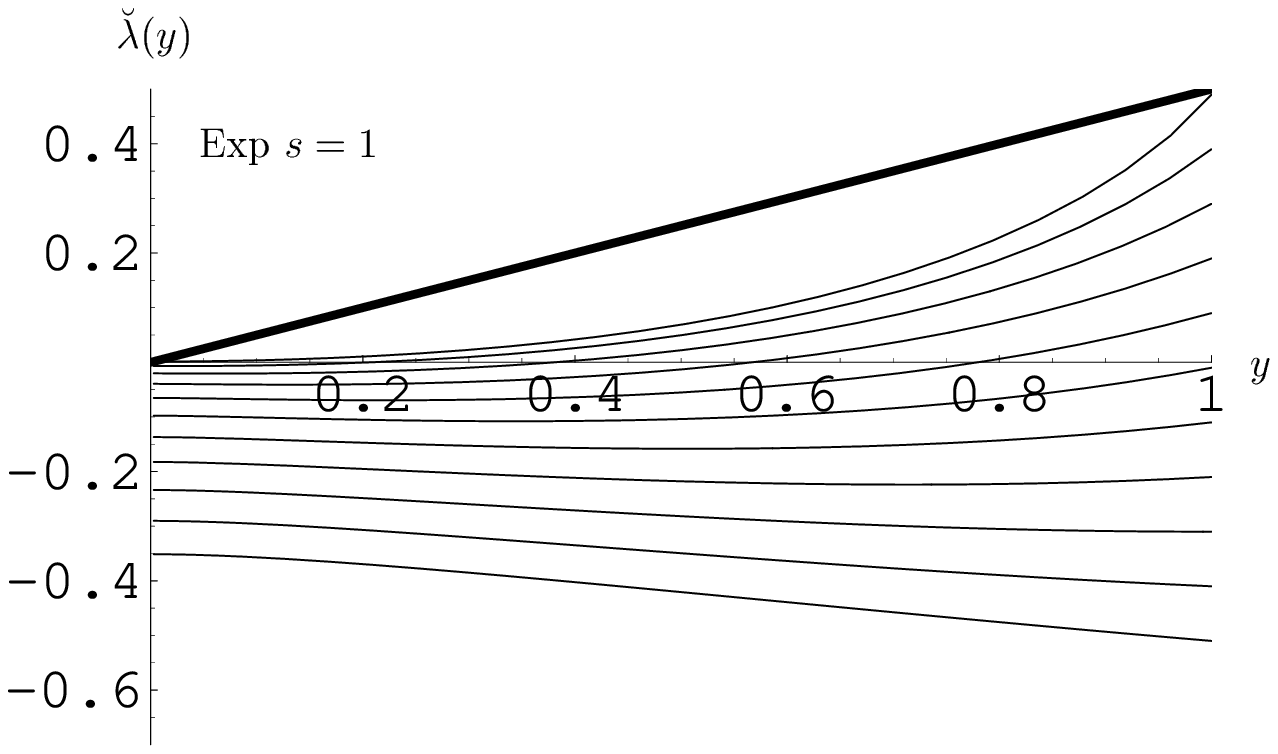}
\epsfxsize=0.48\textwidth
\epsffile{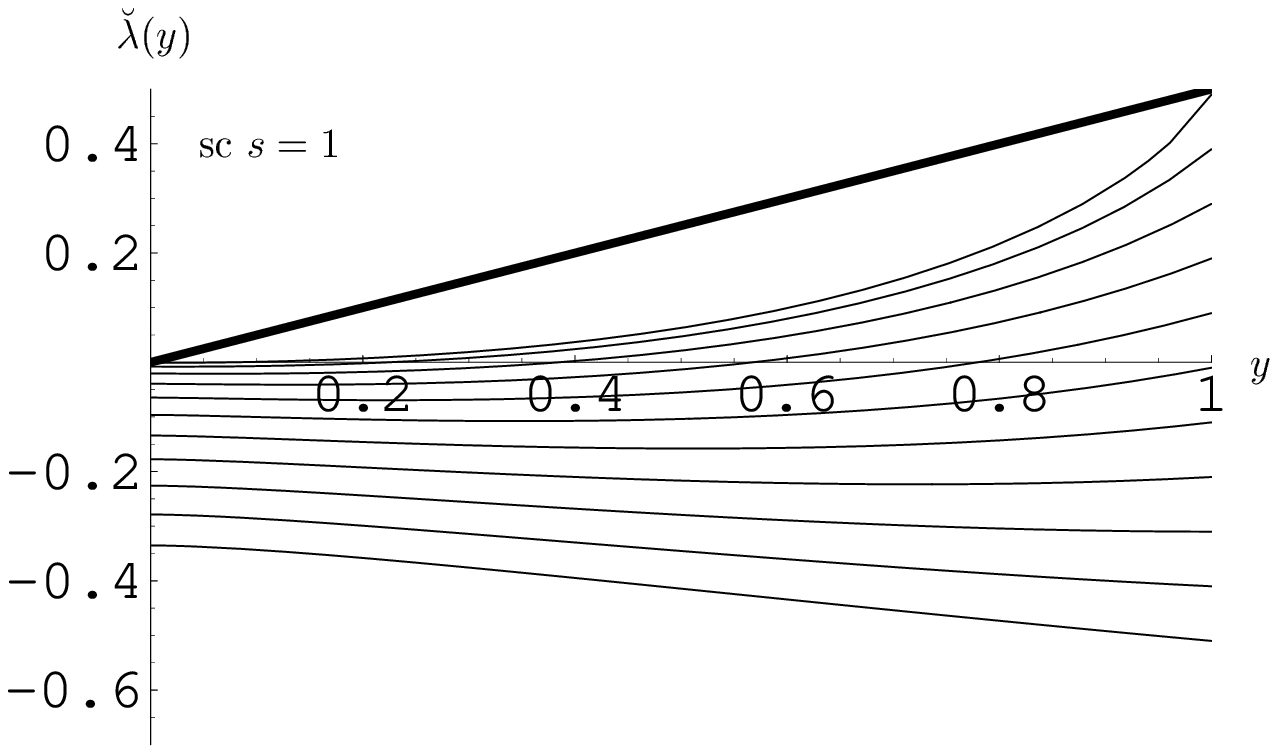}
\end{center}
\parbox[c]{\textwidth}{\caption{\label{drei}{\footnotesize Numerical solutions to the flow equations \rf{4.3} and \rf{4.3a} for the exponential and the sharp cutoff for $\Gh(0)=1$ and shape parameter $s=1$. The bold, straight line indicates the boundary $\lh=y/2$ introduced by the threshold functions. In contrast to the generic case $s=2$ no trajectories terminate on the boundary $\lh=y/2$.}}}
\end{figure}

Looking at FIG. \ref{drei} we see that the choice $s=1$ leads to a non-generic behavior since the termination of the trajectories at $\lh = y/2$ does not occur. In this case all admissible initial conditions imposed at $\hat{y}=1$ give rise to trajectories which can be continued down to $y=0$. They all yield a negative or vanishing $\lh_0$.
\end{subsection}
\begin{subsection}{The Complete System}
Let us now also drop the approximation $G_k=\mbox{const}$ and consider the full flow equations \rf{2.23} and \rf{2.23a}, treating both $\Gh(k)$ and $\lh(k)$ dynamical. We compute the solutions for the initial values 
\be\label{4.5}
\Gh(\hat{\y}=1) = 0.25, \qquad \lh(\hat{\y}=1) \; \mbox{arbitrary}
\ee
The resulting trajectories for the sharp cutoff are shown in FIG. \ref{vier}. Here two classes of solutions immediately become apparent.

Trajectories which end at a negative value of the cosmological constant $\lh(0)<0$ have already appeared in FIGs. \ref{dreia} and \ref{drei}. Later on they will be referred to as of ``Type Ia''. Comparing the trajectories with $\lh(0) < 0$ of FIG. \ref{vier} to those in FIG. \ref{dreia} and \ref{drei} one finds that including a dynamical Newton constant in the flow equation counteracts the effect of focusing $|\lb_0|$ towards smaller values, rather leading to even more negative values of $\lh(0)$. 

The second class of solutions in FIG. \ref{vier}, which will be classified as Type IIIa, is formed by the trajectories ending on the boundary line $\lh = y/2$. For $\lh \rightarrow y/2$  the anomalous dimension $\eta_N$ rapidly diverges. This leads to a vast increase of the Newton constant $\Gh(y)$ preceding the termination of the trajectory. Due to the divergence of $\eta_N$, the RHS of both eq. \rf{2.23} and \rf{2.23a} approaches $-\infty$ as $\lh \rightarrow y/2$. Therefore the tangents of the functions $\Gh(y)$ and $\lh(y)$ turn vertical causing the trajectory to terminate at a finite value $\y_{\rm term}>0$.

Furthermore, FIG. \ref{vier} clearly shows the ``anti-screening'' character of pure quantum gravity, i.e. the monotone decrease of the Newton constant $G_k$ with increasing values of $y$ or $k$ \cite{ERGE}.
\begin{figure}[t]
\renewcommand{\baselinestretch}{1}
\epsfxsize=0.48\textwidth
\begin{center}
\leavevmode
\epsfxsize=0.48\textwidth 
\epsffile{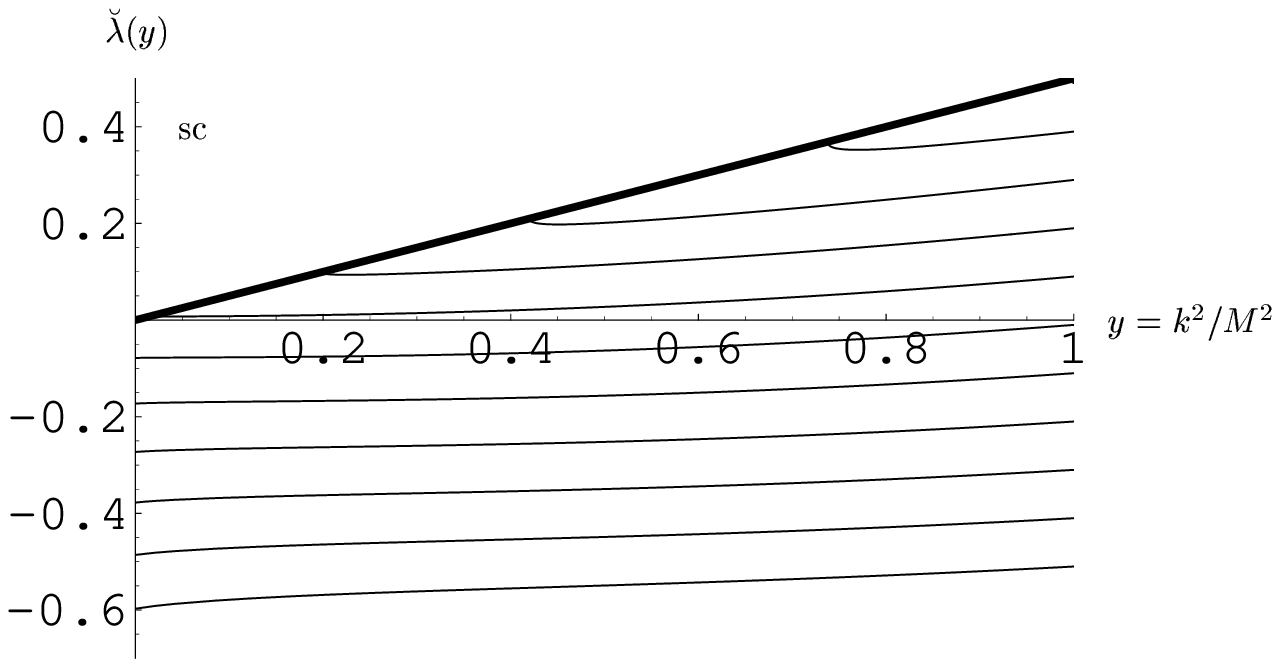}
\epsfxsize=0.48\textwidth
\epsffile{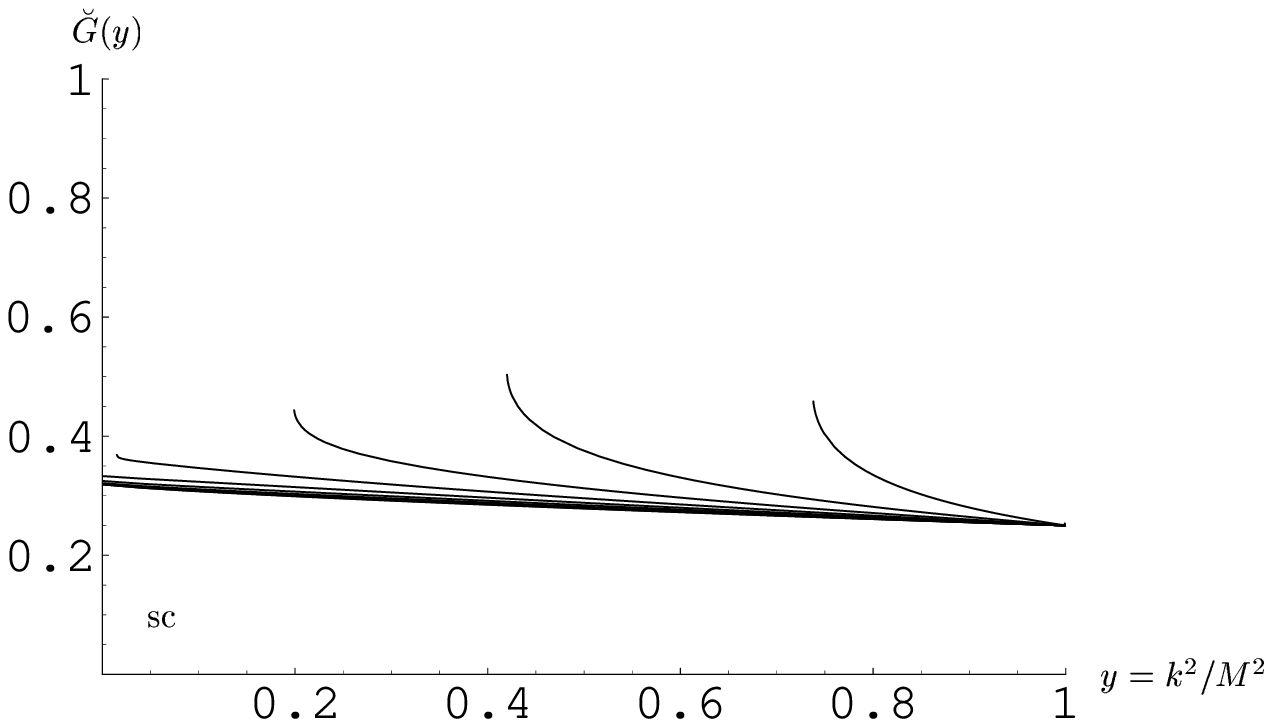}
\end{center}
\parbox[c]{\textwidth}{\caption{\label{vier}{\footnotesize Solution to the full flow equation with $\Gh(\hat{\y}=1) = 0.25$ and various initial values $\lh(\hat{\y}=1)$ for the sharp cutoff with $s=1$. The bold line indicates the boundary at $\lh=y/2$.}}}
\end{figure}

The most important change arising from the inclusion of a running $\Gh(y)$, which can not directly be deduced from the FIGs. \ref{dreia}, \ref{drei} and \ref{vier}, is the modification of the backward evolution when we {\it in}crease $y$ and try to send it to infinity. For the flow equations \rf{4.3} and \rf{4.3a} with $\Gh(y) = \Gh(0)$ kept constant we find that the backward evolution becomes undefined for sufficiently large values of $y > \hat{y}$. All trajectories terminate at the boundary line $\lh = y/2$ at a finite value $y < \infty$ and cannot be continued to $y$``=''$\infty$.

As we shall see in the next section this behavior changes drastically when the running of $\Gh(y)$ is included. This is due to the appearance of a non-trivial fixed point which governs the RG flow of the coupled $\Gh$-$\lh-$system for large $y \gg 1$. As a consequence, all trajectories shown in FIG. \ref{vier} where the running of $\Gh$ was included have a well defined backward evolution and can be continued up to $y$``=''$\infty$.
\begin{figure}[t]
\renewcommand{\baselinestretch}{1}
\epsfxsize=0.48\textwidth
\begin{center}
\leavevmode
\epsffile{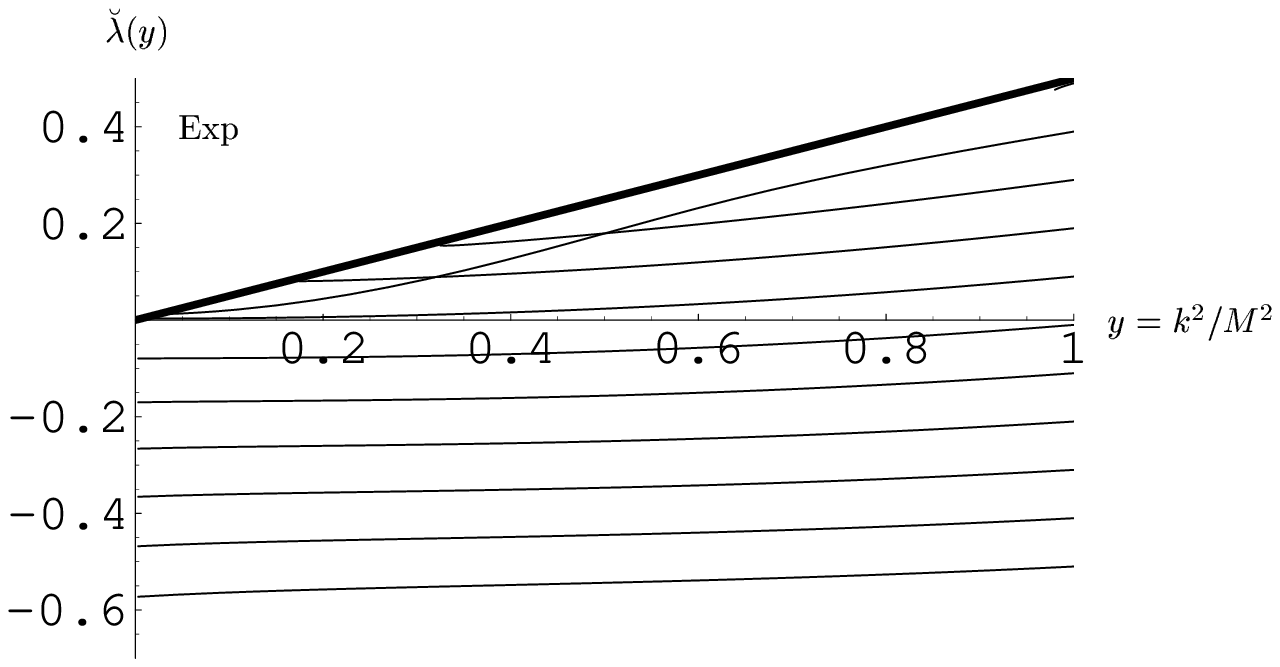}
\epsfxsize=0.48\textwidth
\epsffile{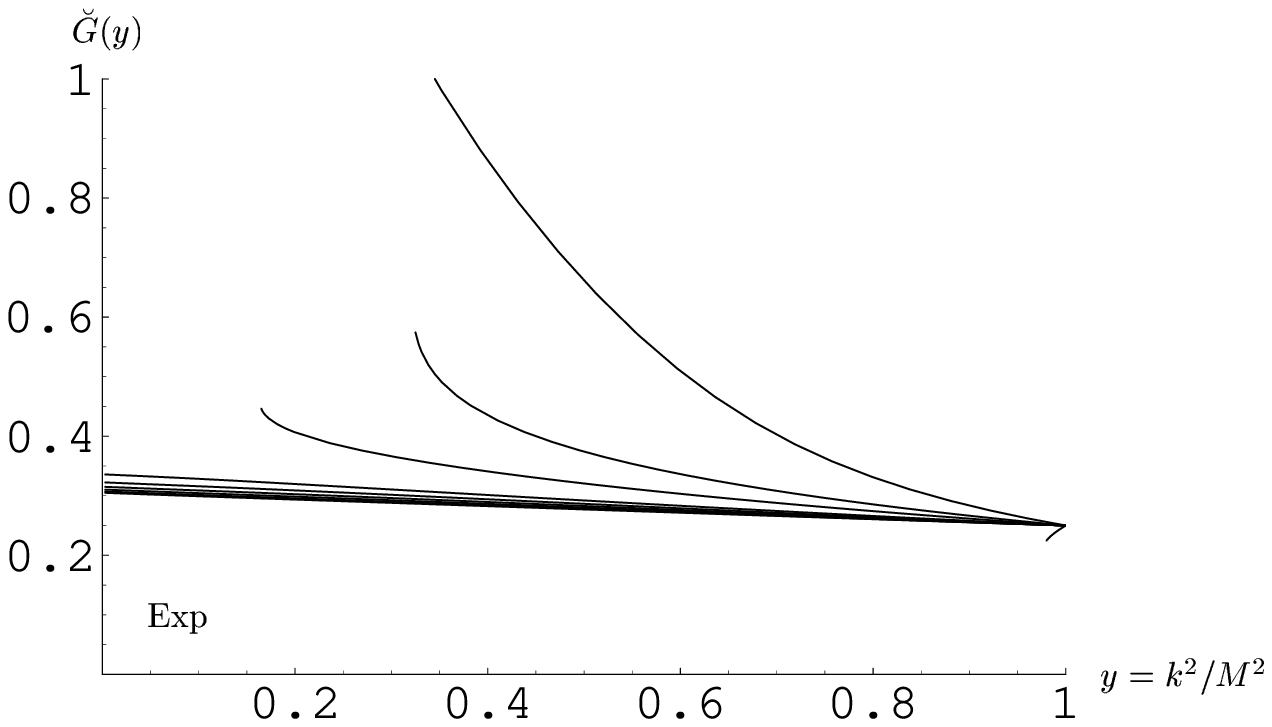}
\end{center}
\parbox[c]{\textwidth}{\caption{\label{viera}{\footnotesize Solution to the full flow equation with $\Gh(\hat{\y}=1) = 0.25$ and various initial values $\lh(\hat{\y}=1)$ for the exponential with $s=1$. The bold line indicates the singularity at $\lh=y/2$.}}}
\end{figure}

Next we solve eqs. \rf{2.23} and \rf{2.23a} using the initial conditions \rf{4.5} and the {\it exponential} cutoff. The results are displayed in FIG. \ref{viera}. Comparing the trajectories in FIGs. \ref{vier} and \ref{viera} one sees that the trajectories obtained using the sharp and the exponential cutoff are very similar except when they get close to the singular line $\lh = y/2$. 

A new phenomenon occurs if we choose initial conditions close to the boundary $\lh = y/2$. Here the trajectory obtained using the sharp cutoff shows no particularities while the one found using the exponential cutoff crosses all the other Type IIIa trajectories which start at lower $\lh(\hat{y})$ and terminates at an ``unnaturally low'' value $y$, see the first diagram of FIG. \ref{viera}. This behavior is due to the divergence of $\eta_N^{\rm Exp}$ on a certain 2-dimensional surface in $\lh$-$\Gh$-$y-$space which is discussed in detail in Appendix A. It leads to a termination of the trajectories slightly before the $\lh = y/2$-line is reached. This difference between the sharp and the smooth cutoff occurs in a region very close to the singularity where the Einstein-Hilbert truncation is unreliable in any case. It is clear that it has no physical significance. For the sharp cutoff the analogous singularity $\eta_N^{\rm sc} \rightarrow \infty$ is located in the (probably unphysical) region with negative $\Gh(\y)$. 
\end{subsection}
\mysection{The complete renormalization group flow}
After having studied the properties of the RG flow below the scale $\M$ in the last section we now investigate the complete RG flow, i.e. the RG flow on the entire $\lambda$-$g-$plane. In this course we first look at the qualitative features of the flow equation, determining its fixed points and their stability properties. Afterwards we construct the full phase portrait of the flow by numerically solving the ``$k$-scaled'' flow equation with the sharp cutoff.
\begin{subsection}{The Fixed Points}
\begin{subsubsection}{General Remarks}
The existence and consequences of a non-trivial fixed point in the Einstein-Hilbert truncation of pure quantum gravity have already been discussed in refs. \cite{oliver,oliver2,souma}. Ref. \cite{souma} uses the exponential shape function \rf{3.3} for a cutoff of TYPE A, establishing  the existence of this fixed point for a wide range of shape parameters $s$. In \cite{oliver} the properties of the fixed point are investigated using the TYPE B cutoff for various smooth shape functions $R^{(0)}$. In this subsection we extend these surveys to the sharp cutoff $R^{(0){\rm sc}}$ within the original TYPE A cutoff scheme and compare our results to those obtained by using $R^{(0){\rm Exp}}$.

In order to investigate the fixed points we now turn to the flow equations written in terms of the $k$-scaled coupling constants $\lambda, g$ of \rf{2.18} and \rf{3.15}. The existence of a fixed point $\lambda^*, g^*$ requires that the $\Fbeta$-functions of the Einstein-Hilbert truncation vanish simultaneously:
\be\label{5.1}
\Fbeta_{\lambda} (\lambda = \lambda^*, g=g^*) =0, \qquad \Fbeta_{g}(\lambda = \lambda^*, g = g^*) = 0 
\ee
Equation \rf{5.1} has two solutions: A trivial one for which $\lambda^*=0, g^*=0$ and a non-trivial solution with $g^*\not=0$. They give rise to a ``Gaussian'' and a ``non-Gaussian'' fixed point, respectively. 

An important property of a fixed point is the anomalous dimension $\eta_N(\lambda^*, g^*)$ at this fixed point. From \rf{2.18} and \rf{2.19} one finds $\eta_N(\lambda^*, g^*)=0$ and $\eta_N(\lambda^*, g^*) = 2-d $ for the ``Gaussian'' and the ``non-Gaussian'' fixed point, respectively. 

In order to investigate their stability properties we linearize the RG flow at the fixed points,
\be\label{5.2}
\partial_t \, \mbox{g}_i \approx \sum_j B_{ij} \, (\mbox{g}_j-\mbox{g}^*_j),  \qquad 
{\bf B} \equiv [B_{ij}] = \left[ \begin{array}{cc} \frac{\partial \, \beta_{\lambda}}{\partial \, \lambda}   & \frac{\partial \, \beta_{\lambda}}{\partial \, g} \\ \frac{\partial \, \beta_g}{\partial \, \lambda} & \frac{\partial \, \beta_g}{\partial \, g} \end{array} \right]
\ee
In our case the generalized couplings ${\rm g}_i$ are given by ${\rm g}_1 = \lambda, \; {\rm g}_2 = g$, and the derivatives defining $B_{ij}$ are taken at $\mbox{g}_i=\mbox{g}^*_i$. At an arbitrary point $(\lambda, g)$, and for any $R^{(0)}$, the partial derivatives of the $\Fbeta$-functions are easily found by making use of the recursion formula \rf{3.2}:
\bea\label{5.4}
\nonumber \frac{\partial \beta_{\lambda}}{\partial \lambda} &=& -(2-\eta_N) + \left( \lambda - \frac{g}{2} (4 \pi)^{1-d/2} \, d \, (d+1) \pt{1}{d/2}{-2\lambda}{} \right) \frac{\partial \, \eta_N}{\partial \, \lambda} \\
\nonumber && + \frac{g}{2} \, (4 \pi)^{1-d/2} \, \left( 4d \, (d+1) \p{2}{d/2}{-2 \lambda}{} -2d \, (d+1) \, \eta_N \, \pt{2}{d/2}{-2\lambda}{} \right) \\[1.5ex]
\nonumber \frac{\partial \beta_{\lambda}}{\partial g} &=& \left( \lambda - \frac{g}{2}\,(4 \pi)^{1-d/2} \, d  (d+1) \, \pt{1}{d/2}{-2\lambda}{} \right) \, \frac{\partial \, \eta_N}{\partial \, g} + \frac{1}{2} \, (4 \pi)^{1-d/2} \cdot \\
\nonumber && \cdot \left( 2d\,(d+1) \p{1}{d/2}{-2  \lambda}{} - 8d \, \p{1}{d/2}{0}{} - d  (d+1) \, \eta_N \, \pt{1}{d/2}{-2  \lambda}{} \right) \\[1.5ex]
\nonumber \frac{\partial \beta_g}{\partial \lambda} &=& \frac{g^2}{1-g\,B_2(\lambda)} \left( B_1^{~\prime} (\lambda) + \eta_N \, B_2^{~\prime} (\lambda) \right) \\[1.5ex]
\frac{\partial \beta_g}{\partial g} &=& d-2+ \left( 2 + \frac{g \, B_2(\lambda)}{1-g\,B_2(\lambda) } \right) \eta_N 
\eea
Here the derivatives of $\eta_N$ are given by
\be\label{5.5}
\nonumber \frac{\partial \, \eta_N}{\partial \, g} = \left( \frac{1}{g} + \frac{ B_2(\lambda) }{1-g\, B_2(\lambda) }\right) \eta_N, \quad
\frac{\partial \, \eta_N}{\partial \, \lambda} = \frac{g}{ 1-g\,B_2(\lambda) } \, \left( B_1^{~\prime} (\lambda)  + \eta_N \, B_2^{~\prime} (\lambda) \, \right)
\ee
and $B_1^{~\prime}(\lambda)$ and $B_2^{~\prime}(\lambda)$ are the derivatives of $B_1(\lambda)$ and $B_2(\lambda)$ with respect to their arguments:
\bea\label{5.6}
B_1^{~\prime} (\lambda) &=& \frac{1}{3} (4 \pi)^{1-d/2} \left( 2d \, (d+1) \p{2}{d/2-1}{-2\lambda}{} - 24d \, (d-1) \p{3}{d/2}{-2 \lambda}{} \right) \hspace*{1.5cm} \\[1.5ex]
\nonumber B_2^{~\prime} (\lambda) &=& - \, \frac{1}{6} \, (4 \pi)^{1-d/2} \left( 2d \, (d+1) \, \pt{2}{d/2-1}{-2\lambda}{} - 24d \, (d-1) \pt{3}{d/2}{-2\lambda}{} \right) 
\eea
Since these equations make no use of the fixed point values $\lambda^*$ and $g^*$ they can be used to investigate both the trivial and the non-trivial fixed point. The eigenvalues and right eigenvectors of ${\bf B}$, evaluated at the corresponding fixed point, then determine its critical exponents and scaling fields, respectively. Since, generically, ${\bf B}$ is not symmetric, its eigenvalues are not real and the eigenvectors are not orthogonal in general. We  define the stability coefficients $\theta^{\rm I}$, I=1,2, as the negative eigenvalues of ${\bf B}$ satisfying the equation ${\bf B} V^{\rm I} = - \theta^{\rm I} V^{\rm I}$, where $V^{\rm I}$ are the right eigenvectors of ${\bf B}$.
\end{subsubsection}
\begin{subsubsection}{The Trivial Fixed Point}
Substituting $\lambda^* = 0, g^* = 0$ into \rf{5.2} the stability matrix simplifies to
\be\label{5.7}
{\bf B}_{\rm GFP} = \left[ \begin{array}{cc} \hspace{0.6cm}-2 \hspace*{0.6cm}   & (4 \pi)^{1-d/2} \, d\,(d-3) \, \p{1}{d/2}{0}{} \\ \hspace{0.6cm} 0 \hspace*{0.6cm} & d-2 \end{array} \right]
\ee
Diagonalizing \rf{5.7} then leads to {\it two real} stability coefficients with their corresponding right eigenvectors:
\bea\label{5.8}
\nonumber \theta_1 = 2 \qquad &{\rm with}& \qquad V^1 = \left( \begin{array}{c} 1 \\ 0 \end{array}\right) \\[1.5ex]  
\theta_2 = 2-d \qquad &{\rm with}& \qquad V^2 = \left( \begin{array}{c}  (4 \pi)^{1-d/2} \, (d-3) \, \p{1}{d/2}{0}{} \\ 1 \end{array} \right) 
\eea

The results \rf{5.8} can be used to write down the linearized renormalization group flow of the coupling constants $\lambda_k, g_k$ in the vicinity of the Gaussian fixed point:
\bea\label{5.9a}
\lambda_k &=& \alpha_1 \, \frac{M^{2} }{k^2} + \alpha_2 \, (4 \pi)^{1-d/2} \, (d-3) \, \p{1}{d/2}{0}{} \, \frac{k^{d-2}}{M^{d-2} } + \cdots \\
\label{5.9b}
g_k &=& \alpha_2 \, \frac{k^{d-2}}{ M^{d-2}} + \cdots
\eea
Here $\alpha_1$ and $\alpha_2$ are constants of integration allowing to adjust the solution to given initial conditions. 

The equations \rf{5.9a} and \rf{5.9b} show that, in $d=4$, the $V^2$-direction of the Gaussian fixed point is attractive for $k\rightarrow 0$, leading to a vanishing $g_k$, while the $V^1$-direction is repulsive. The behavior of $\lambda_k$ crucially  depends on the sign of $\alpha_1$. For $\alpha_1<0$, $\alpha_1>0$ and $\alpha_1=0$, the trajectories start to the left, to the right, or on the $V^2$-axis. They will be referred to as trajectories of Type Ia, IIa and IIIa, respectively. The corresponding renormalization group behavior of $\lambda_k$ is summarized in TABLE \ref{one}. 
\begin{table}
\renewcommand{\baselinestretch}{1}
\begin{tabular}{ccc} 
\hspace*{0.5cm}Type\hspace*{0.5cm} & \hspace*{0.8cm}sign of $\alpha_1$ \hspace*{0.8cm}& \hspace*{1cm} asymptotic behavior \hspace*{1cm} \\ \hline 
Ia & $\alpha_1<0$ & $\alpha_1$-contribution dominates in \rf{5.9a}; \\
&& $\lambda_k \rightarrow -\infty$ in the limit $k\rightarrow 0$. \\[2ex]
IIa & $\alpha_1=0$ & $\alpha_2$-contribution dominates in \rf{5.9a}; \\
&& $\lambda_k$ is proportional to $k^{d-2}$: $\lambda_k \rightarrow 0$ in the limit $k\rightarrow 0$. \\[2ex]
IIIa & $\alpha_1>0$ &  $\alpha_1$-contribution dominates in \rf{5.9a}; \\
&& $\lambda_k \rightarrow +\infty$ in the limit $k\rightarrow 0$. \\  
\end{tabular}
\caption{\label{one}Renormalization group flow of $\lambda_k$ in the vicinity of the trivial fixed point, depending on the sign of $\alpha_1$.} 
\end{table}

For the dimensionful coupling constants the equations \rf{5.9a} and \rf{5.9b} read 
\bea\label{5.10}
\nonumber G_k &=& G_0 + \cdots\\[1.5ex]
\lb_k &=& \lb_0 + (4 \pi)^{1-d/2}\, (d-3) \, \p{1}{d/2}{0}{} \, G_0 \,  k^d + \cdots
\eea
Here we chose $\M=m_{\rm Pl}$ by setting $\alpha_2=1$, and we identified $\lb_0 = \alpha_1 \, m^2_{\rm Pl}$. From \rf{5.10} it is easy to see that both $G_k$ and $\lb_k$ run towards  constant but non-zero values in the limit $k \rightarrow 0$, unless we set $\alpha_1=0$ by hand. But since there is no compelling reason for $\alpha_1$ to be zero we see that $\lb_0$ depends on the free parameter $\alpha_1$ and therefore on the trajectory chosen. Hence the Gaussian fixed point does not determine the value of the cosmological constant $\lb_0$ in the infrared.
\end{subsubsection}
\begin{subsubsection}{The Non-trivial Fixed Point}
The existence of a non-trivial fixed point with $g^*\not=0$ implies $d-2+\eta_N(\lambda^*,g^*)=0$ in order for $\Fbeta_g$ to vanish. Using equation \rf{2.19} this relation can be solved for $g^*$ as a function of $\lambda^*$:
\be\label{5.11}
g^*(\lambda^*) =  \frac{d-2}{(d-2) \, B_2(\lambda^*) - B_1(\lambda^*)}
\ee
This equation can be used to eliminate the $g$-dependence of $\Fbeta_\lambda$ at the non-trivial fixed point, i.e. we have to solve $\Fbeta_\lambda(\lambda^*, g^*(\lambda^*))=0$ for $\lambda^*$. Due to its complicated structure this equation can only be solved numerically. For the $\Fbeta$-functions with sharp cutoff \rf{3.15} the numerical evaluation yields a non-Gaussian fixed point at
\be\label{5.12}
\lambda^* = 0.330, \qquad g^* = 0.403
\ee

In order to discuss its properties, we take the general stability matrix \rf{5.4} and substitute the condition for the non-trivial fixed point, $\eta_N(\lambda^*, g^*) = 2-d$. Using the sharp cutoff we find the following matrix entries:
\bea\label{5.13}
\nonumber \frac{\partial \, \beta_{\lambda}}{\partial \, \lambda} \bigg|_{\rm NGFP}^{\rm sc} &=& - d + \left( \lambda^* -  (4 \pi)^{1-d/2}\,  \frac{g^*}{2} \, \frac{d \, (d+1)}{\Gamma(d/2+1)} \right) \, \frac{g^*}{1 - g^*  B_2(\lambda^*)^{\rm sc} } \, \cdot\\
\nonumber && \cdot \frac{1}{3} \, (4 \pi)^{1-d/2} \left( \frac{2d \, (d+1)}{\Gamma(d/2-1)} \, \frac{1}{ 1-2\lambda^* } - \frac{12d \, (d-1)}{\Gamma(d/2)} \, \frac{1}{(1-2\lambda^* )^2} \right)\\
\nonumber && + (4 \pi)^{1-d/2}\,  g^* \, \frac{}{} \, \frac{2d \, (d+1)}{\Gamma(d/2)} \, \frac{1}{1- 2 \lambda^*} \\[1.5ex]
\nonumber \frac{\partial \, \beta_{\lambda}}{\partial \, g} \bigg|_{\rm NGFP}^{\rm sc}  &=& \left( 2-d \right) \left( \lambda^* - (4 \pi)^{1-d/2}\, \frac{g^*}{2} \, \frac{d \, (d+1)}{\Gamma(d/2+1)} \right) \left( \frac{1}{g^*} + \frac{ B_2(\lambda^*)^{\rm sc} }{1 - g^* B_2(\lambda^*)^{\rm sc} } \right) \\
\nonumber && \! \! \! \! + \frac{(4 \pi)^{1-d/2}}{2} \left( - \, \frac{2d \, (d+1)}{\Gamma(d/2)} \, \ln(1-2\lambda^*) + 2d \, (d-3) \, \varphi_{d/2} - \frac{d \, (d+1) \, (2-d)}{\Gamma(d/2+1)} \right) \\[1.5ex]
\nonumber \frac{\partial \, \beta_g}{\partial \, g}  \bigg|_{\rm NGFP}^{\rm sc} &=& \frac{(4 \pi)^{1-d/2} }{3} \, \frac{g^{*^2}}{1 - g^* \, B_2(\lambda)^{\rm sc} } \, \left( \frac{2d \, (d+1)}{\Gamma(d/2-1)} \, \frac{1}{ 1-2\lambda^* } - \frac{12d \, (d-1)}{\Gamma(d/2)} \, \frac{1}{(1-2\lambda^* )^2} \right) \\[1.5ex]
\frac{\partial \, \beta_g}{\partial \, g}  \bigg|_{\rm NGFP}^{\rm sc} &=& \left( 2-d \right) \left( 1 + \frac{g^* \, B_2(\lambda^*)^{\rm sc} }{1-g^* \, B_2(\lambda^*)^{\rm sc} } \right) 
\eea
$B_1(\lambda)^{\rm sc} $ and $B_2(\lambda)^{\rm sc}$ are given by equation \rf{3.17}. 
Substituting the numerical values of the non-trivial fixed point \rf{5.12} and diagonalizing the stability matrix leads to the following pair of {\it complex} stability coefficients:
\bea\label{5.14}
\nonumber \theta^1 \equiv& \theta^{\prime} + i \theta^{\prime \prime} &= 1.941 + i \, 3.147\\
\theta^2 \equiv& \theta^{\prime} - i \theta^{\prime \prime} &= 1.941 - i \, 3.147
\eea
 
In analogy to the equations \rf{5.9a} and \rf{5.9b} the  solution to the linearized flow equation in the vicinity of the non-trivial fixed point then reads
\bea\label{5.14a}
\left( \begin{array}{c} g(t) - g^* \\ \lambda(t) - \lambda^* \end{array} \right)= \alpha_1 \sin(- \theta^{\prime \prime}\,t) \, e^{- \theta^{\prime} t} \, \mbox{Re}(V^1) + \alpha_2 \cos(- \theta^{\prime \prime}\,t) \, e^{- \theta^{\prime} t} \, \mbox{Im}(V^1) 
\eea
Here one sees that due to the positive real-part of $\theta^I$ the non-trivial fixed point is UV attractive in both directions of the $g$-$\lambda-$plane, i.e. attractive for $t = \ln(k/\widehat{k}) \rightarrow \infty$. Furthermore, the non-zero imaginary part of the stability coefficient causes the trajectories to spiral into the non-Gaussian fixed point when $t \rightarrow \infty$.
 
We emphasize that the nonzero imaginary part of the stability coefficients is not an artifact of our singular cutoff but appears for the smooth exponential cutoff as well, as is shown in TABLE \ref{Fixpunktdaten}. 

Let us now turn to the investigation of the cutoff scheme dependence of the results obtained above. A change of the cutoff function $R^{(0)~{\rm sc}}$ generally leads to a change in the RG flow. By definition, universal quantities, i.e. quantities that should in principle be observable, are cutoff scheme independent in an exact treatment. However, an artificial scheme dependence can arise due to the approximations one has to make in all practical calculations. Analyzing the cutoff dependence of universal quantities therefore provides a useful tool for judging the quality of the truncation. One expects that a truncation which yields a good approximation of the exact RG flow leads to universal quantities which are fairly independent of the cutoff scheme used.
\begin{figure}[t]
\renewcommand{\baselinestretch}{1}
\epsfxsize=0.49\textwidth
\begin{center}
\leavevmode
\epsffile{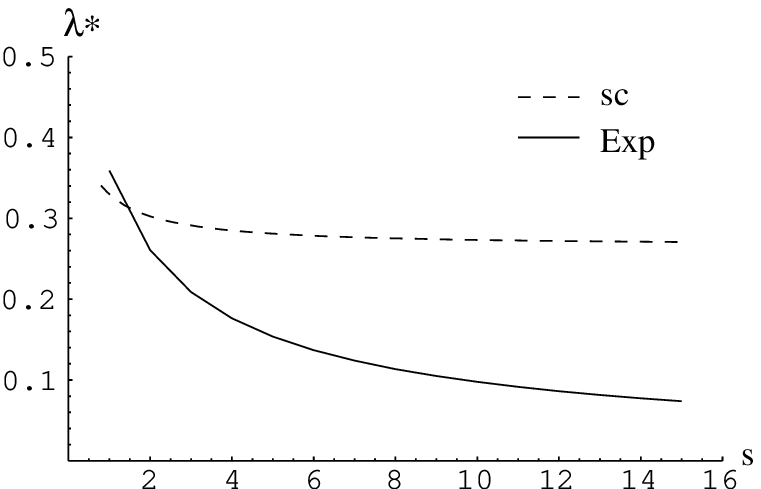}
\epsfxsize=0.48\textwidth
\epsffile{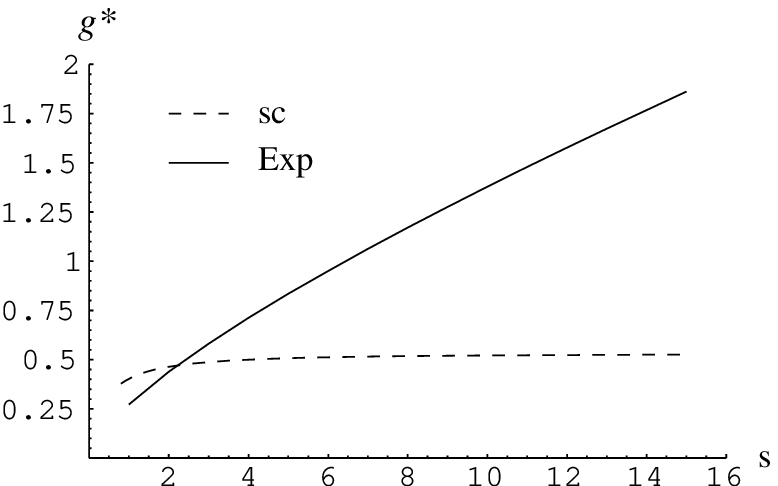}
\end{center} 
\epsfxsize=0.49\textwidth
\begin{center}
\leavevmode
\epsffile{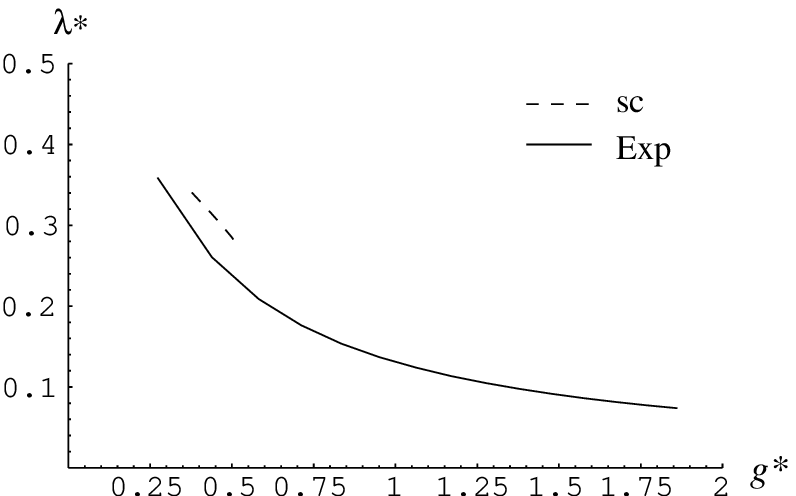}
\epsfxsize=0.48\textwidth
\epsffile{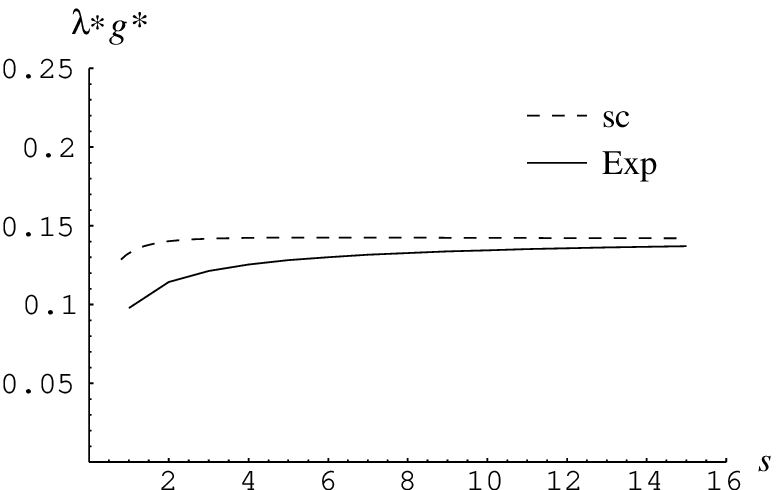}
\end{center} 
\parbox[c]{\textwidth}{\caption{\label{sechs}{\footnotesize The location of the non-trivial fixed point depending on the shape parameter $s$. While $\lambda^*$ and $g^*$ show a fairly strong cutoff scheme dependence, their product $\lambda^* g^*$ is constant with a remarkable precision, approaching the same ``plateau'' values of $\lambda^* g ^*$ for large values of $s$. In the third diagram $\lambda^*$ is plotted against $g^*$. Both of the two lines start at $s=0.8$ and end at $s=15$. From the different lengths of the lines for the sharp and the exponential cutoff one sees that the sharp cutoff leads to a much smaller $s$-dependence than its exponential counterpart.}}}
\end{figure}

Natural candidates for testing the quality of the Einstein-Hilbert truncation are the stability coefficients $\theta^{\prime}$ and $\theta^{\prime \prime}$ \cite{sumi}. Furthermore, one can argue that the product $\lambda^* g^*$ should be universal, too \cite{oliver,nin}. The argument is as follows: The functions $k \mapsto g_k, \lambda_k$ and their UV limits $g^*$ and $\lambda^*$ depend on the cutoff operator $R_k$ and are not directly observable therefore. While $k$ and, as a consequence, $G_k$ and $\bar{\lambda}_k$ at a prescribed value of $k$ cannot be measured separately, we may invert the function $k \mapsto G_k$ and insert the result $k = k(G)$ into $\bar{\lambda}_k$. This leads to a relationship between the Newton constant and the cosmological constant which, at least in principle, could be tested experimentally: $\bar{\lambda}=\bar{\lambda}(G)$. In general this relation depends on the renormalization group trajectory chosen. But in the fixed point regime all trajectories approach $\bar{\lambda}_k = \lambda^* k^2$ and $G_k = g^* / k^2$ which gives rise to $\bar{\lambda}(G) = g^* \lambda^*/G$ for $\bar{\lambda} \gg m_{\rm Pl}^2$ and $G \ll m_{\rm Pl}^{-2}$. Assuming that $\bar{\lambda}$ and $G$ have the status of observable quantities, this relation shows that $g^*\lambda^*$ should be observable, and hence $R_k$-independent.

Therefore we also include the quantity $\lambda^* g^*$ when we plot the numerical values for $g^*$ and $\lambda^*$ obtained from the exponential and the sharp cutoff. The results are shown in FIG. \ref{sechs}. 

Here one finds that the non-trivial fixed point exists in the entire region under investigation, $0.8<s<15$. While the numerical values for $\lambda^*(s)$ and $g^*(s)$ vastly differ for the various values of $s$ and between the two families of cutoff functions one finds that the product $\lambda^*(s) g^*(s)$ is almost $s$-independent and that its values for both types of cutoff functions are very similar. For the ``plateau'' values of $\lambda^* g^*$ we find with the different functions $R^{(0)}$ employed in this paper: 
\be\label{5.14c}
\lambda^*g^* = 0.14 \; \mbox{for TYPE A, sc, $\alpha = 1$}, \qquad \lambda^*g^* = 0.14 \; \mbox{for TYPE A, Exp, $\alpha = 1$} 
\ee
The analogous results found in \cite{oliver} using the TYPE B cutoff scheme are:
\be\label{5.14d}
\lambda^*g^* = 0.12 \; \mbox{for TYPE B, Exp, $\alpha = 1$}, 
\qquad  \lambda^*g^* = 0.14 \; \mbox{for TYPE B, Exp, $\alpha = 0$} 
\ee 
We see that these values are very close to our present findings. In particular, the differences between the TYPE A and TYPE B cutoff is comparable to the difference between the gauge fixing parameters $\alpha = 0$ and $\alpha=1$ \cite{oliver,souma}. This supports the assumption that the product $\lambda^* g^*$ has a universal meaning. The rather small numerical deviations of the plateau height can be attributed to using only a very simple truncation. 
\begin{table}
\renewcommand{\baselinestretch}{1}
\begin{tabular}{c|ccccc|ccccc}
& \multicolumn{5}{c}{sharp cutoff} & \multicolumn{5}{c}{exponential cutoff} \\
$s$ & $\lambda^*$ & $g^*$ & $\lambda^* g^*$ & $\theta^{\prime}$ & $\theta^{\prime \prime}$
& $\lambda^*$ & $g^*$ & $\lambda^* g^*$ & $\theta^{\prime}$ & $\theta^{\prime \prime}$
\\ \hline
0.8 & 0.340 & 0.378 & 0.129 & 2.141 & 3.438 & 0.390 & 0.233 & 0.091 & 1.376 & 4.710 \\[1.2ex] 
1   & 0.330 & 0.403 & 0.133 & 1.941 & 3.147 & 0.359 & 0.272 & 0.098 & 1.422 & 4.307 \\[1.2ex]
5   & 0.281 & 0.507 & 0.143 & 1.348 & 2.743 & 0.154 & 0.834 & 0.128 & 1.499 & 3.224 \\[1.2ex] 
10  & 0.273 & 0.521 & 0.142 & 1.294 & 2.654 & 0.098 & 1.378 & 0.144 & 1.518 & 3.111 \\[1.2ex] 
30  & 0.268 & 0.529 & 0.142 & 1.270 & 2.592 & 0.044 & 3.149 & 0.140 & 1.562 & 3.064 \\
\end{tabular}
\caption{\label{Fixpunktdaten} \footnotesize Comparison between the numerical results for the non-trivial fixed point using the sharp and exponential cutoff with different values for the shape parameter $s$.}
\end{table} 

In order to round up our survey of the properties of the non-trivial fixed point in $d=4$, we now treat $\varphi_1$ and $\varphi_2$ as two independent, positive numbers obeying no further constraints. The resulting numerical values for $\lambda^*, g^*, \theta^{\prime}$ and $\theta^{\prime \prime}$ are shown in FIG. \ref{sieben}. 
\begin{figure}[bt]
\renewcommand{\baselinestretch}{1}
\epsfxsize=0.49\textwidth
\begin{center}
\leavevmode
\epsffile{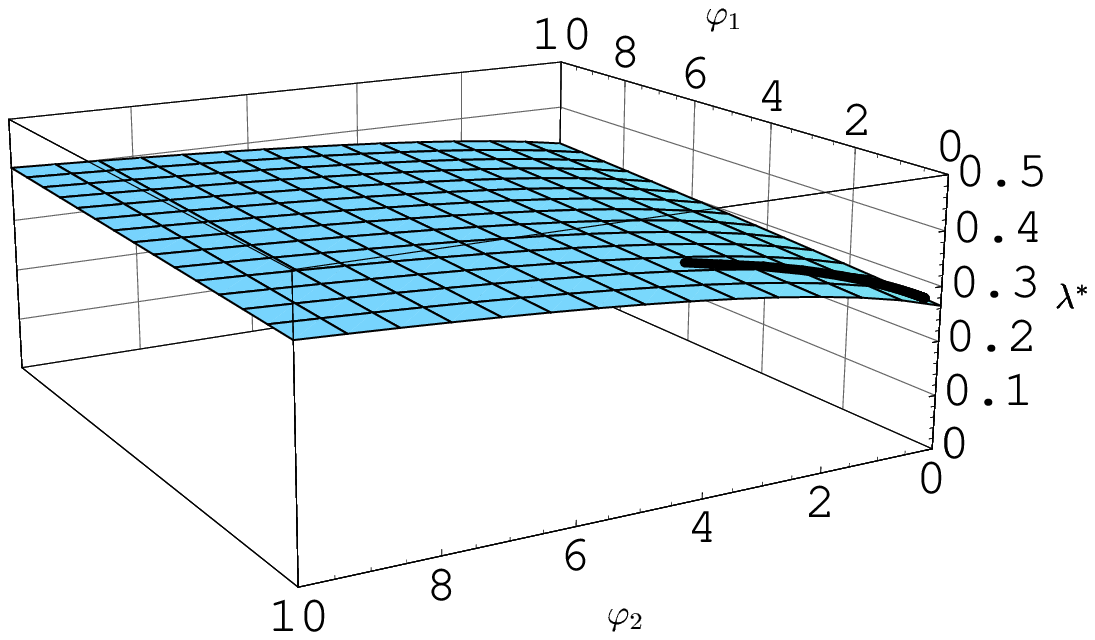}
\epsfxsize=0.48\textwidth
\epsffile{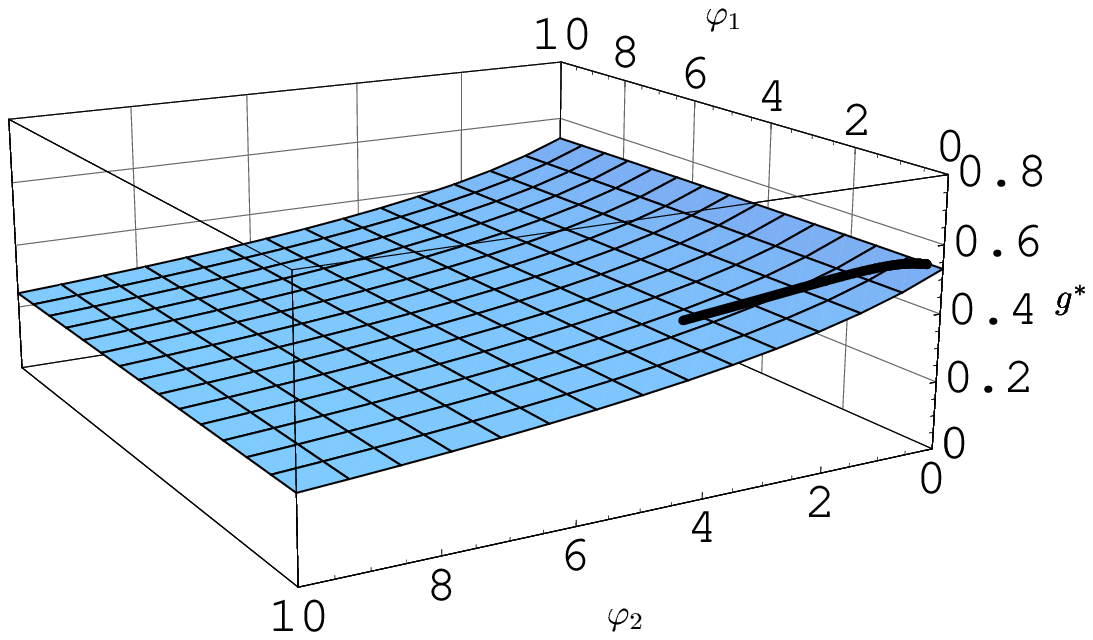}
\end{center}
\begin{center}
\epsfxsize=0.48\textwidth
\leavevmode
\epsffile{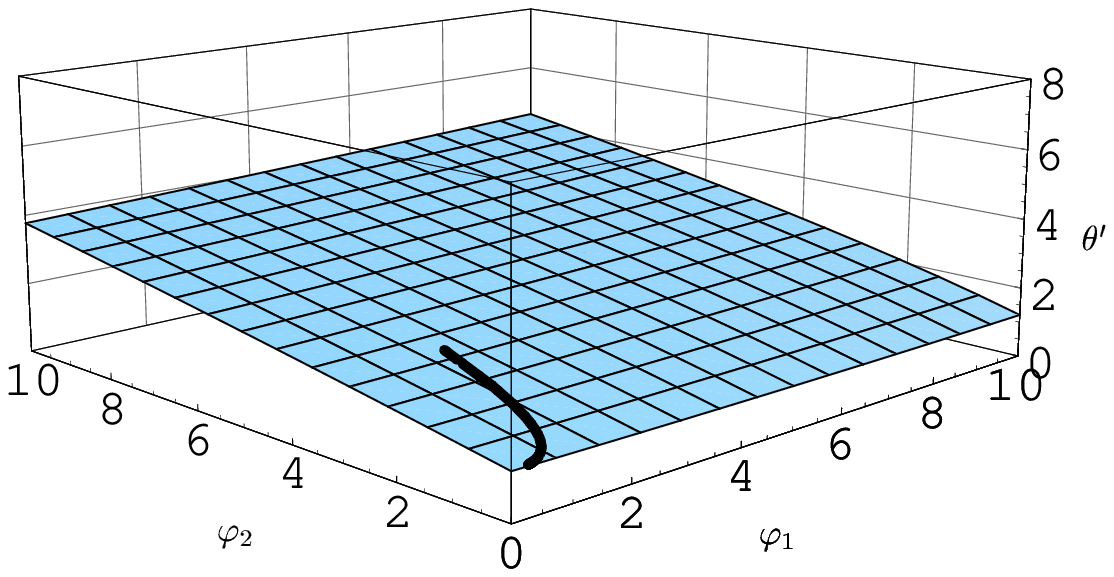}
\epsfxsize=0.48\textwidth
\epsffile{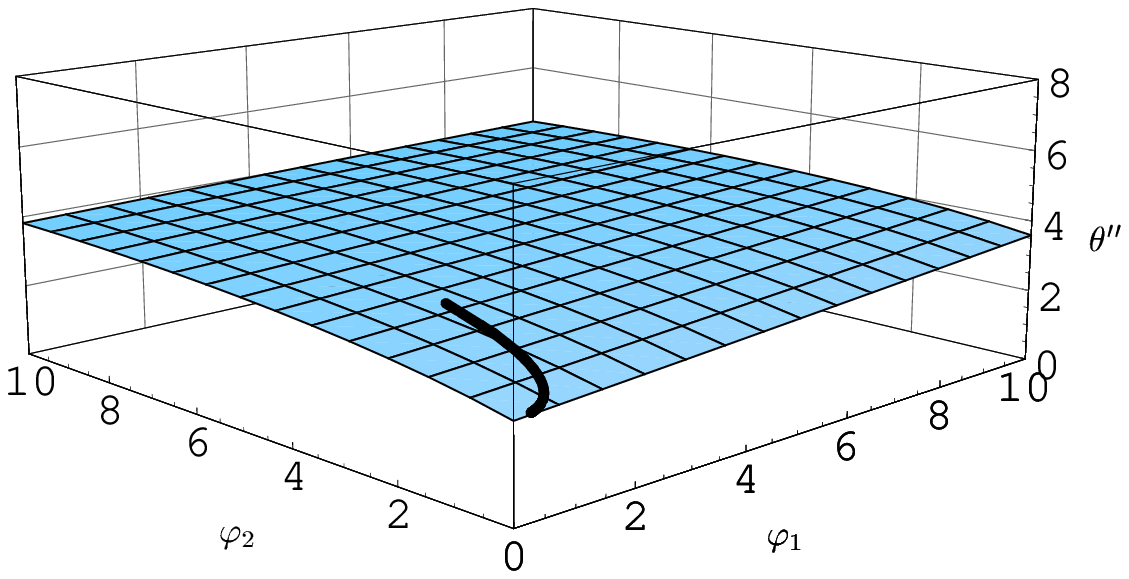}
\end{center}
\parbox[c]{\textwidth}{\caption{\label{sieben}{\footnotesize Numerical results for the location and stability coefficients of the non-trivial fixed point, treating $\varphi_1$ and $\varphi_2$ as independent. The solid line corresponds to the results obtained by considering $\varphi_1$ and $\varphi_2$ as a function of the shape parameter $s$. For all values of $\varphi_1$ and $\varphi_2$ the non-Gaussian fixed point is seen to have the same qualitative features.}}}
\end{figure}
 
This analysis clearly shows the existence of the non-trivial fixed point for the complete region under investigation, $0<\varphi_1, \varphi_2 < 10$. Even though the magnitude of the stability coefficients varies by about a factor of three, as an effect of our truncation, the qualitative properties of the fixed point are the same for all values of $\varphi_1$ and $\varphi_2$. The fixed point always possesses a pair of complex-conjugate stability coefficients whose real-part is positive, ${\rm Re}(\theta^I) \equiv \theta^\prime > 0$, i.e. the fixed point is always UV-attractive. This provides us with further evidence for the existence of a non-trivial fixed point in the full theory \cite{oliver,oliver2,souma}. 
\end{subsubsection}
\end{subsection}
\begin{subsection}{The Phase Portrait}
Motivated by the very good agreement between the exponential and the sharp cutoff found when solving the $\M$-scaled flow equation in Section III and analyzing the non-trivial fixed point, we now investigate the full $\lambda$-$g-$parameter space, using the $\Fbeta$-functions \rf{3.15} with the technically much more convenient sharp cutoff. 
\begin{figure}[t]
\renewcommand{\baselinestretch}{1}
\epsfxsize=8.5cm
\centerline{\epsfbox{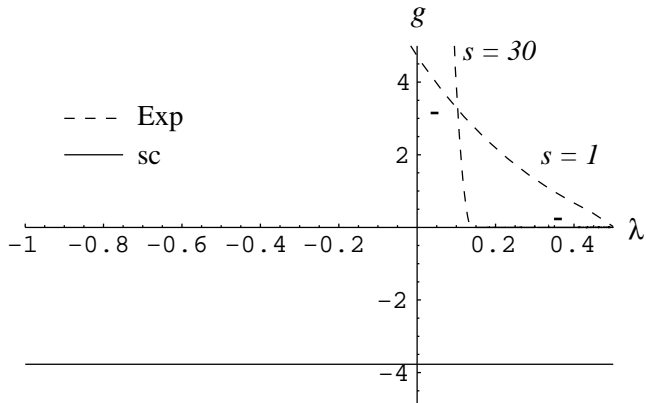}}
\parbox[c]{\textwidth}{\caption{\label{acht}{\footnotesize The points in the parameter space where $\eta_N$ diverges for the sharp and exponential cutoff. The bars indicate the location of the non-trivial fixed point for the exponential cutoff with shape parameters $s=1$ and $s=30$.}}}
\end{figure}

But before presenting the numerical solutions, we summarize the properties of the parameter space that can be directly read off from \rf{3.15}. As has been noted in the context of \rf{4.4}, the flow equation is singular for $\lh \ge y/2$. In terms of the $k$-scaled coupling constant $\lambda_k = \lh(y)/\y$ this singularity occurs for $\lambda \ge 1/2$, resulting in a boundary of the $\lambda$-$g-$parameter space given by the line $\lambda=1/2$. 

The vanishing of $\Fbeta_g(\lambda, g)$ for $g=0$ and arbitrary $\lambda$ leads to a separation of the phase space into two decoupled regions with positive and negative coupling $g$, respectively. Trajectories starting in one of these regions will never cross the separation line $g=0$. 

The characteristics of the trivial and the non-trivial fixed point have already been discussed in the previous subsection. One expects that the non-Gaussian fixed point dominates the RG flow for large $t=\ln(k/\hat{k})$, leading to a spiraling in of the trajectories on the point $(\lambda^*, g^*)$. The stability axis $V^2$ of the Gaussian fixed point will provide a separation between the trajectories which run towards $\lambda_k = -\infty$ and $\lambda_k = +1/2$, respectively.

In terms of the coordinates $\lambda, g$ it is also easy to visualize the singularity of the anomalous dimension $\eta_N$ encountered in Section III. The function $\eta_N(\lambda, g)$ diverges at those points $(g, \lambda)$ at which the denominator of equation \rf{2.19} vanishes. The function $g(\lambda)$ defined by $1 - g(\lambda) B_2(\lambda)=0$ is shown in FIG. \ref{acht}. For the sharp cutoff one finds that this singularity occurs on a $s$-independent horizontal line at $g=- 6 \pi/5$. For the exponential cutoff the divergence of $\eta_N^{\rm Exp}$ leads to a ($s$-dependent) line in the first quadrant of the $\lambda$-$g$--plane. 
\begin{figure}[t]
\renewcommand{\baselinestretch}{1}
\epsfxsize=0.99 \textwidth
\centerline{\epsfbox{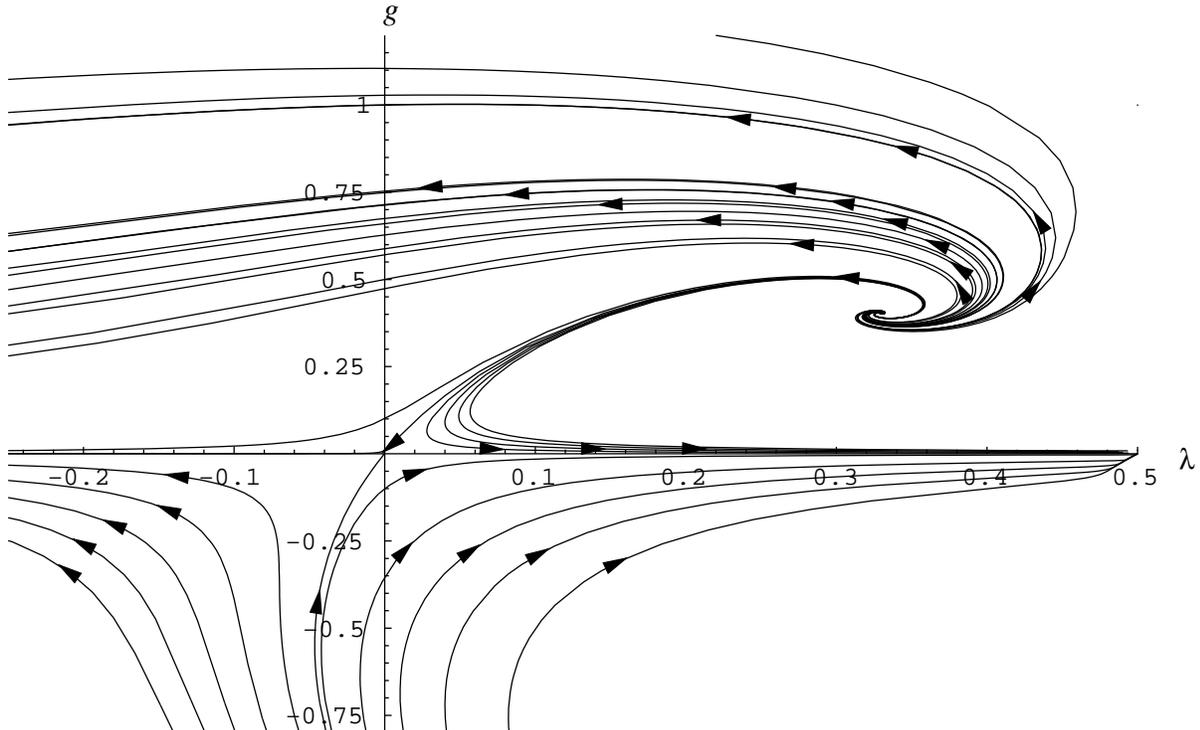}}
\parbox[c]{\textwidth}{\caption{\label{neun}{Part of the parameter space with its RG flow. The arrows along the trajectories point in the direction of the renormalization group flow, i.e. towards decreasing values of $k$. The flow pattern is dominated by a non-Gaussian fixed point in the first quadrant and a trivial one at the origin.}}}
\end{figure}

We shall now derive the full phase portrait of the Einstein-Hilbert truncation by numerically solving the autonomous differential equation for the ``$k$-scaled'' quantities $g_k$ and $\lambda_k$. In the first step we focus on the part of the $\lambda$-$g$--plane where the RG flow is governed by the two fixed points found in the previous subsection.

The resulting flow diagram is shown in FIG. \ref{neun}. 

This figure clearly shows the separation between the trajectories with positive and negative coupling $g$. 

The most interesting feature of this diagram is the interplay between the non-Gaussian and the Gaussian fixed point in the positive coupling region. In the limit $t \rightarrow \infty$ the non-Gaussian fixed point completely dominates the flow of all trajectories in this region, leading to their spiraling in on the point $(\lambda^*, g^*)$. This result exactly matches the behavior expected from the complex stability coefficients found in the previous subsection. The behavior of the trajectories in the vicinity of the trivial fixed point confirms TABLE \ref{one}: Trajectories starting to the left of the separation line $(\alpha_1<0)$ are running towards $\lambda \rightarrow -\infty$ while those to the right $(\alpha_1>0)$ terminate at the border singularity $\lambda = 1/2$ at finite values of $k$. 

The trajectory separating these two regions is of special interest and will be called ``the'' separatrix. It connects the non-trivial fixed point in the UV to the trivial fixed point for $k \rightarrow 0$. As a result, this trajectory leads to a {\it vanishing} renormalized cosmological constant $\lb_0$. 

The trajectories running to the left of the separatrix also possess a well defined limit $k \rightarrow 0$. They lead to negative values of the renormalized cosmological constant $\lb_0$.

\begin{figure}[t]
\renewcommand{\baselinestretch}{1}
\epsfxsize=0.99 \textwidth
\centerline{\epsfbox{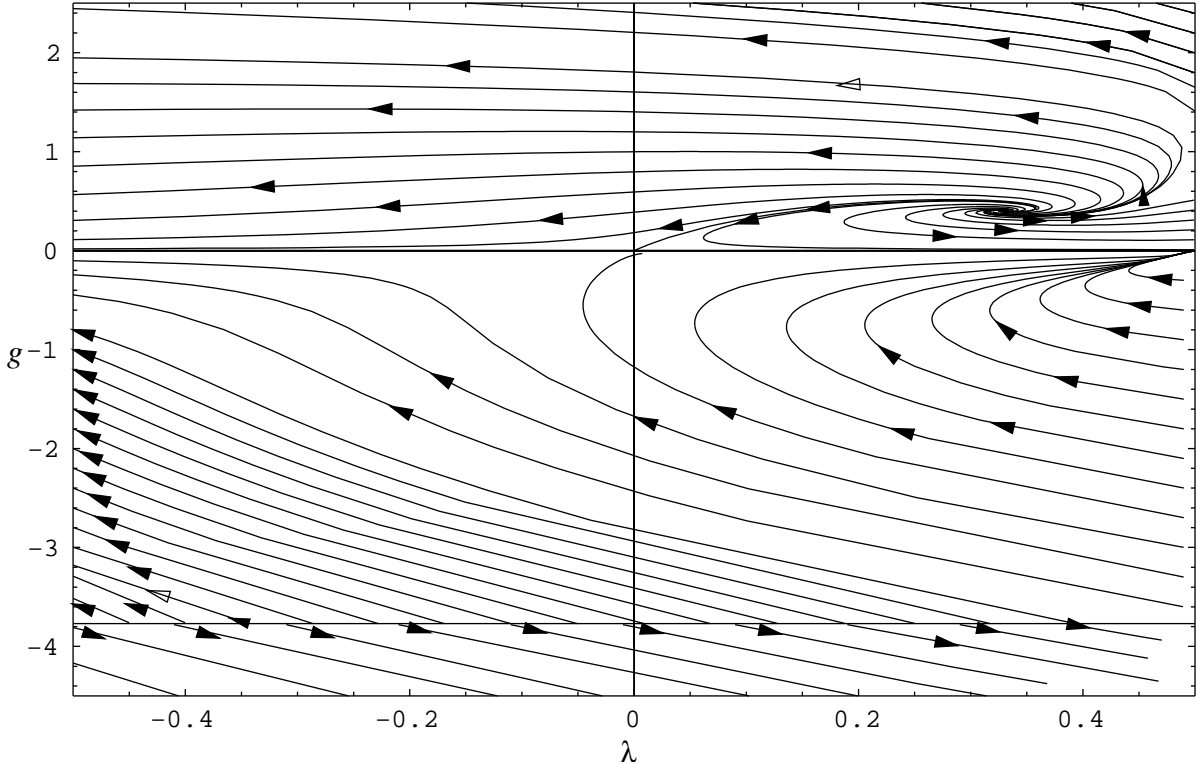}}
\parbox[c]{\textwidth}{\caption{\label{zehn}{Full phase structure of the Einstein-Hilbert truncated theory. On the bold horizontal line $\eta_N^{\rm sc}$ diverges.}}}
\end{figure}

In the next step we now extend our numerical survey to the full parameter space. The resulting renormalization group flow is shown in FIG. \ref{zehn}. Here it becomes obvious that the negative-$g$ region also contains a trajectory which separates the regions with trajectories running towards $\lambda \rightarrow -\infty$ and $\lambda \rightarrow 1/2$. Furthermore one sees that the singularity of $\eta_N^{\rm sc}$ at $g=-6 \pi / 5$ results in a separation between trajectories showing a screening and anti-screening behavior of the Newton constant $g_k$ in the IR. This resembles the behavior found for the exponential cutoff in Section III.
\begin{figure}[t]
\renewcommand{\baselinestretch}{1}
\epsfxsize=12cm
\centerline{\epsfbox{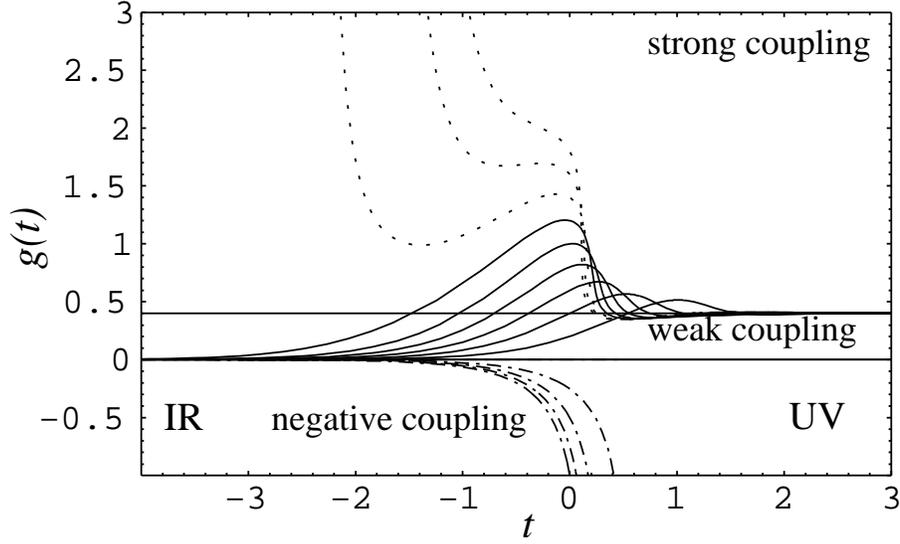}}
\parbox[c]{\textwidth}{\caption{\label{elf}{\footnotesize Projecting the full renormalization group flow onto the $g$-axis leads to a separation between the trajectories in the strong, weak and negative coupling region.}}}
\end{figure}

In a first step of classifying the trajectories found in FIG. \ref{zehn}, we project the renormalization group trajectories of the {\it full system} onto the $g$-axis. The results are displayed in FIG. \ref{elf}. According to their limit for $k \rightarrow 0$ or $t=\ln(k/\widehat{k}) \rightarrow -\infty$ three different classes of trajectories can be distinguished:
\begin{enumerate}
\item Trajectories with $\lim\limits_{t \to -\infty} g(t) \rightarrow \infty$. They form the ``strong coupling region''.
\item Trajectories with $\lim\limits_{t \to -\infty} g(t) = 0$. They form the ``weak coupling region''.
\item Trajectories with $\lim\limits_{t \to -\infty} g(t) < 0$. They form the ``negative coupling region''. 
\end{enumerate}
Note the oscillating behavior of $g(t)$ before the trajectories adopt their asymptotic value. This is caused by the nonzero imaginary part of the stability coefficients found for the non-trivial fixed point\footnote{A plot of $g(t)$ similar to FIG. \ref{elf} has been given in ref. \cite{souma}, see fig. 2 there. In this reference the running of $\lambda$ has been disregarded, however. As a consequence, no oscillations were found.}. 
\begin{figure}[t]
\renewcommand{\baselinestretch}{1}
\epsfxsize=0.99 \textwidth
\centerline{\epsfbox{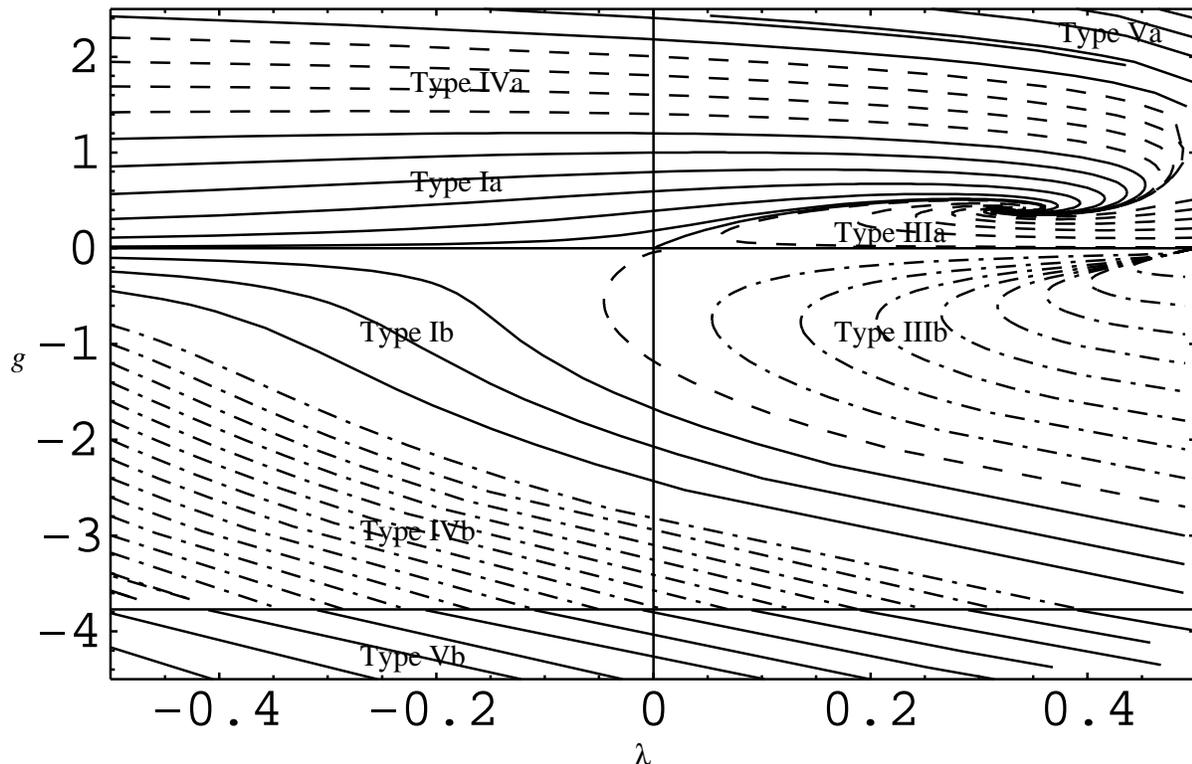}}
\parbox[c]{\textwidth}{\caption{\label{zw"olf}{\footnotesize Classification of the {\it full} Einstein-Hilbert truncated theory space. The trajectories IIa and IIb end at the trivial fixed point and separate the regions Ia and IIIa, Ib and IIIb, respectively. They are not labeled explicitly.}}}
\end{figure}

In a second step we classify the trajectories shown in FIG. \ref{zehn} according to their starting and end points. FIG. \ref{zw"olf} shows the resulting phase space regions, which are distinguished by a different kind of dashing of their trajectories. 

The characteristics of each region are summarized in TABLE \ref{three} which contains the classification of all trajectories occurring in the Einstein-Hilbert truncation. 
\begin{table}
\renewcommand{\baselinestretch}{1} \small \normalsize
\caption{\label{three}Classification of all RG trajectories occurring in the Einstein-Hilbert truncation.}
\begin{tabular}{cccc} 
\raisebox{-2mm}{Type} & \raisebox{-2mm}{UV-behavior} & \raisebox{-2mm}{IR-behavior} & \raisebox{-2mm}{coupling region} \\[1.5ex] \hline 
Ia & NGFP $(\lambda^*, g^*)$ & o.k. \parbox{1.4cm}{\[ \begin{array}{c} \lambda_k \rightarrow -\infty \\ g_k \rightarrow 0^+ \end{array}  \]} & weak coupling \\ 
\parbox{2cm}{\begin{tabular}{c} IIa \\ separatrix \end{tabular}} & NGFP $(\lambda^*, g^*)$ & GFP \parbox{1.4cm}{\[ \begin{array}{c} \lambda_k \rightarrow 0^+ \\ g_k \rightarrow 0^+  \end{array}  \]} & weak coupling \\ 
IIIa & NGFP $(\lambda^*, g^*)$ & Sing. \parbox{1.2cm}{\[ \begin{array}{c} \lambda = 1/2 \\ g > 0 \end{array}  \]} & weak coupling \\[1.5ex]
IVa & NGFP $(\lambda^*, g^*)$ & Sing. \parbox{1.2cm}{\[ \begin{array}{c} \lambda_k \rightarrow -\infty \\  g_k \rightarrow \infty \end{array}  \]} & strong coupling \\
Va & Sing. \parbox{2cm}{\[ \begin{array}{c} \lambda = 1/2 \\ g > 0 \end{array}  \]} & Sing. \parbox{1.2cm}{\[ \begin{array}{c} \lambda_k \rightarrow -\infty \\ g_k \rightarrow \infty \end{array}  \]} & strong coupling \\[1.5ex]
Ib & \hspace*{1mm} Sing. \parbox{2cm}{\[ \begin{array}{c} \lambda = 1/2 \\ g < 0 \end{array}  \]} & o.k. \parbox{1.2cm}{\[ \begin{array}{c} g_k \rightarrow 0^- \\ \lambda_k \rightarrow -\infty \end{array}  \]} & negative coupling \\
IIb & \hspace*{1mm} Sing. \parbox{2cm}{\[ \begin{array}{c} \lambda = 1/2 \\ g < 0 \end{array}  \]} & GFP \parbox{1.4cm}{\[ \begin{array}{c} \lambda_k \rightarrow 0^- \\ g_k \rightarrow 0^- \end{array}  \]} & negative coupling \\
IIIb & \hspace*{1mm} Sing. \parbox{2cm}{\[ \begin{array}{c} \lambda = 1/2 \\ g < 0 \end{array}  \]} & Sing. \parbox{1.2cm}{\[ \begin{array}{c} \lambda = 1/2 \\ g < 0 \end{array}  \]} & negative coupling \\
IVb & \hspace*{1mm} Sing. \parbox{2cm}{\[ \begin{array}{c} \lambda \\ g = -6 \, \pi/5 \end{array}  \]} & o.k. \parbox{1.2cm}{\[ \begin{array}{c} \lambda_k \rightarrow -\infty \\ g_k \rightarrow 0^- \end{array}  \]} & negative coupling \\
Vb & \hspace*{1mm} Sing. \parbox{2cm}{\[ \begin{array}{c} \lambda \\ g = -6\, \pi/5 \end{array}  \]} & Sing. \parbox{1.2cm}{\[ \begin{array}{c} \lambda = 1/2 \\ g < -6 \, \pi/5 \end{array}  \]} & negative coupling \\
\end{tabular}
\end{table}
TABLE \ref{three} is organized as follows: The first column labels the type of the trajectory as it is marked in the phase space diagram FIG. \ref{zw"olf}. (Only the {\it single} trajectories of Type IIa and IIb separating the regions Ia and IIIa, and Ib and IIIb, respectively, are not marked explicitly in this diagram.) The columns labeled ``UV-'' and ``IR-behavior'' indicate the characteristic features of the trajectories, ``UV'' referring to the endpoint of the trajectories for $k \rightarrow \infty$ and ``IR'' relating to $k \rightarrow 0$. These limits do not exist for all the classes. The aborting of the trajectory at a finite value of $k$ in either the UV or the IR is indicated by ``Sing.''. The values of $\lambda, g$ given in the table indicate where the corresponding RG trajectories end. 

In the column ``UV-behavior'' the label ``NGFP'' means that the trajectory runs into the non-Gaussian fixed point. The trajectories labeled with ``Sing.'' either end at the boundary line $\lambda = 1/2$ where we distinguish between the two regions $g>0$ and $g<0$ or in the singular line caused by the divergence of $\eta_N^{\rm sc}$ at $g = -6 \pi /5$ and arbitrary values of $\lambda$. Note that the limit $k \rightarrow \infty$ exists only for the trajectories running into the ``NGFP''.

In the IR, trajectories which possess an IR-limit $k\rightarrow 0$ are indicated by the note ``o.k.''. They lead to negative values of the renormalized cosmological constant $\lb_0$ with positive $(g_k \rightarrow 0^+)$ or negative $(g_k \rightarrow 0^-)$ Newton constant, respectively. The label ``GFP'' indicates that the corresponding trajectory ends at the Gaussian fixed point, yielding a vanishing cosmological constant $\lb_0$. Singular behavior, i.e. the termination of the trajectory at a finite value of $k$, appears when the trajectory reaches the boundary $\lambda = 1/2$ in the positive $(g >0)$ or negative $(0 > g > -6 \pi/5$, $g < -6 \pi/5)$ coupling region. The possibility of $g_k$ diverging at a finite value of $k$ is indicated by $g_k \rightarrow \infty, \lambda_k \rightarrow -\infty$.

The column ``coupling region'' finally specifies to which coupling region, as defined in FIG. \ref{elf}, the trajectory belongs.

The Newtonian regime consists of the trajectories in the weak coupling region. They are the only ones yielding a finite, positive Newton constant $G_k$ in the IR. In this respect it is important to note that only the trajectories of the Types Ia and IIa can be continued down to $k = 0$, leading to a finite value of $G_0$ and a negative or vanishing cosmological constant, respectively. Since for these two classes the limit $k \rightarrow \infty$ is also well-defined, these solutions could possibly be used in order to define a fundamental quantum theory of gravity with a vanishing or a negative renormalized cosmological constant, respectively. The trajectories of Type IIIa, which run towards positive values of $\lb$, terminate at the singularity $\lambda = 1/2$ and do not give rise to well-defined renormalized parameters $\lb_0$ and $G_0$ at $k=0$.   
\end{subsection}
\begin{subsection}{Crossover Behavior}
The phase diagram FIG. \ref{neun} shows that the trajectories Ia, IIa and IIIa which are relevant for the Newtonian limit of quantum gravity cross over from the non-trivial fixed point in the UV to the basin of attraction of the trivial fixed point in the IR. In this subsection we first concentrate on the classes Ia and IIa. They lead to finite $k \rightarrow 0$ limits of the coupling constants. Since these trajectories extend down to $k = 0$ they allow us to identify the scale $\M$ with the Planck-scale $m_{\rm Pl} = G_0^{-1/2}$. 

Due to the trivial and the non-trivial fixed point governing the RG flow of these trajectories in the limits $k \rightarrow 0$ and $k \rightarrow \infty$, respectively, the trajectories of these classes obey simple scaling laws in these regimes. 

In the UV one finds the following scale dependence of the coupling constants, valid for both types of trajectories: $\lambda_k \rightarrow \mbox{const}, g_k \rightarrow \mbox{const}$ $\Leftrightarrow$ $G_k, \Gh \propto k^{-2}, \quad \lb_k, \lh \propto k^2$. 

In the IR the trivial fixed point leads to the scaling laws given by the equations \rf{5.9a} and \rf{5.9b}. For Type Ia trajectories they are $g_k \propto k^2 \Leftrightarrow G_k, \Gh \propto k^0, \quad \lambda_k \propto k^{-2} \Leftrightarrow \lb_k, \lh \propto k^0$, and for the Type IIa they read $g_k \propto k^2 \Leftrightarrow G_k, \Gh \propto k^0, \quad \lambda_k \propto k^{2} \Leftrightarrow \lb_k, \lh \propto k^4$.  

Next we determine the scale $\kh$, at which the transition between these regimes takes place. Thanks to the identification $M=m_{\rm Pl}$ the related dimensionful quantity $k = \kh \, m_{\rm Pl}$ has a clear physical interpretation. We solve equation \rf{3.15} with the ``final condition'' $\Gh(k=0)=1$, together with $\lh(k=0) = -0.01$ and $\lh(k=0) = 0$ for a typical trajectory of Type Ia and IIa, respectively. This leads to the results displayed in FIGs. \ref{dreizehn} and \ref{vierzehn}.
\begin{figure}[tp]
\renewcommand{\baselinestretch}{1}  
\epsfxsize=0.49\textwidth
\begin{center}
\leavevmode
\epsffile{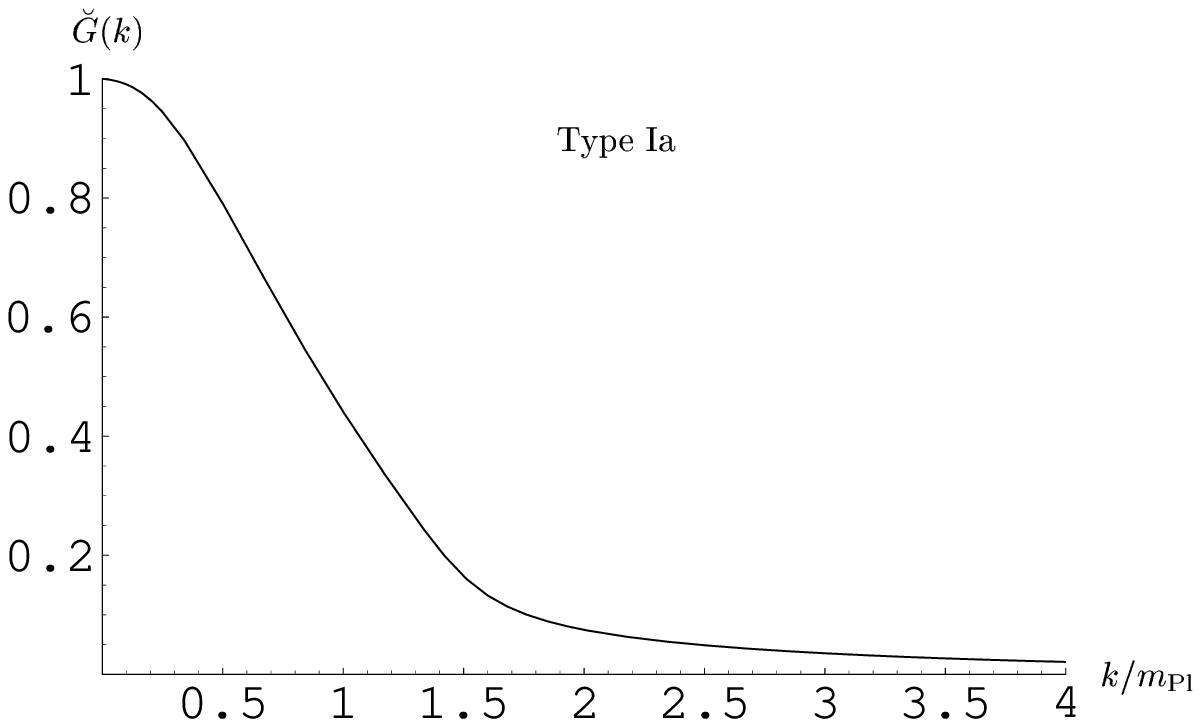}
\epsfxsize=0.49\textwidth
\epsffile{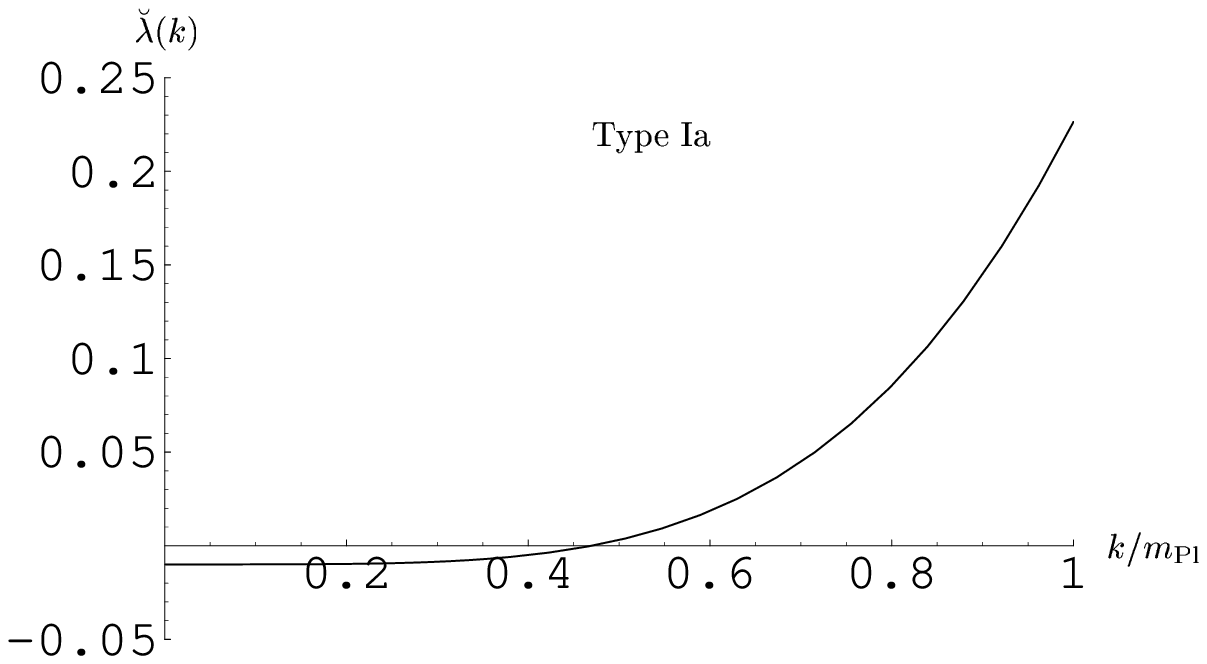}
\end{center}
\vspace*{1.5ex}
\begin{center}
\leavevmode
\epsfxsize=0.49\textwidth 
\epsffile{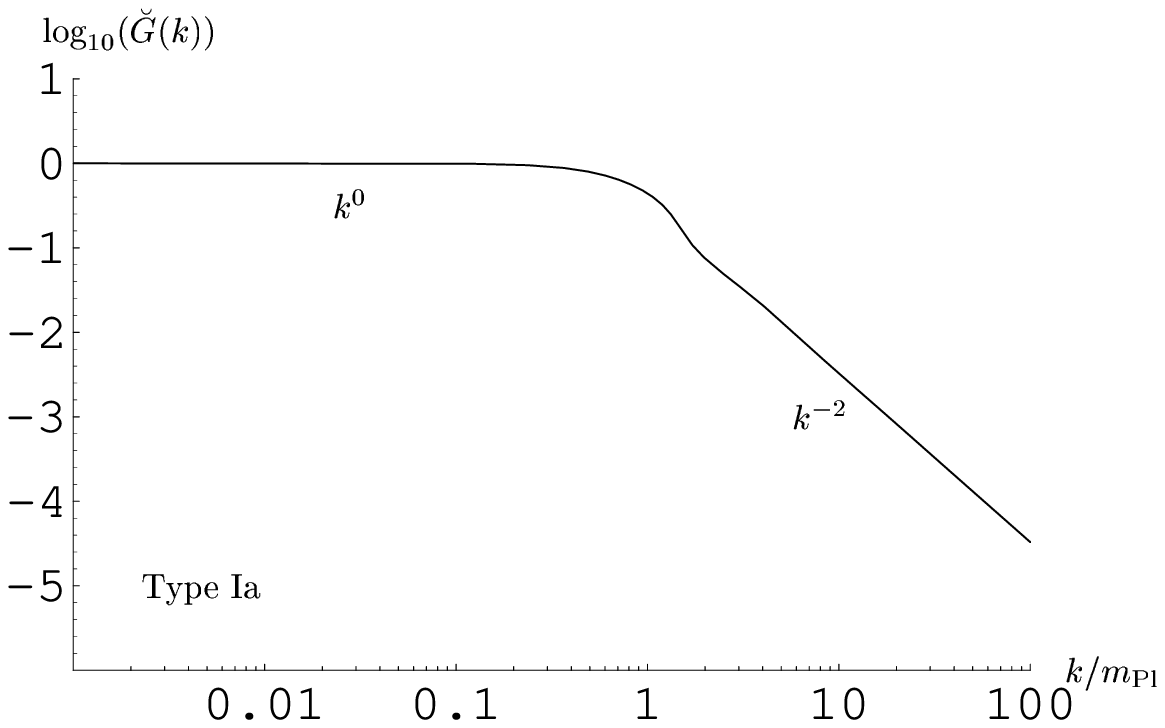}
\epsfxsize=0.49\textwidth
\epsffile{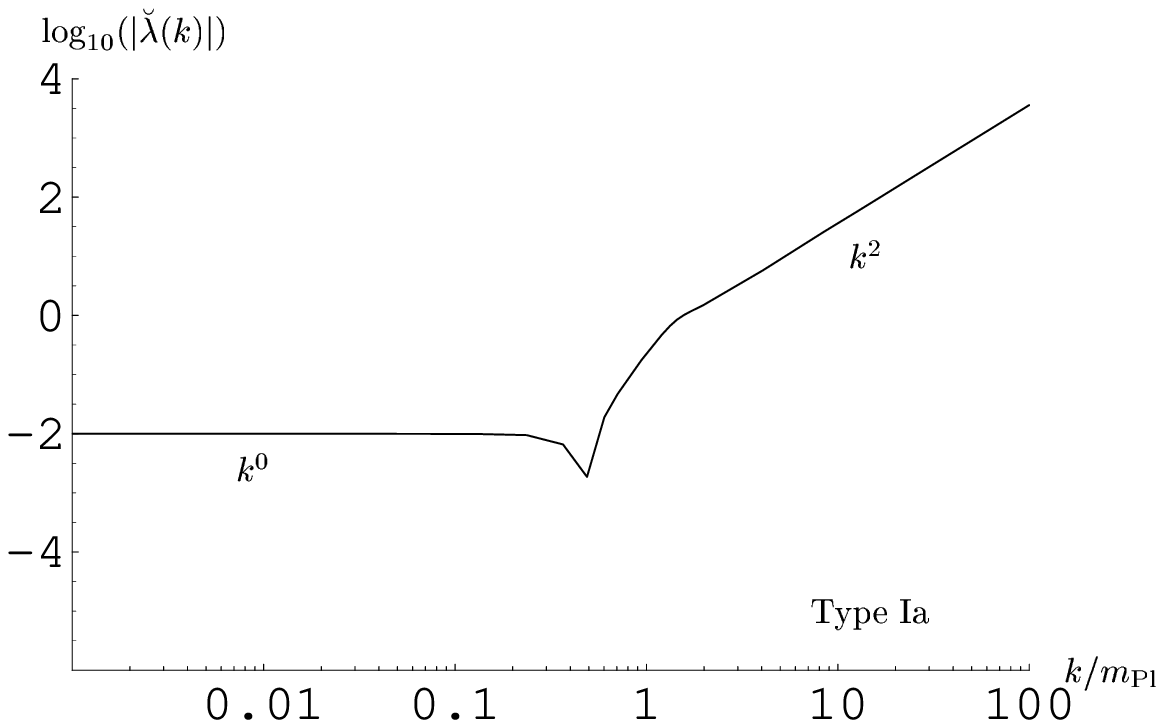}
\end{center}
\vspace*{1.5ex}
\begin{center}
\leavevmode
\epsfxsize=0.49\textwidth 
\epsffile{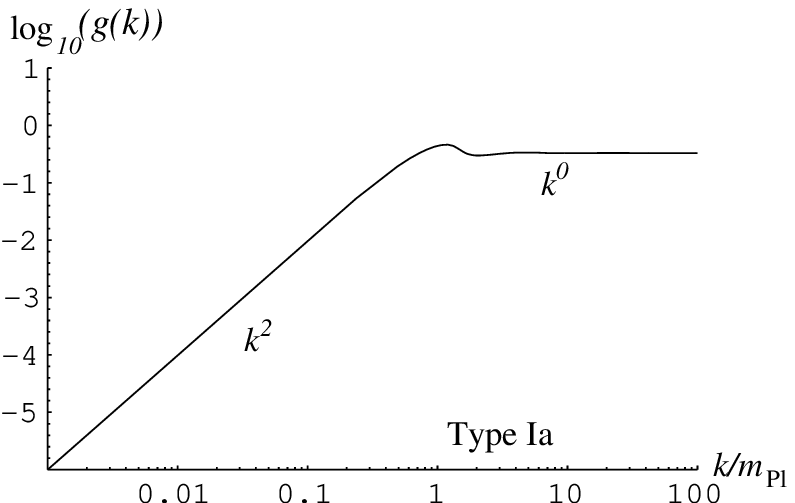}
\epsfxsize=0.49\textwidth
\epsffile{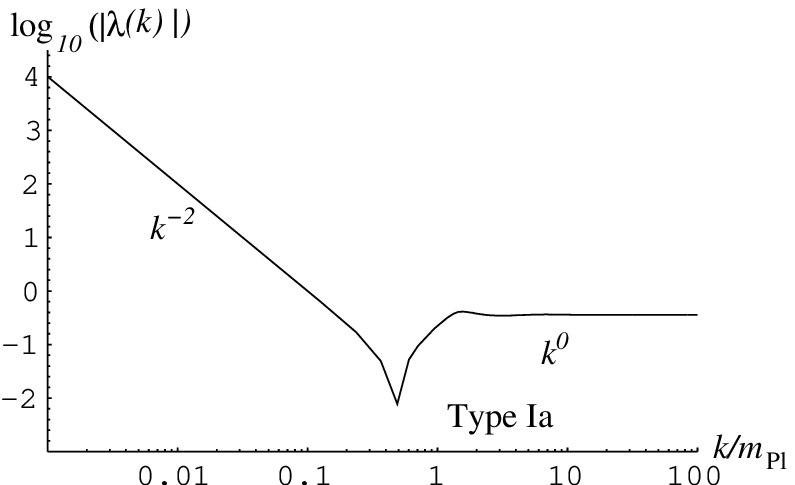}
\end{center}
\parbox[c]{\textwidth}{\caption{\label{dreizehn}{\footnotesize Scaling laws for Typ Ia trajectories. For $\kh < 0.1$ and $\kh > 10$ the flow is governed by the scaling laws of the Gaussian and the non-Gaussian fixed point, respectively. The transition between these scaling laws happens at $k \approx m_{\rm Pl}$. The cusp appearing in the double logarithmic diagrams for the modulus of the cosmological constant is caused by $\lh(\kh)$ and $\lambda_k$ becoming negative at finite $\kh$.}}}
\end{figure}
\begin{figure}[tp]
\renewcommand{\baselinestretch}{1}  
\epsfxsize=0.49\textwidth
\begin{center}
\leavevmode
\epsffile{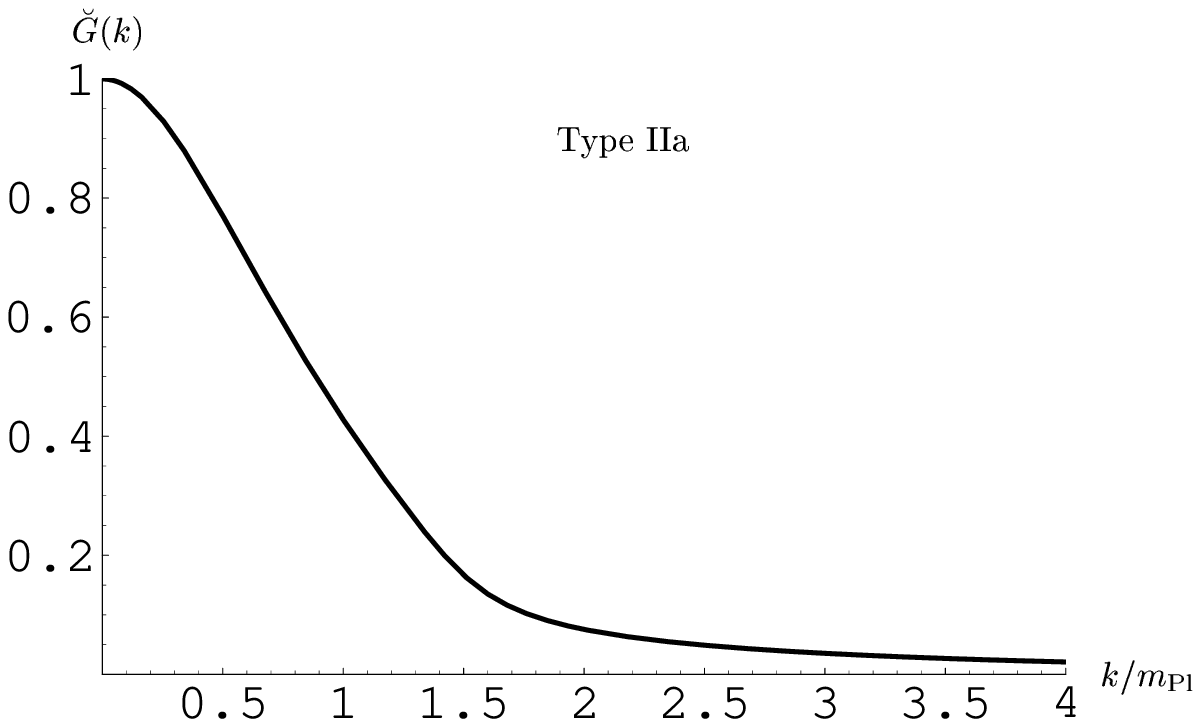}
\epsfxsize=0.49\textwidth
\epsffile{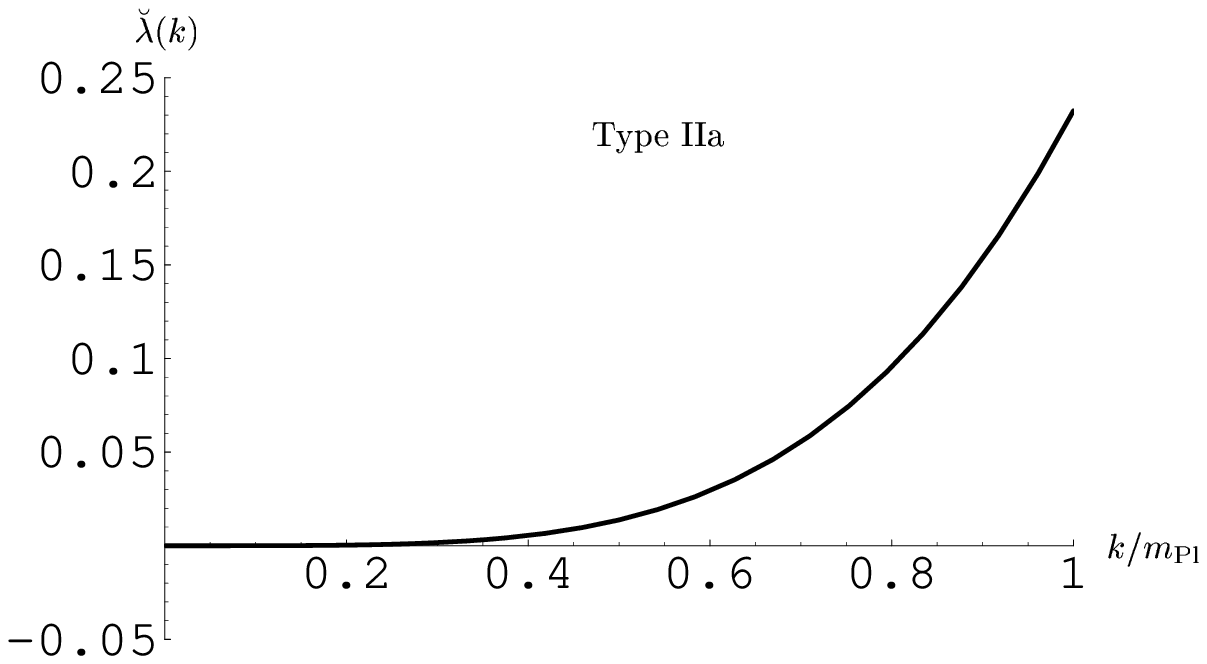}
\end{center}
\vspace*{1.5ex}
\begin{center}
\leavevmode
\epsfxsize=0.49\textwidth 
\epsffile{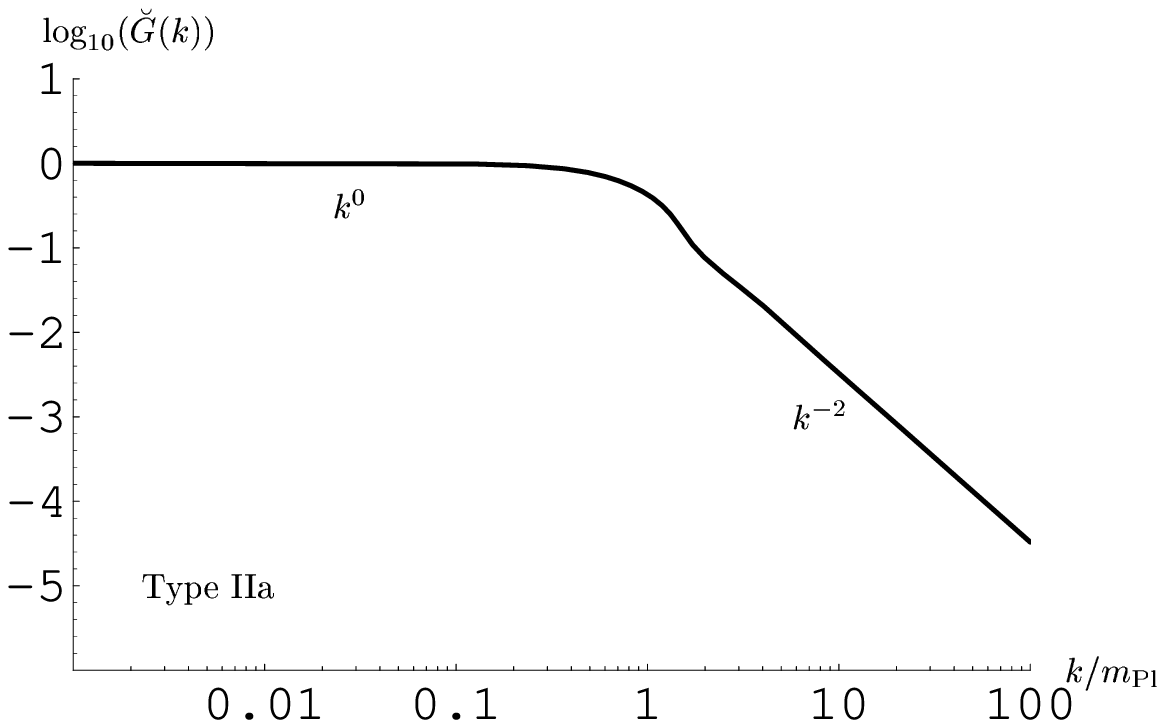}
\epsfxsize=0.49\textwidth
\epsffile{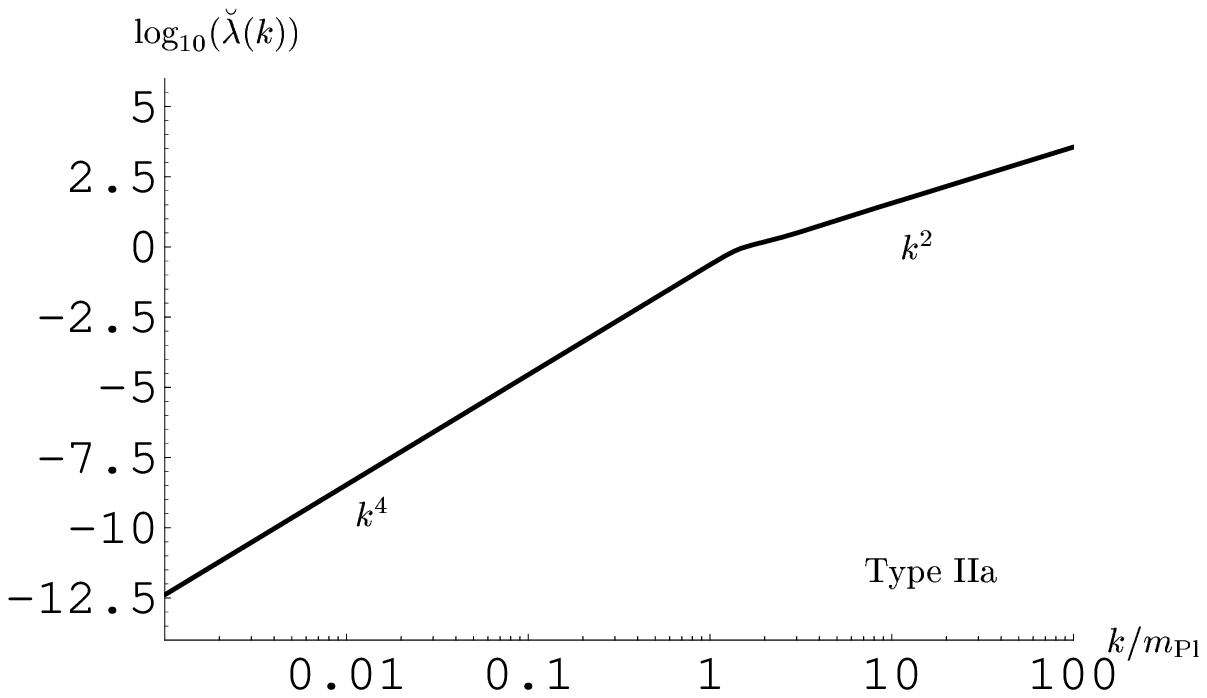}
\end{center}
\vspace*{1.5ex}
\begin{center}
\leavevmode
\epsfxsize=0.49\textwidth 
\epsffile{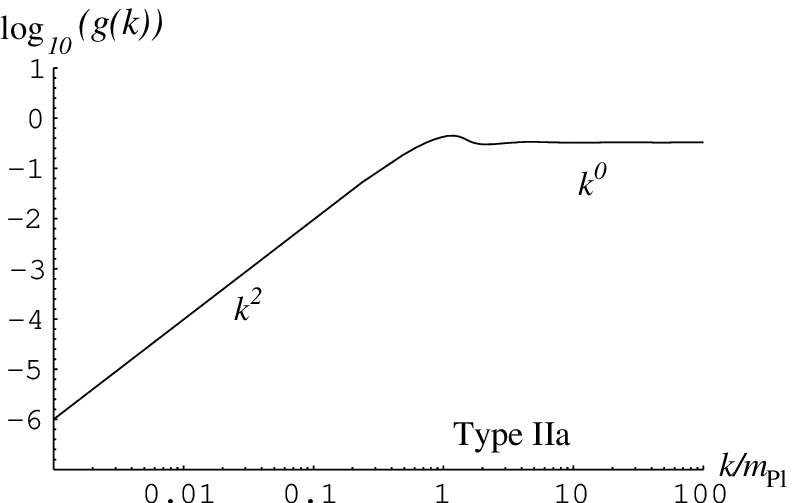}
\epsfxsize=0.49\textwidth
\epsffile{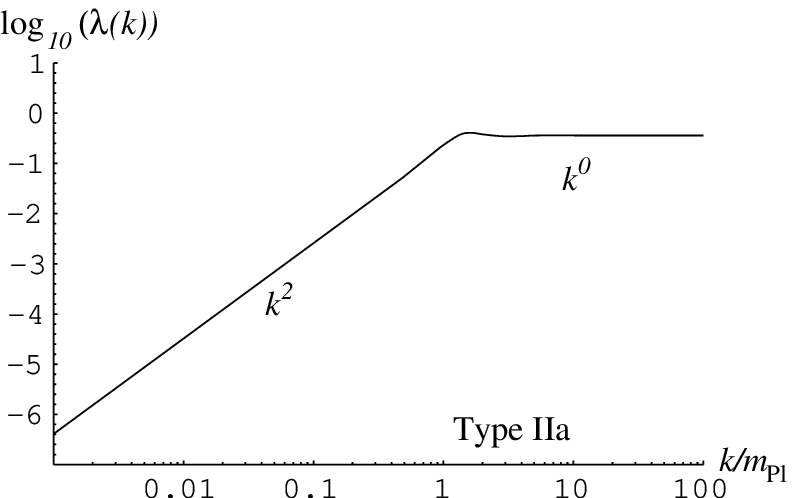}
\end{center}
\parbox[c]{\textwidth}{\caption{\label{vierzehn}{\footnotesize Scaling laws for the trajectory IIa, the separatrix. For $\kh < 0.1$ and $\kh > 10$ the flow is governed by the scaling laws of the Gaussian and the non-Gaussian fixed point, respectively. The change between   these power laws happens at $k \approx m_{\rm Pl}$.}}}
\end{figure}
The main conclusion is that the change in the scaling laws arising from the two fixed point regions is located rather close to the Planck-scale, $k \equiv m_{\rm Pl}$. For $\kh \lesssim 0.1$ and $\kh \gtrsim 10$ the respective scaling laws of the trivial and non-trivial fixed point already dominate the running of the coupling constants. 

This picture is confirmed by the diagrams displaying the anomalous dimension $k \mapsto \eta_N(\lambda_k, g_k)$ along the trajectories Ia, IIa and IIIa shown in FIG. \ref{f"unfzehn}. In the region governed by the non-trivial fixed point one has $\eta_N \approx -2$ while in the IR-region $\eta_N$ vanishes. (Recall that $\eta_N^*=-2$ and $\eta_N^*=0$ at the non-Gaussian and the Gaussian fixed point in $d=4$, respectively.) For the Type IIIa the termination of the trajectory is accompanied by a steep decrease of $\eta_N$, caused by the divergence of $B_1(\lambda)^{\rm sc}$ at constant  $B_2(\lambda)^{\rm sc}$ for $\lambda$ approaching the boundary line $\lambda = 1/2$. (This can easily be checked from equation \rf{3.17}.)
\begin{figure}[t]
\renewcommand{\baselinestretch}{1}  
\epsfxsize=0.48\textwidth
\begin{center}
\leavevmode
\epsffile{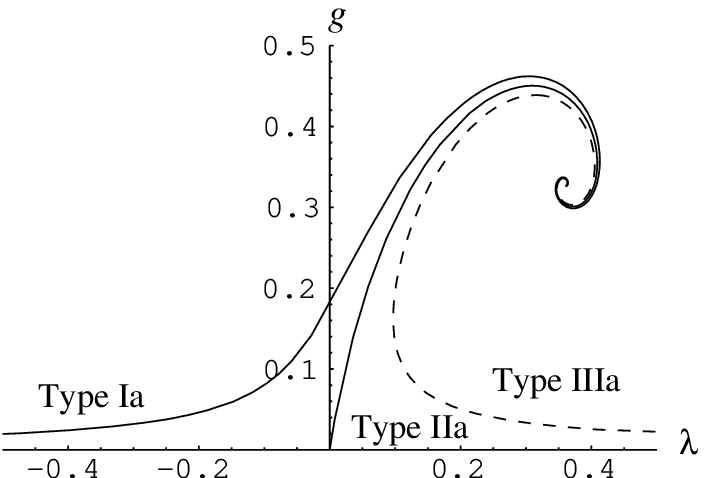}
\epsfxsize=0.48\textwidth
\epsffile{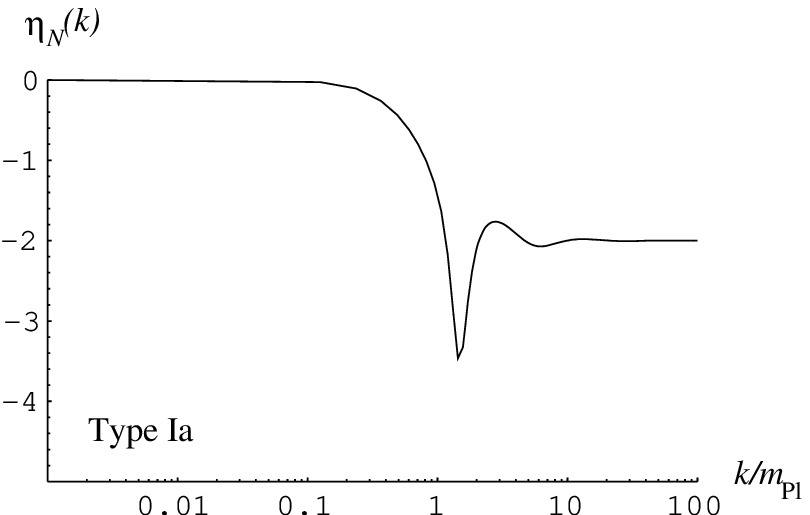}
\\
\leavevmode
\epsfxsize=0.48\textwidth 
\epsffile{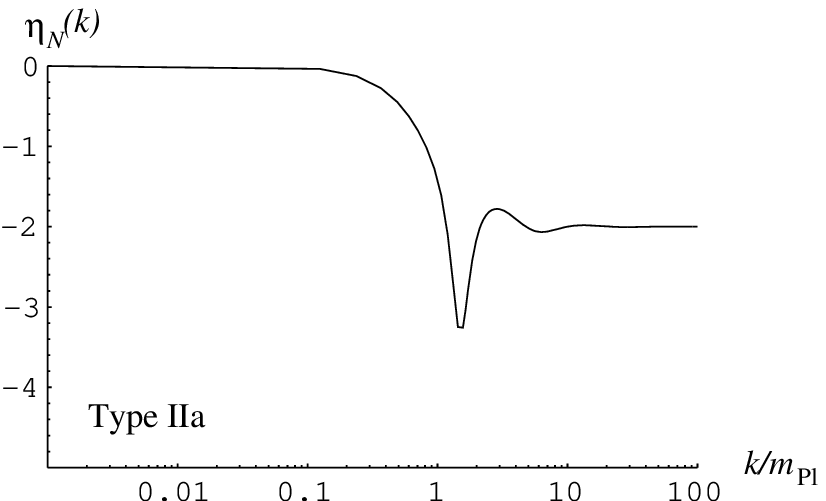}
\epsfxsize=0.48\textwidth
\epsffile{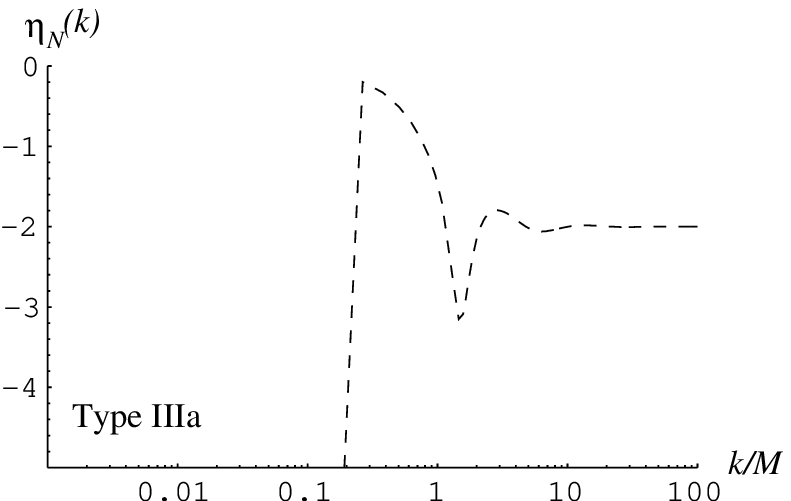}
\end{center}
\parbox[c]{\textwidth}{\caption{\label{f"unfzehn}{\footnotesize Anomalous dimension $\eta_N$ along the Type Ia, IIa and IIIa trajectories shown in the first diagram.}}}
\end{figure}
The vast decrease of $\eta_N$ for the trajectories IIIa suggests that the Einstein-Hilbert truncation may not be sufficient to describe the RG flow close to the boundary line $\lambda = 1/2$. 
\begin{figure}[t]
\renewcommand{\baselinestretch}{1}
\epsfxsize=12cm
\centerline{\epsfbox{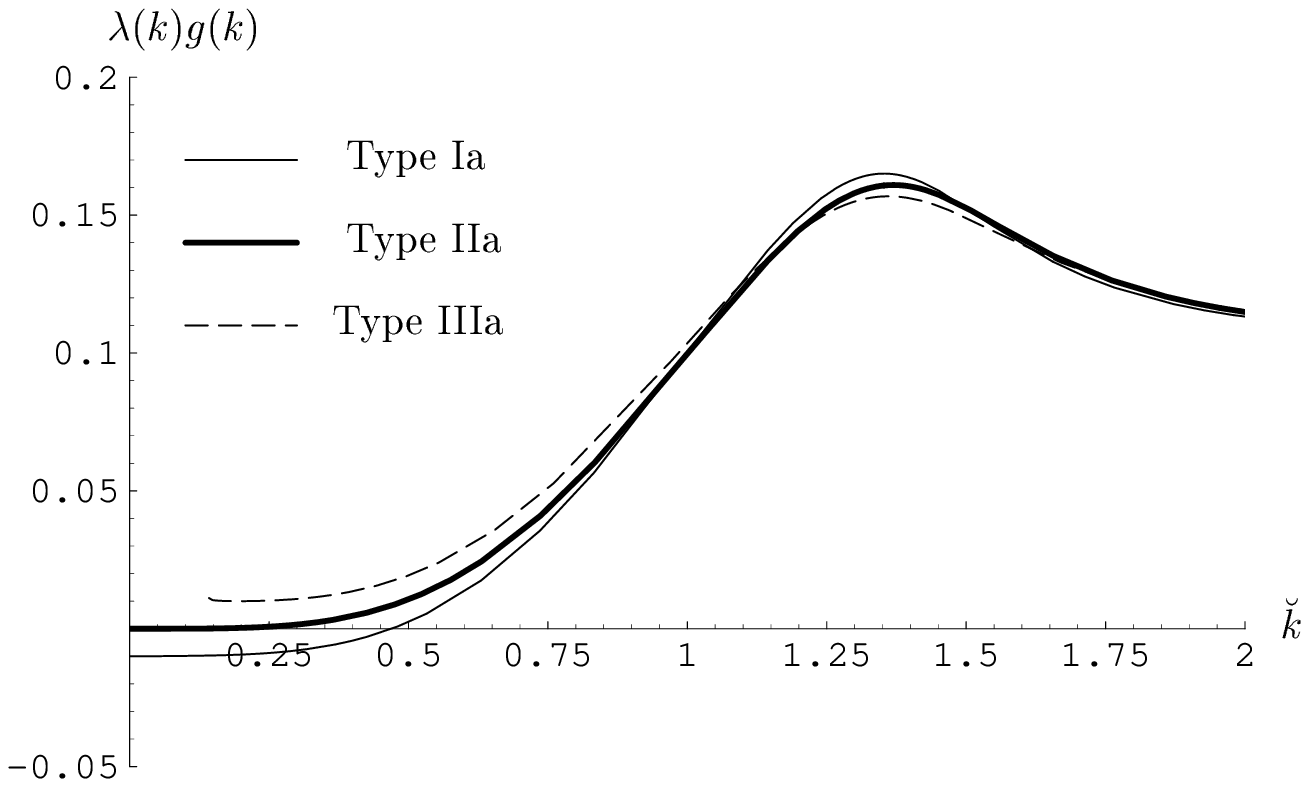}}
\parbox[c]{\textwidth}{\caption{\label{zwanzig}{\footnotesize Product $\lambda(k) g(k)$ for the representative trajectories Ia, IIa and IIIa shown in FIG. \ref{f"unfzehn}. The parameter $\kh$ is given by $\kh= k/m_{\rm Pl}$ for Type Ia and IIa, and $\kh=k/\M$ for Type IIIa.}}}
\end{figure}

Motivated by the important role played by the product $\lambda^* g^*$ we also plot $\kh \mapsto \lambda(\kh) g(\kh)$ for those trajectories which were considered in FIG. \ref{f"unfzehn}. The results are shown in FIG. \ref{sechzehn}. 

Here we observe that for small values of $\kh$, $\kh \lesssim 1$ say, the value of $\lambda(\kh) g(\kh)$ along the separatrix provides a clear separation of the trajectories Ia and IIIa. In the region above $\kh \approx 1$ the trajectories cross. In this region the $\lambda g$-value along the separatrix does not provide a separator between the trajectories Ia and IIIa. This behavior is caused by the non-zero imaginary part of the stability coefficients of the non-Gaussian fixed point. It leads to the observed oscillations of the trajectories in the intermediate region. In this respect we point out that for a non-Gaussian fixed point which by chance has $\mbox{Im}(\theta^I)=0$, the quantity $\lambda(\kh) g(\kh)$  along the separatrix would provide a good separator between the trajectories running in the regions Ia and IIIa for all values of $\kh$. This remark will become important when we compare our results to those found in lattice calculations in Section VI.   
\end{subsection}
\begin{subsection}{The Non-Trivial Fixed Point in other Dimensions}
The $\Fbeta$-functions \rf{2.18} and the entries of the stability matrix ${\bf B}$ in \rf{5.4} have been derived for arbitrary $d$ and allow for an investigation of the non-trivial fixed point in any dimension $d$. 
\begin{figure}[t]
\renewcommand{\baselinestretch}{1}
\epsfxsize=0.49\textwidth
\begin{center}
\leavevmode
\epsffile{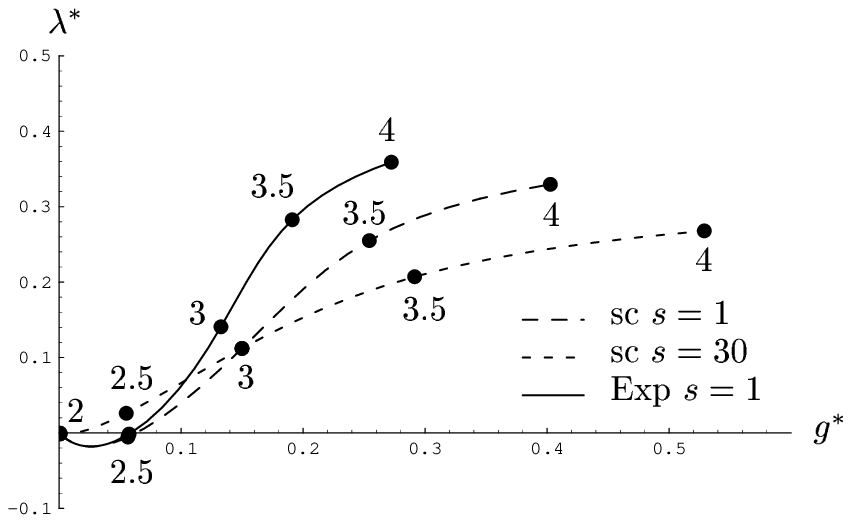}
\epsfxsize=0.48\textwidth
\epsffile{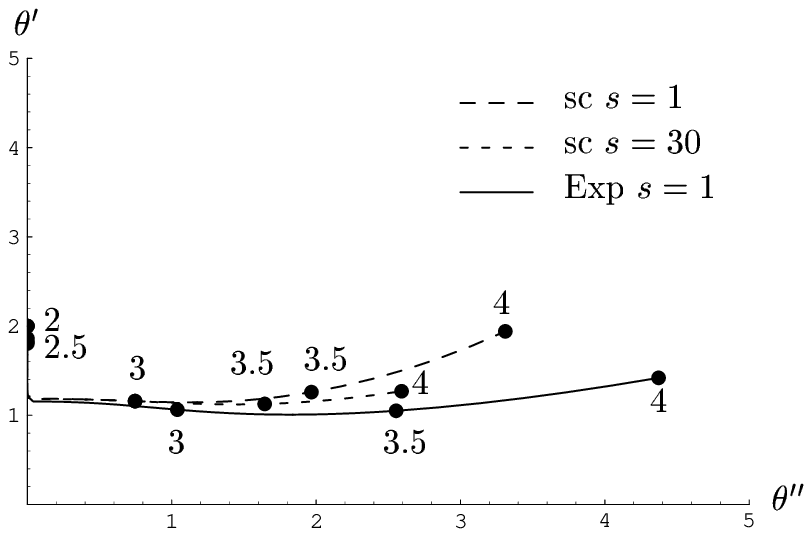}
\end{center}
\begin{center}
\leavevmode
\epsfxsize=0.6\textwidth
\epsffile{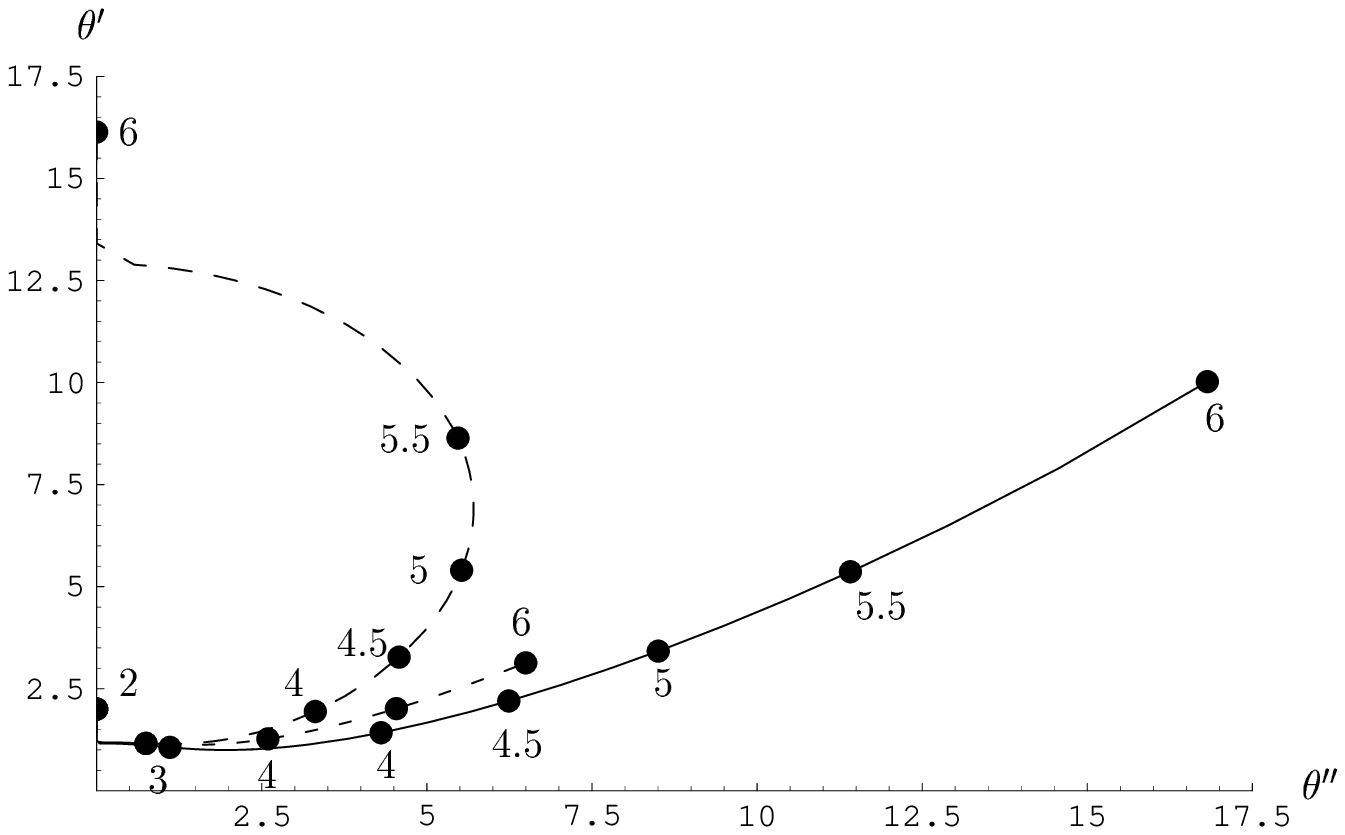}
\end{center}
\parbox[c]{\textwidth}{\caption{\label{sechzehn}{\footnotesize Comparison of the numerical values of $\lambda^*,g^*, \theta^{\prime}$ and $\theta^{\prime \prime}$ for different cutoff functions in dependence of the dimension $d$. The upper line shows that for $2+ \epsilon \le d \le 4$ the cutoff scheme dependence of the results is rather small. The lower diagram shows that increasing $d$ beyond about 5 leads to a significant difference in the results for $\theta^{\prime}, \theta^{\prime \prime}$ obtained with the different cutoff schemes.}}}
\end{figure}
For $2+\epsilon \le d \le 4$, the existence of the non-trivial fixed point for the $\Fbeta$-function with the exponential cutoff \rf{3.3} has already been demonstrated in \cite{oliver,oliver2,souma}. 

FIG. \ref{sechzehn} shows the numerical values of $\lambda^*,g^*$, $\theta^{\prime}$ and $\theta^{\prime \prime}$ obtained by using both the exponential and the sharp cutoff. The $s$-dependence of the sharp cutoff is again defined by the relation $\rf{5.14b}$ which trivially extends to any $d$. 

The results in FIG. \ref{sechzehn} show that the scheme dependence of both the (nonuniversal) location and the (universal) critical indices at the fixed point increase steadily with increasing $d$. But up to dimensionalities between about $4.5$ or $5$, say, one still finds a relatively good agreement between the results obtained with the different cutoff functions. Beyond this point the universal quantities $\theta^{\prime}$ and $\theta^{\prime \prime}$ become seriously scheme dependent. 

It is reassuring to see that there seem to be no qualitative differences between $d=4$ and $d=2+\epsilon$. We interpret this as another confirmation of the reliability of the Einstein-Hilbert truncation in 4 dimensions, at least at a qualitative level.
\begin{figure}[t]
\renewcommand{\baselinestretch}{1}
\epsfxsize=0.49\textwidth
\begin{center}
\leavevmode
\epsffile{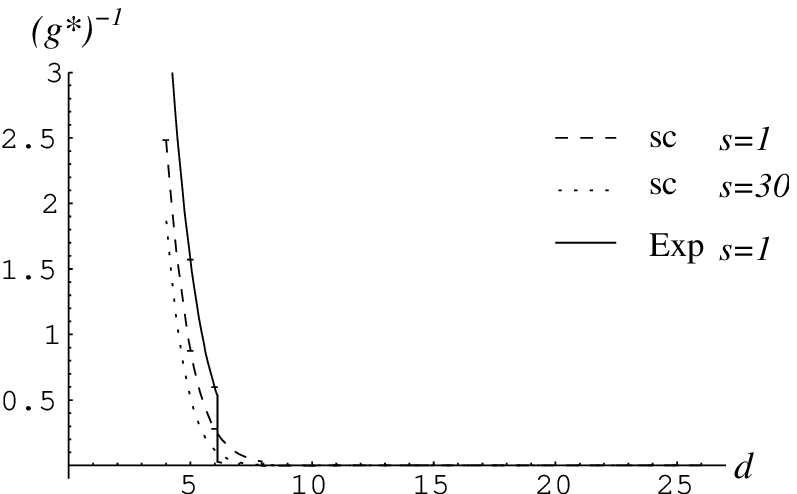}
\epsfxsize=0.48\textwidth
\epsffile{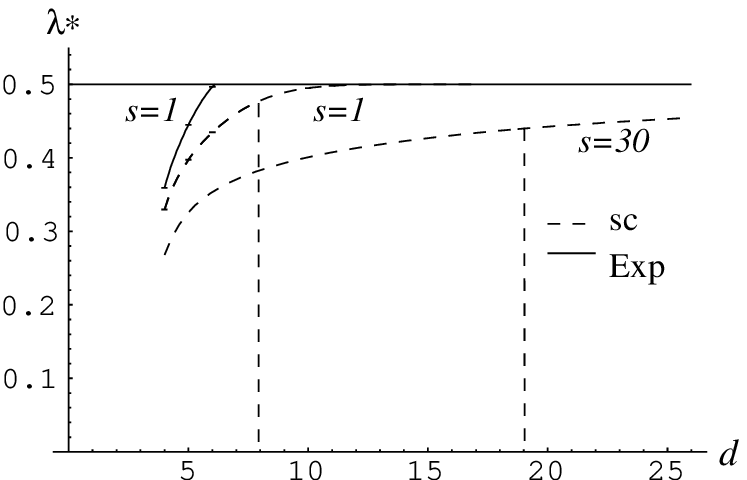}
\end{center}
\parbox[c]{\textwidth}{\caption{\label{siebzehn}{\footnotesize Numerical values $\lambda^*$ and $(g^*)^{-1}$ for the non-trivial fixed point in $d>4$. The branching occurring for the sharp cutoff indicates the existence of a second fixed point in the negative coupling region.}}}
\end{figure}

Looking at the cutoff dependence of $\lambda^*,g^*$ in the region $d \gg 4$, one finds that the trend of the growing scheme dependence of the fixed point properties continues. FIG. \ref{siebzehn} shows that the value of $g^*$ vastly increases for increasing $d$, and that $\lambda^*$ approaches the singularity at $\lambda = 1/2$. Once $\lambda^*=1/2$ is reached, the non-trivial fixed point disappears at a certain critical dimension $d_c$ which strongly depends on the cutoff function. In FIG. \ref{siebzehn} one finds:
\bea\label{5.15}
\nonumber d_c \approx 6 && \quad \mbox{for the exponential cutoff with $s=1$.}\\
\nonumber d_c \approx 17 && \quad \mbox{for the sharp cutoff with $s=1$.}\\
d_c > 26 && \quad \mbox{for the sharp cutoff with $s=30$.}
\eea

The first diagram of FIG. \ref{siebzehn} shows quite impressively that $d=4$ seems still to lie on the safe side of a rather pronounced ``phase-transition'' at $d \approx 5$. At this point, $g^*$ suddenly jumps from $g^*<1$ to $g^* \gg 1$, so that we must expect the truncation to become problematic at $d \approx 5$.
\begin{figure}[t]
\renewcommand{\baselinestretch}{1}
\epsfxsize=0.49\textwidth
\begin{center}
\leavevmode
\epsffile{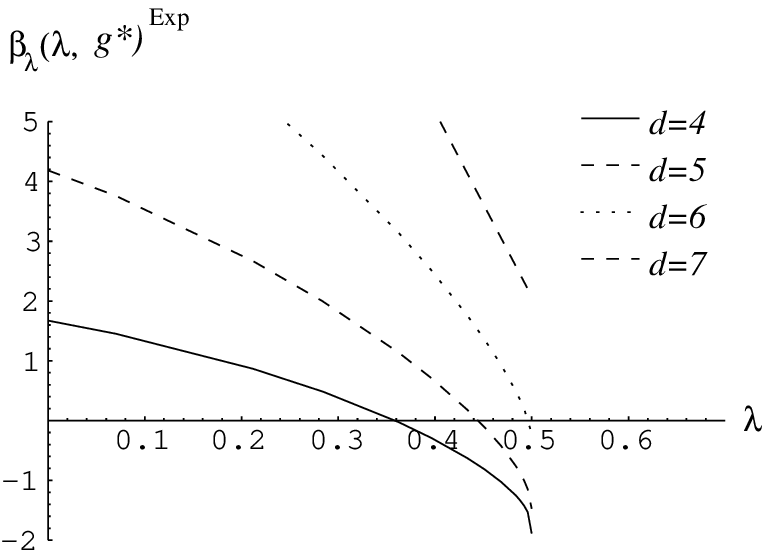}
\epsfxsize=0.48\textwidth
\epsffile{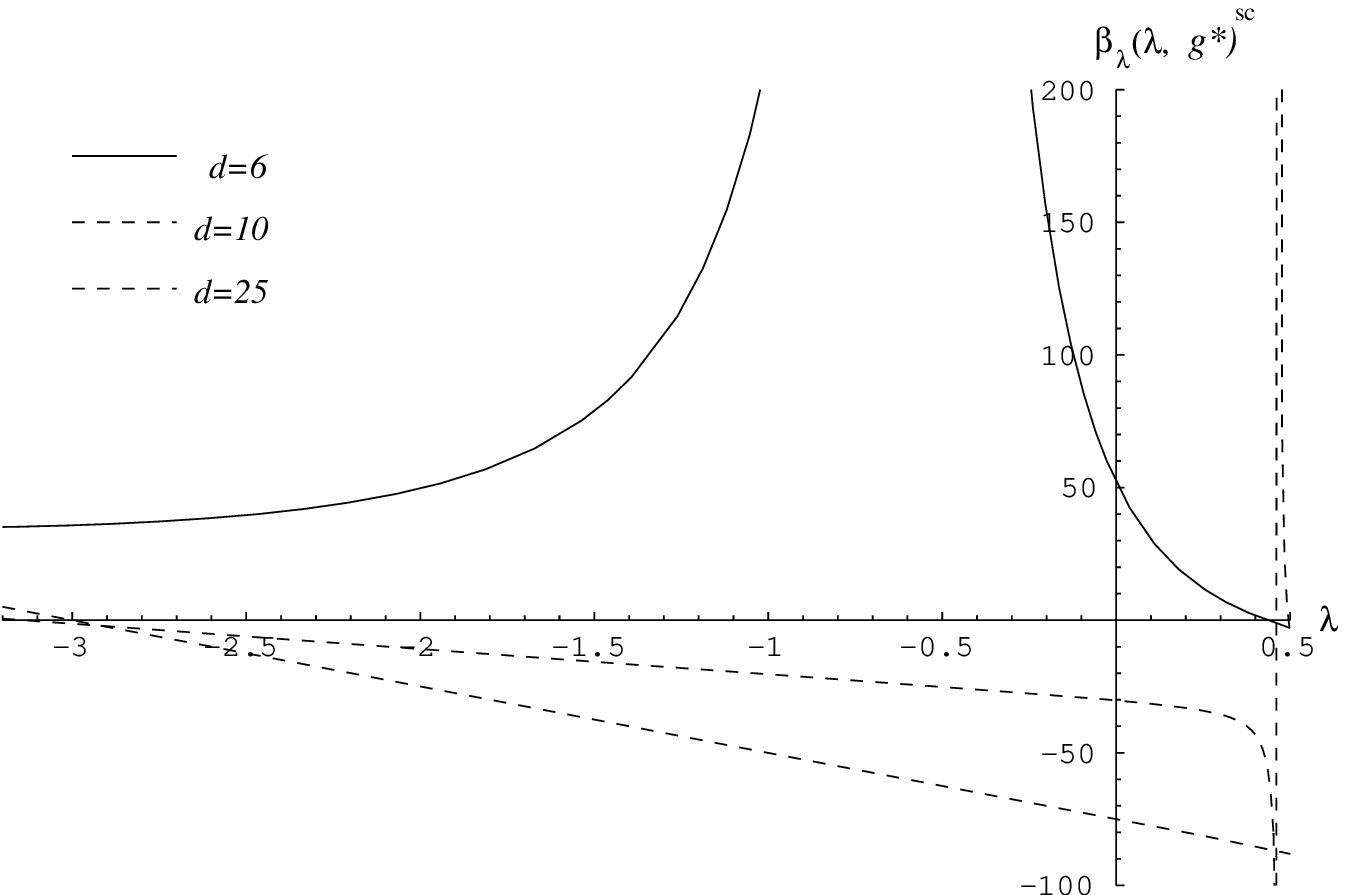}
\end{center}
\parbox[c]{\textwidth}{\caption{\label{achzehn}{\footnotesize $\Fbeta_{\lambda}(\lambda, g^*(\lambda))$ as a function of $\lambda$ for the exponential and the sharp cutoff with shape parameter $s=1$ for selected dimensions $d$. The zero of this function is $\lambda^*$. The left diagram demonstrates the escape of the non-trivial fixed point into the boundary $\lambda =1/2$ for the exponential cutoff. The right diagram shows the appearance of a second zero and the vanishing of the familiar non-trivial fixed point for the sharp cutoff.}}}
\end{figure}
  
In order to better understand the disappearance of the non-Gaussian fixed point, we plot in FIG. \ref{achzehn} the function $\lambda \mapsto \Fbeta_\lambda(\lambda, g^*(\lambda))$, whose zero is $\lambda^*$, for both the exponential and sharp cutoff and for various dimensions $d$. The function $g^*(\lambda)$ thereby is given by equation \rf{5.11}. One finds that, for the exponential cutoff, $\Fbeta_\lambda(\lambda, g^*(\lambda))^{\rm Exp}$ is positive definite for $d>d_c$ which implies that there can be no non-trivial fixed point. For the sharp cutoff, $\Fbeta_{\lambda}(\lambda, g^*(\lambda))$ develops a second zero in the negative coupling region before the old non-trivial fixed point in the positive region escapes through the $\lambda = 1/2$ boundary. The appearance of the second zero is indicated by the branching shown in the $\lambda^*$-$d$--diagram of FIG. \ref{achzehn}. For $d>d_c$ only the new fixed point in the negative coupling region remains. 

The properties of these fixed points for selected dimensions are shown in TABLE \ref{four} and TABLE \ref{fife} for the sharp and the exponential cutoff, respectively. 
\begin{table}[tp]
\renewcommand{\baselinestretch}{1}
\begin{tabular}{c|ccccc|cccc}
& \multicolumn{5}{c}{ non-trivial fixed point } & \multicolumn{4}{c}{ fixed point in the negative coupling region} \\
$d$ & $\lambda^*$ & $g^*$ & $g^*\lambda^*$ & $\theta^{\prime}$ & $\theta^{\prime \prime}$
& $\lambda^*$ & $g^*$ & $\theta^{\prime}$ & $\theta^{\prime \prime}$ \\ \hline
3  & \hspace*{0mm} 0.112  & 0.150 &  0.017   & 1.159   & 0.746 &   -    &   -   &   -   &   - \\[1.5ex]   
4  & \hspace*{0mm} 0.330  & 0.403 &  0.133   & 1.941   & 3.311 &   -    &   -   &   -   &   -             \\[1.5ex] 
5  & \hspace*{0mm} 0.398  & 1.143 &  0.454   & 5.401   & 5.526 &   -    &   -   &   -   &   -             \\[1.5ex] 
   &                      &   &      & 10.89   &       &        &       &       &       \\[-1ex]
\raisebox{1.5ex}[-1.5ex]{6}  & \raisebox{1.5ex}[-1.5ex]{\hspace*{0mm} 0.435}  & \raisebox{1.5ex}[-1.5ex]{\hspace*{0mm} 1.554} & \raisebox{1.5ex}[-1.5ex]{3.576}    & 16.14  &   \raisebox{1.5ex}[-1.5ex]{0}   &   \raisebox{1.5ex}[-1.5ex]{-}    &   \raisebox{1.5ex}[-1.5ex]{-}   &   \raisebox{1.5ex}[-1.5ex]{-}   &  \raisebox{1.5ex}[-1.5ex]{-}            \\ [1.5ex] 
   &                      &   &      & 10.15   &       &        &       &  8.167&       \\[-1ex]
\raisebox{1.5ex}[-1.5ex]{10} & \raisebox{1.5ex}[-1.5ex]{\hspace*{0mm} 0.495  } & \raisebox{1.5ex}[-1.5ex]{324.1    } & \raisebox{1.5ex}[-1.5ex]{\hspace*{0mm} 160.4} & 922.7  &   \raisebox{1.5ex}[-1.5ex]{0}    & \raisebox{1.5ex}[-1.5ex]{-3.138 } & \raisebox{1.5ex}[-1.5ex]{-2232.} & 10.12 & \raisebox{1.5ex}[-1.5ex]{0} \\[1.5ex] 
   &                      &     &    & 11.07   &       &        &       &  9.074&       \\[-1ex]
\raisebox{1.5ex}[-1.5ex]{11}  & \raisebox{1.5ex}[-1.5ex]{\hspace*{0mm} 0.498  }  & \raisebox{1.5ex}[-1.5ex]{\hspace*{0mm} 476.8}  & \raisebox{1.5ex}[-1.5ex]{957.4}   & 2480.   & \raisebox{1.5ex}[-1.5ex]{0} & \raisebox{1.5ex}[-1.5ex]{-3.042}  &\raisebox{1.5ex}[-1.5ex]{-6075.}  & 11.05& \raisebox{1.5ex}[-1.5ex]{0}  \\ [1.5ex] 
   &                      &      &   &   &       &        &       &  23.00&       \\[-1ex]
\raisebox{1.5ex}[-1.5ex]{ 25 } &   \raisebox{1.5ex}[-1.5ex]{-} &   \raisebox{1.5ex}[-1.5ex]{-} & \raisebox{1.5ex}[-1.5ex]{-} &   \raisebox{1.5ex}[-1.5ex]{ - } & \raisebox{1.5ex}[-1.5ex]{-}  & \raisebox{1.5ex}[-1.5ex]{-3.000 }  &\raisebox{1.5ex}[-1.5ex]{$-5 \cdot 10^{10}$ }  & 25.00 & \raisebox{1.5ex}[-1.5ex]{0}  \\[1.5ex] 
   &                      &       &   &  &       &        &       &  24.00 &       \\[-1ex]
\raisebox{1.5ex}[-1.5ex]{26} &   \raisebox{1.5ex}[-1.5ex]{-} &  \raisebox{1.5ex}[-1.5ex]{-} &  \raisebox{1.5ex}[-1.5ex]{-}   &   \raisebox{1.5ex}[-1.5ex]{-}  &  \raisebox{1.5ex}[-1.5ex]{-}  & \raisebox{1.5ex}[-1.5ex]{-3.000 }  &\raisebox{1.5ex}[-1.5ex]{$-2 \cdot 10^{11}$ }  & 26.00& \raisebox{1.5ex}[-1.5ex]{ 0 } \\ 
\end{tabular}
\caption{\label{four} \footnotesize Numerical values for the non-trivial fixed point and the fixed point appearing in the negative coupling region for the $\Fbeta$-functions with sharp cutoff with $s=1$ in selected dimensions $d$. Here `-' denotes that no fixed point of the corresponding type exists for this dimension. A pair of real-valued stability coefficients, appearing for $d>6$, is indicated by giving the two corresponding $\theta$-values in the $\theta^{\prime}$-column. For $d \ge 6$ the results found here are certainly not reflecting properties of the full theory and should be seen as a demonstration of the limitations of the Einstein-Hilbert truncation in higher dimensional spacetimes.}
\end{table}    
\begin{table}[tph]
\renewcommand{\baselinestretch}{1}
\begin{tabular}{c|ccccc}
& \multicolumn{5}{c}{non-trivial fixed point for exponential cutoff with $s=1$} \\
$d$ & $\lambda^*$ & $g^*$ & $\lambda^* g^*$ & $\theta^{\prime}$ & $\theta^{\prime \prime}$ \\ \hline
3  &	0.140  & 0.133   &  0.019 & 1.063   & 1.109  \\
4  & 0.359  & 0.272   &  0.098 & 1.422   & 4.307  \\ 
5  & 0.445  & 0.636   &  0.283 & 3.419   & 8.503  \\ 
6  & 0.497  & 1.673   &  0.831 & 10.02   & 16.82  \\ 
\end{tabular}
\caption{\label{fife} \footnotesize Numerical values for the non-trivial fixed point with exponential cutoff in selected dimensions $d$. For $d>6$ no non-trivial fixed point exists.}
\end{table} 

The data shown in TABLE \ref{four} and \ref{fife} suggests that the Einstein-Hilbert truncation should produce reliable results up to $d=4$ or maybe $d=5$. Beyond that the results shown in TABLE \ref{four} are untrustworthy and should be understood as an illustration of the effects arising from an improper truncation. For the non-trivial fixed point this unreliability is indicated by the fixed point lying very close to the boundary line $\lambda = 1/2$ and at large values $g^* \gg 1$. In this region of the $\lambda$-$g-$plane, the termination of the trajectories and the steep decrease in $\eta_N$ indicate that the Einstein-Hilbert truncation might be insufficient in describing the RG flow. The appearance of the second zero of the $\Fbeta$-function with sharp cutoff and the absence of this new zero for the smooth exponential cutoff point towards a strong scheme dependence even at the {\it qualitative} level. This is a typical symptom of an insufficient truncation.

Summarizing the results of this subsection we find that the Einstein-Hilbert truncation is most likely insufficient to describe the RG flow for $d \gtrsim 5$. This limitation is due to operators like $\int d^dx \sqrt{g} R^2$ and higher powers of the curvature scalar which are not included in the truncation. Based on the canonical dimensions of their coupling constants one expects an increasing importance of these terms in the UV as $d$ is increased. Therefore it is likely that in order to properly describe quantum gravity in higher dimensional spacetimes a more refined truncation will be needed. 
\end{subsection}
%
%
\mysection{Discussion and Conclusion}
\setcounter{equation}{0} 
In this paper we studied the exact renormalization group equation for the effective average action of pure quantum gravity in its original formulation \cite{ERGE}. The RG flow of Newton's constant and the cosmological constant was investigated nonperturbatively within the Einstein-Hilbert truncation of the theory space. The new contributions of the present paper are the introduction of a sharp cutoff and a comprehensive numerical analysis of the RG trajectories in 4 spacetime dimensions.

As in ref. \cite{ERGE}, we used a cutoff action $\Delta_kS$ whose general structure is of ``TYPE A'' in the terminology of ref. \cite{oliver}\footnote{This cutoff should not be confused with the new one introduced in ref. \cite{oliver} whose general structure is of the ``TYPE B'' which appears naturally when the transverse-traceless decomposition of the metric is used.}. It contains a shape function $R^{(0)}$ for which we considered both a smooth exponential ansatz and, as a singular limit, a sharp step function of infinite height. While in other applications sharp cutoffs often lead to inaccurate or even undefined or divergent results we found that the flow equations for $g$ and $\lambda$ remain perfectly well defined in the sharp cutoff limit. Moreover, all RG trajectories which we computed with both the smooth and the sharp cutoff turned out virtually identical. Thus having confidence in the technically much more convenient sharp cutoff we employed it for a complete classification and computation of the RG trajectories on the $g$-$\lambda-$parameter space. Our main results are summarized in FIGs. \ref{neun} and \ref{elf} and in TABLE \ref{three}.

The most prominent feature of the RG flow resulting from the Einstein-Hilbert truncation is a non-Gaussian fixed point which acts as an UV-attractor for the trajectories Ia, IIa, IIIa and IVa on the $g>0$ half plane. It is an extremely important question whether this fixed point is a truncation artifact or whether it is also present in the full theory. In the latter case 4-dimensional Quantum Einstein Gravity is likely to be ``asymptotically safe'' \cite{weinbergcc}, i.e. nonperturbatively renormalizable by taking the infinite cutoff limit at this fixed point. Einstein gravity would then have the status of a fundamental rather than merely an effective theory, and it could be valid at arbitrarily small distance scales.

In refs. \cite{oliver,oliver2} this question was investigated in detail. The reliability of the Einstein-Hilbert truncation was tested both by analyzing the scheme dependence within this truncation \cite{oliver} and by adding a higher-derivative invariant to it \cite{oliver2}. The picture suggested by these investigations is that, in 4 dimensions, the RG flow in the vicinity of the non-Gaussian fixed point is very well described by the Einstein-Hilbert truncation. Hence we have very good reasons to believe that the fixed point actually should exist in the full theory. Only when one lowers (!) the cutoff and leaves the region of asymptotic scaling more complicated operators such as $R^2$, $R^3$, $\cdots$ are generated.

The results of the present paper provide further evidence supporting the hypothesis of the non-Gaussian fixed point. In order to judge the reliability of the truncation, we checked the cutoff scheme dependence of various universal quantities $(\theta^{\prime}, \theta^{\prime \prime}, g^* \lambda^*)$ which are expected to be scheme independent in an exact treatment. Quite remarkably, the results found here for a sharp cutoff of TYPE A are rather similar to those found in \cite{oliver} with a smooth cutoff of TYPE B. Typically the scheme dependence within the family of sharp cutoffs (parameterized by $\varphi_1$ and $\varphi_2$) is of about the same magnitude as the differences between the sharp and the smooth cutoff, and between the TYPE A and TYPE B structure. In particular the product $\lambda^*g^*$ was found to be scheme independent with a very surprising precision. Here the sharp cutoff leads to even slightly more stable results than  the smooth one, as is shown in the last diagram of FIG. \ref{sechs}.

We also investigated how the scheme dependence of the fixed point data varies with the dimensionality $d$. With increasing $d$ the quality of the Einstein-Hilbert truncation deteriorates and it certainly becomes insufficient at about $d=6$ where the existence or nonexistence of the non-trivial fixed point depends on the cutoff chosen. It seems that in $d=4$, at least qualitatively, the conditions are still very similar to $d=2+\epsilon$ with $0 < \epsilon \ll 1$. All admissible cutoffs lead to the presence of the non-trivial fixed point.  They agree on a quantitative $(\lambda^* g^*)$ or at least semi-quantitative level  $(\theta^{\prime}, \theta^{\prime \prime})$. 

While in \cite{oliver,oliver2} only the linearized flow near the fixed points is discussed, the numerical investigations of the present paper allow us to follow the trajectories emanating from the non-Gaussian fixed point all their way down from the UV to the IR. In particular we were able to determine the scale $k_{\rm asym}$ where the asymptotic scaling behavior governed by the UV fixed point comes to an end. For the trajectories which can be continued to $k=0$ we found that $k_{\rm asym} \approx m_{\rm Pl} \equiv G(k=0)^{-1/2}$. Below this scale the RG flow is governed by a trivial fixed point at the origin of the $\lambda$-$g-$plane. The (Planck) scale $k_{\rm asym}$ is analogous to the mass scale $\Lambda_{\rm QCD}$ in QCD in the sense that it marks the lower border of the asymptotic scaling region.

Probably the analogy between QCD and gravity goes even further. For $k \gg \Lambda_{\rm QCD}$, thanks to asymptotic freedom, the effective average action of Yang-Mills theory has a very simple local structure, and simple truncations such as $\Gamma_k = 1/4 \, \cz_k \int d^dx (F_{\mn}^a)^2$ are sufficient \cite{QCD}. Only when $k$ is close to, or much smaller than $\Lambda_{\rm QCD}$ its structure becomes very complicated and nonlocal. Also gravity seems to be asymptotically free \cite{ERGE}, and if the results of \cite{oliver,oliver2} and the present paper point in the right direction, gravity, too, can be described by a simple local truncation above $k_{\rm asym} \approx m_{\rm Pl}$. Only when we lower $k$ down to the Planck scale, $\Gamma_k$ will contain higher order local invariants ($R^n, \cdots)$ as well as nonlocal invariants \cite{nonloc,olivercond}. Presumably the use of the Einstein-Hilbert truncation is much more problematic at $k \approx m_{\rm Pl}$ than at $k \gg m_{\rm Pl}$. In fact, the aborting trajectories of Type IIIa and the unbounded increase of $|\eta_N|$ which precedes their termination most probably hint at an insufficiency of the Einstein-Hilbert truncation {\it in the infrared}. It must be emphasized that there are no analogous consistency problems above $k_{\rm asym}$.

It is likely that in the IR a proper treatment of the Type IIIa trajectories (or rather their counterparts in the exact theory) requires much more sophisticated truncations. It is plausible to speculate that those truncations will encode the strong IR-quantum effects which have been discussed in refs. \cite{oliver}. They might be the key to a dynamical resolution of the cosmological constant problem.

The above picture of quantum gravity being ``simple'' above $m_{\rm Pl}$ and ``complicated'' below $m_{\rm Pl}$ contradicts the general prejudice that far below $m_{\rm Pl}$ gravity is well described by the ``simple'' Einstein-Hilbert action and becomes very ``complicated'' for all $k \gtrsim m_{\rm Pl}$. We think that this prejudice might turn out wrong for (at least) two complementary reasons: (i) Ordinary perturbation theory predicts that $\Gamma_k$ becomes ``complicated'' (higher operators are generated) when $k$ approaches the Planck scale {\it from below}. While this is probably correct, it is an unjustified extrapolation that $\Gamma_k$ continues to be ``complicated'' for $k \gg m_{\rm Pl}$. In this regime the ``simplicity'' due to asymptotic freedom sets in, but this cannot be discovered by ordinary perturbation theory. (ii) The classic experimental tests of general relativity all refer to the theory without a cosmological constant. From this side nothing is known about the IR-properties of the (quantum) theory with $\lb \not = 0$ which could be rather ``complicated'' without contradicting any experimental fact.

The weak coupling region of the $\lambda$-$g-$plane contains the trajectories of Type Ia, leading to $\lb_0 < 0$, and of Type IIIa, terminating at the boundary $\lambda = 1/2$. The two classes of trajectories are separated by a single trajectory of Type IIa, the separatrix. It runs from the non-trivial to the trivial fixed point and leads to a vanishing renormalized cosmological constant $\lb_0 = 0$ at its endpoint.

We saw that the Ia and IIa trajectories can be continued down to $k=0$ without any problem and that $|\eta_N|$ remains bounded along these trajectories. This might indicate that in this sector, characterized by a renormalized cosmological constant $\lb_0 \le 0$, the Einstein-Hilbert truncation is reliable even in the IR, at least at a qualitative level. With the present calculational techniques it would be very hard to check this conjecture on a theoretical basis. However, as far as the case $\lb_0=0$ is concerned, the phenomenological success of classical general relativity (without a cosmological constant) is an excellent confirmation of the Einstein-Hilbert action in the IR.

The Einstein-Hilbert action thus being a good approximation to $\Gamma_k$ both in the extreme UV and the extreme IR, it is plausible to assume that the separatix which connects the two fixed points does not change much when we go from the truncated to the full theory. It is then possible to define the theory called ``Quantum Einstein Gravity'' (with zero renormalized cosmological constant) by means of this specific RG trajectory. 
 
It is natural to interpret the separatrix as a kind of phase transition line. It is in fact similar to the critical line running into the Gaussian fixed point in scalar theories. In the latter case this line separates the symmetric from the spontaneously broken phase, i.e. trajectories with positive and negative $\mbox{(mass)}^2$, respectively. In the case of gravity the line separates the trajectories which, in the IR, lead to a positive or negative cosmological constant, respectively. On the side with $\lb>0$ the RG trajectories terminate at a certain finite value of $k$. Whether this is an artifact of the truncation or a real physical effect cannot be decided on the basis of the present analysis. 

In this context it is intriguing to compare our results to those obtained by numerical simulations within Regge's simplicial formulation of gravity \cite{hamber,ambbook}. These studies indicate that simplicial quantum gravity in four dimensions exhibits a phase transition in the bare coupling $G$ between the following two phases: a strong coupling phase in which the geometry is smooth at large distance scales with $\langle g_{\mn} \rangle \approx c \delta_{\mn}$, and a weak coupling phase in which the geometry is degenerate, $\langle g_{\mn} \rangle =0$  (branched polymer-like phase). Interestingly enough, in the strong coupling phase one finds a small {\it negative} average curvature.

It is tempting to identify the strong coupling phase with our Type Ia trajectories which ultimately go to negative $\lb_0$ and negative curvature $R=4 \lb$, and the weak coupling phase with the type IIIa trajectories which have no limit $k\rightarrow 0$ and cannot describe large, nearly flat universes therefore. Hamber \cite{hamber} parameterizes the bare couplings by a quantity ${\bf k}$ (not to be confused with our cutoff) which reads, in our notation,
\benn
{\bf k}= \left[ 8 \pi G \lb \right]^{-1/2} = \left[ 8 \pi g \lambda \right]^{-1/2}
\eenn
Strong (weak)coupling means ${\bf k}<{\bf k}_c$ $({\bf k}>{\bf k}_c)$ for a certain critical value ${\bf k}_c$. This matches precisely with our findings in Section V. Depending on whether the product $g \lambda$ takes on negative or positive values in the IR, either the Type Ia or the Type IIIa trajectories are realized. This can be read off from FIG. \ref{zwanzig} provided one ignores the oscillations and the crossing of the trajectories which are caused by the imaginary part of the critical exponent $\theta$. It remains to be seen if this identification is correct.

A quantity which in principle might lend itself to a quantitative comparison is the exponent $\theta$. In \cite{hamber} the analogous exponent is assumed to be real, $\theta=\theta^{\prime}=1 /{\mathbf \nu}$, with ${\mathbf \nu}$ the conventionally defined critical exponent used in the theory of critical phenomena. The fit to the data yields a value close to ${\mathbf \nu} = 1/3$ so that we should expect $\theta^\prime \approx 3$. The typical values we found in Section V are somewhat smaller. However in \cite{oliver2} a detailed investigation of the fixed point with a generalized truncation and a quantitatively probably more reliable smooth cutoff has been performed. The results suggest that an improved calculation indeed could stabilize close to $\theta^\prime \approx 3$. It is unclear, however, how the imaginary part of $\theta$ should be properly taken into account in this comparison.

\vspace{1.2cm}
\noindent
Acknowledgement: We would like to thank A.Bonanno, W.Dittrich, O.Lauscher and C.Wetterich for helpful discussions.

\begin{appendix}
\begin{section}{Singularity of $\eta_N$ for the exponential cutoff}
When investigating the complete system \rf{2.23} and \rf{2.23a} in Section IV we omitted discussing some peculiar properties of the RG flow derived with the exponential cutoff. These properties will be analyzed in this appendix.

The key to understanding this behavior is the fact that there exists a line on the $\lambda$-$g-$plane along which $\eta_N^{\rm Exp}(\lambda, g)$ diverges. This line is shown in FIG. \ref{acht}. Switching from the variables $(\lambda,g,k)$ to $(\lh, \Gh, y)$ leads to an analogous divergence of $\eta_N^{\rm Exp}(\lh, \Gh, y\equiv \kh^2)$ on a 2-dimensional curved surface in the 3-dimensional $\lh$-$\Gh$-$y-$space. We can visualize this surface by intersecting it with some plane $\Gh = \mbox{const}$, which then leads to a line of singularities on this $\lh$-$y-$plane. For a given value of $\Gh$, we parameterize this line as $y \mapsto \lh_{\rm sing}(y)$. The result for $\Gh=0.5$ is shown in FIG. \ref{Aeins}.
\begin{figure}[htb]
\epsfxsize=0.60\textwidth
\renewcommand{\baselinestretch}{1}
\begin{center}
\leavevmode
\epsffile{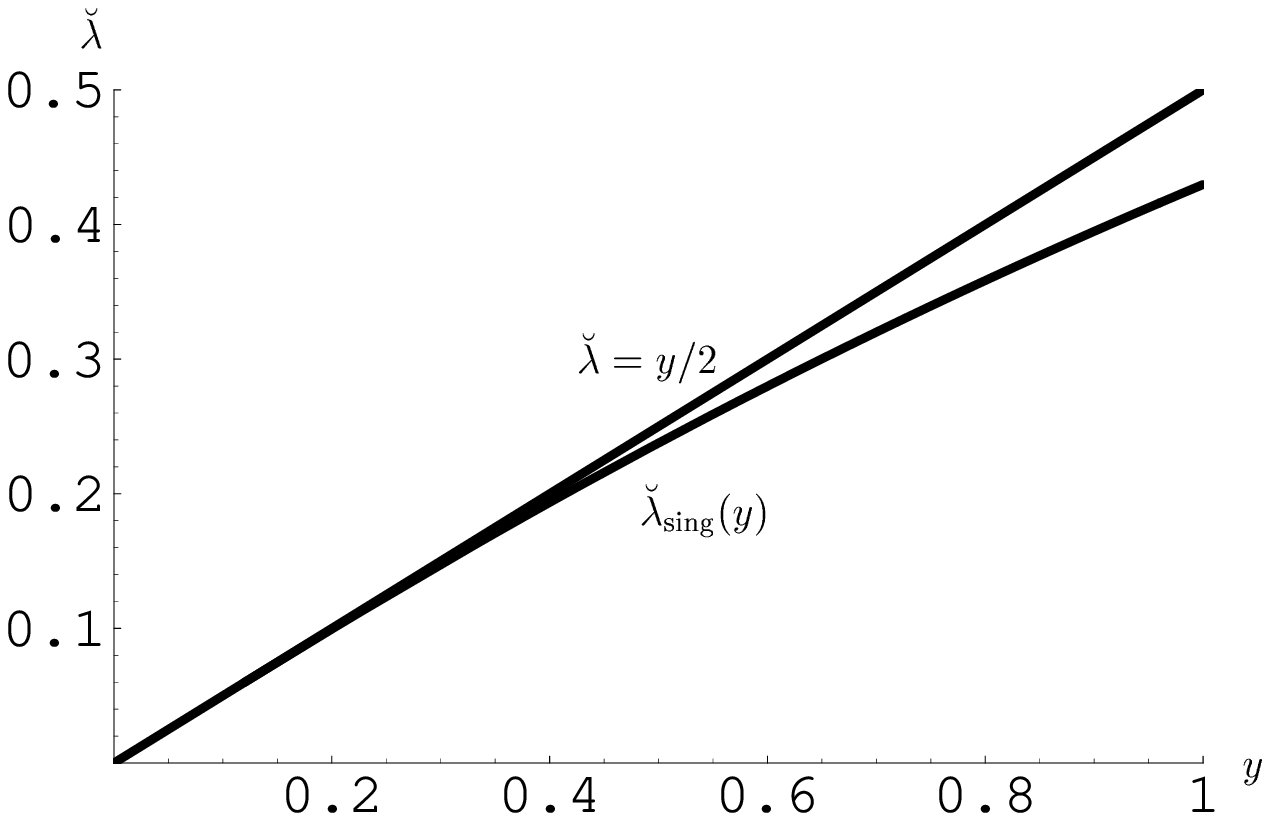}
\end{center}
\parbox[c]{\textwidth}{\caption{\label{Aeins}{\footnotesize Line of $\eta_N^{\rm Exp}$-singularities for fixed $\Gh = 0.5$. This line separates regions with $\eta_N^{\rm Exp} > 0$ above it and with $\eta_N^{\rm Exp} < 0$ below it.}}}
\end{figure}
One finds that the line $\lh_{\rm sing}(y)$ is located below the boundary $\lh = y/2$, approaching it as $y \rightarrow 0$. This line separates the $\lh$-$y-$parameter space (for $\Gh$ fixed) into a region below $(\lh < \lh_{\rm sing}(y))$ and above $(\lh_{\rm sing}(y) < \lh < y/2)$ it. In these regions one finds $\eta_N^{\rm Exp} < 0$ and $\eta_N^{\rm Exp}>0$, respectively. From a purely mathematical point of view, one can pose admissible initial conditions for the RG equations in both regions.
\begin{figure}[htb]
\epsfxsize=0.48\textwidth
\renewcommand{\baselinestretch}{1}
\begin{center}
\leavevmode
\epsffile{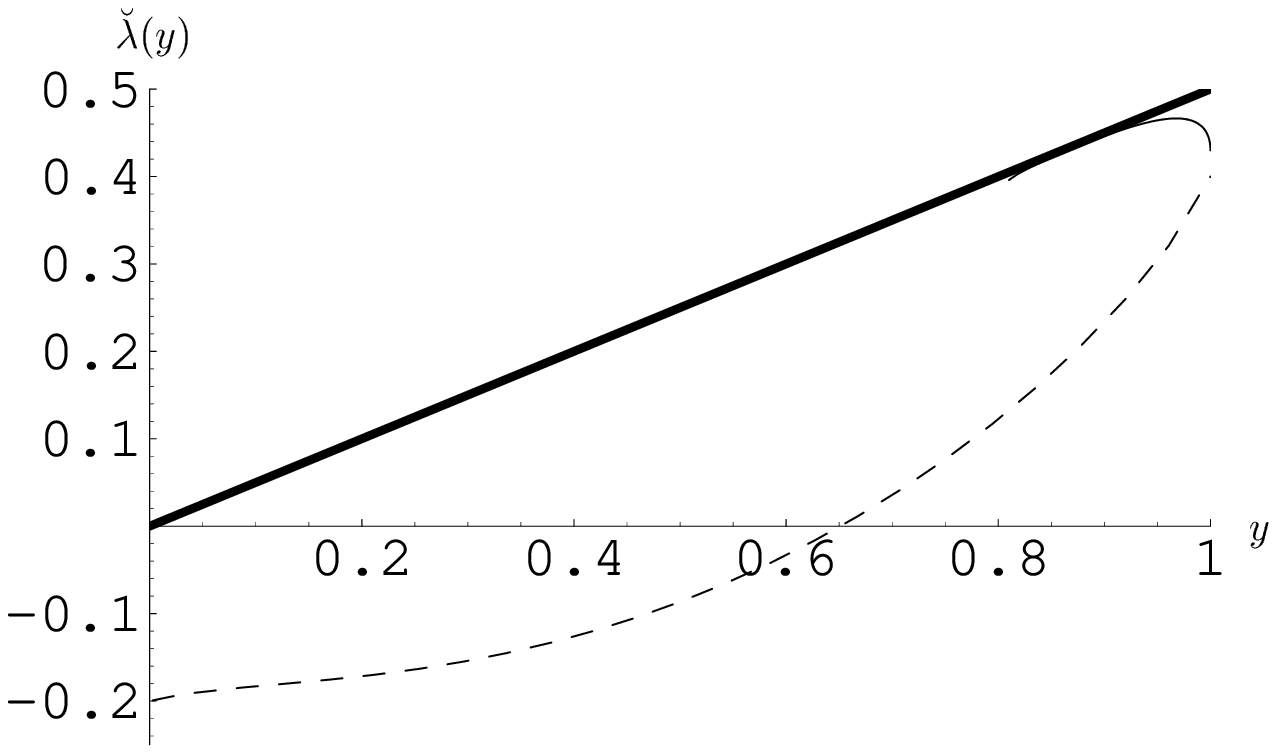}
\epsfxsize=0.48\textwidth
\epsffile{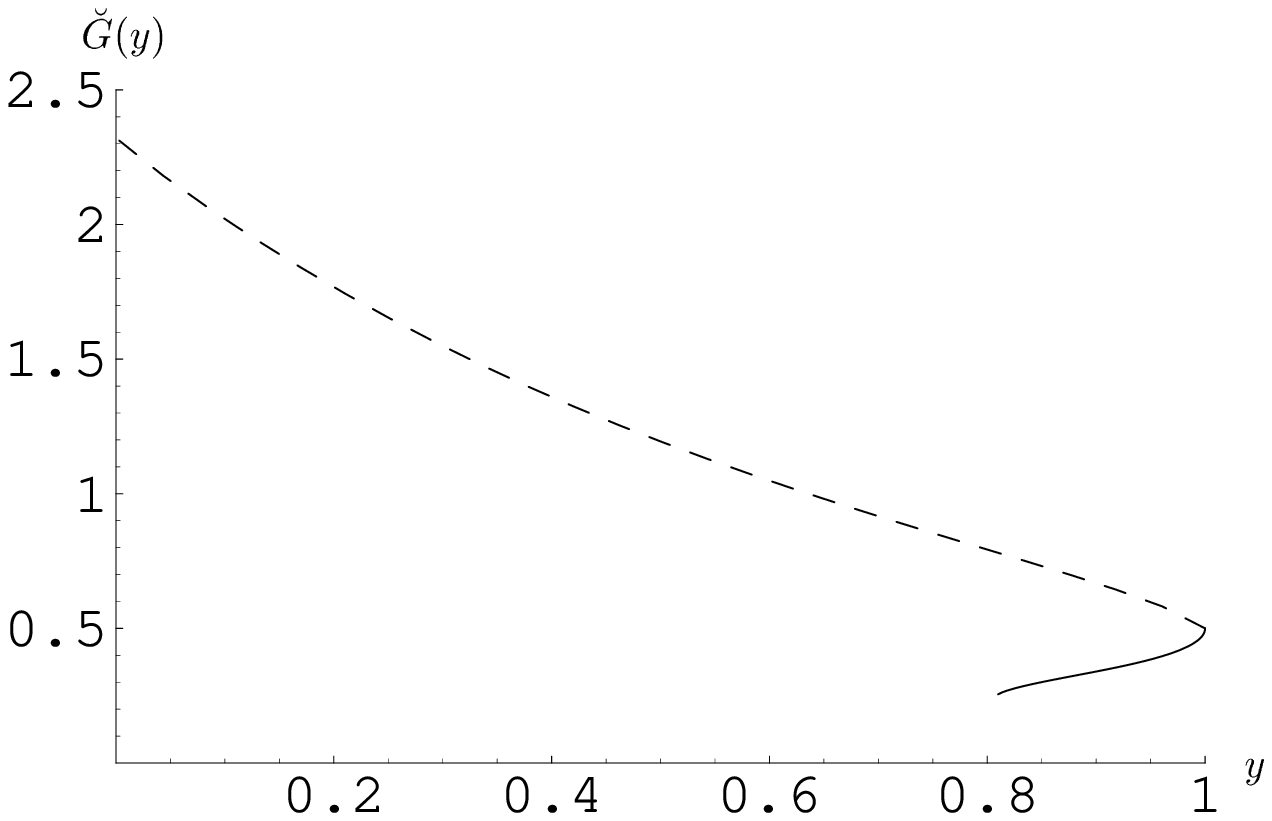}
\end{center}
\parbox[c]{\textwidth}{\caption{\label{Azwei}{\footnotesize Typical trajectories starting in the regions $\lh < \lh_{\rm sing}(\hat{y})$ (dashed line) and $\lh_{\rm sing}(\hat{y}) < \lh < \hat{y}/2$ (solid line). While the trajectory below $\lh_{\rm sing}(\hat{y})$ shows the typical behavior discussed in Section IV, the solution above $\lh_{\rm sing}(\hat{y})$ yields a screening behavior of the Newton constant.}}}
\end{figure}

Two typical trajectories starting below and above $\lh_{\rm sing}(\hat{y})$ are shown in FIG. \ref{Azwei}. The most probably unphysical trajectories starting in the narrow region $\lh_{\rm sing}(\hat{y}) < \lh(\hat{y}) < \hat{y}/2$ terminate at the boundary $\lh = y/2$. They would lead to a screening behavior of the Newton constant, i.e. $\Gh(y)$ increases with increasing $y = k^2/M^2$.

The trajectories starting with $\lh(\hat{y}) < \lh_{\rm sing}(\hat{y})$, i.e. in the region where $\eta_N^{\rm Exp} < 0$, lead to the RG trajectories discussed in Section IV.C which are physically relevant and show the expected antiscreening. According to their initial values, these trajectories are solutions of Type Ia, IIa or IIIa. 

Here it is interesting to note that trajectories starting close to the singularity of $\eta_N$, i.e. $\lh(\hat{y}) \lesssim \lh_{\rm sing}(\hat{y})$ can be continued down to ``unnaturally low'' values $y$. The corresponding trajectories cross the ones starting at lower values $\lh(\hat{y})$ in the $\lh$-$y-$plane before they terminate at the boundary. An example is the trajectory with the largest $\lh(\hat{y})$ in the first plot of FIG. \ref{vier}. Since all trajectories of this type terminate, this mechanism cannot be used to change the Type IIIa character of a trajectory.

We want to point out that the region above the singular line at $\lh_{\rm sing}(y)$ is probably unphysical. This assumption is supported by the fact that our investigation with the sharp cutoff did not yield any comparable behavior in the positive coupling region. The trajectories calculated with the smooth and the sharp cutoff, respectively, agree quite well for $y > y_{\rm term}$. Only close to their endpoint qualitative differences appear, such as the earlier termination at $\lh_{\rm sing}(y)$ rather than $\lh = y/2$. This scheme dependence is a typical symptom showing that the truncation becomes unreliable close to the boundary of parameter space.
\end{section}
\end{appendix}

\end{document}